\documentclass[useAMS,usenatbib]{mn2e}
\usepackage{epsfig}
\usepackage{mathtools}

\newcommand{\ci}{[C\,{\sc i}]}
\newcommand{\cii}{[C\,{\sc ii}]}

\title{Modelling \ci~emission from turbulent molecular clouds}
\author[]
{Simon C. O. Glover$^{1}$\thanks{E-mail: glover@uni-heidelberg.de},  Paul C. Clark$^{1,2}$, Milica Micic$^{3}$, \& Faviola Molina$^{1,4}$ \\  
\\ $^{1}$ Universit\"at Heidelberg, Zentrum f\"ur Astronomie, Institut f\"ur Theoretische Astrophysik, \\ \hspace{.1in} Albert-Ueberle-Strasse 2,  69120 Heidelberg, Germany \\
$^{2}$ School of Physics and Astronomy, Cardiff University, Cardiff CF24 3AA, UK \\
$^{3}$ Astronomical Observatory, Volgina 7, 11060 Belgrade, Republic of Serbia \\
$^{4}$ University of Chile, Faculty of Physical and Mathematical Sciences, \\
\hspace{.1in} Department of Computer Science. Beauchef 851, Santiago, Chile \\ 
}

\begin{document}
\maketitle

\begin{abstract}
We use detailed numerical simulations of a turbulent molecular cloud to study the
usefulness of the \ci~609~$\mu$m and 370~$\mu$m fine structure emission 
lines as tracers of cloud structure. Emission from these lines is observed throughout
molecular clouds, and yet they have attracted relatively little theoretical attention.
We show that the widespread \ci~emission results from the fact that the clouds are
turbulent. Turbulence creates large density inhomogeneities, allowing radiation to penetrate 
deeply into the clouds. As a result, \ci~emitting gas is found throughout the cloud.
We examine how well \ci~emission traces the cloud structure, and show that the 
609~$\mu$m line traces column density accurately over a wide range of values.
For visual extinctions greater than a few, \ci~and $^{13}$CO both perform well, 
but \ci~performs better at $A_{\rm V} \leq 3$.
We have also studied the distribution of \ci~excitation temperatures. We show that these 
are typically smaller than the kinetic temperature, indicating that 
the carbon is subthermally excited. We discuss how best to estimate the excitation 
temperature and the carbon column density, and show that the latter
tends to be systematically underestimated. Consequently, estimates of the atomic 
carbon content of real GMCs could be wrong by up to a factor of two.
\end{abstract}

\begin{keywords}
galaxies: ISM -- ISM: clouds -- ISM: molecules --  stars: formation
\end{keywords}

\section{Introduction}
Giant molecular clouds (GMCs) are the sites where the majority of Galactic star formation occurs, and hence it
is important to understand their properties if we are to understand how star formation proceeds within our Galaxy.
The two most abundant chemical species in molecular clouds are molecular hydrogen (H$_{2}$) and atomic 
helium, and between them they are responsible for the vast majority of the mass of the clouds. However, neither 
of these species can easily be observed within molecular clouds, as the gas temperature is too low to
excite their internal energy levels. Observational studies of GMCs are therefore forced to focus on less abundant
tracers that can be observed at typical GMC temperatures. The most popular such tracer is carbon monoxide 
(CO).\footnote{Unless otherwise stated, when we refer to CO in this paper, we mean $^{12}$CO,  the most abundant  isotopologue.}

Unfortunately, CO is not without its problems as a molecular gas tracer. At high column densities, the
problem is one of opacity: the CO rotational transitions start to become optically thick, breaking the
relationship between CO emission and column density \citep[see e.g.][]{pin08}. Fortunately, this problem
can be avoided to a large extent by observing less common isotopic variants of CO, such as $^{13}$CO or 
C$^{18}$O, which remain optically thin up to much higher column densities. At low column densities,
on the other hand, the problem is one of abundance: in low extinction regions, CO is strongly photodissociated
by the interstellar radiation field (ISRF), making it very difficult to observe any CO emission from these
regions \citep[see e.g.][]{vdb88,gold08}, let alone to learn anything about the thermal state or kinematics of this gas.

It is therefore worthwhile examining other possible tracers of the gas within molecular clouds. One particularly 
interesting possibility is neutral atomic carbon. Spin-orbit coupling splits the atomic carbon ground state into three
fine structure levels with total angular momenta $J = 0, 1, 2$, and forbidden transitions between these three
levels give rise to two prominent fine structures lines: the \ci~$^{3}P_{1} \rightarrow \mbox{}^{3}P_{0}$ transition
at 609~$\mu$m (hereafter written simply as the $1 \rightarrow 0$ transition)
and the \ci~$^{3}P_{2} \rightarrow \mbox{}^{3}P_{1}$ transition at 370~$\mu$m (hereafter the $2 \rightarrow 1$
transition). The energy
difference between the $^{3}P_{1}$ and $^{3}P_{0}$ levels is only $\Delta E_{10} / k \simeq 24$~K, 
while between the $^{3}P_{2}$ and $^{3}P_{0}$ levels it is $\Delta E_{20} / k \simeq 60$~K. Therefore,
at typical molecular cloud temperatures of 10--20~K, both of the fine structure lines can be excited. 

So far, however, neutral atomic carbon has attracted much less attention than CO as a molecular cloud
tracer. There are several reasons for this. Observationally, the \ci~lines are harder to study than
the low $J$ rotational transitions of CO: they typically have lower brightness temperatures, and are also
situated in a wavelength regime where the effects of atmospheric absorption can be highly significant.
Theoretically, it was originally thought that neutral atomic carbon would be abundant only
at the edges of clouds \citep{langer76}, limiting its usefulness as a tracer of the bulk of the cloud.
Subsequent observations have shown that this expectation is incorrect, and that \ci~609~$\mu$m
emission is widespread within clouds \citep{frerking89,little94,sch95,kra08}, and yet the
perception of \ci~as a tracer of cloud surfaces has proved difficult to shift.

In the past few years, however, there has been renewed interest in the prospects of \ci~as a 
molecular gas tracer. The CHAMP+ and FLASH instruments on the APEX 
telescope\footnote{http://www.apex-telescope.org/} have allowed both \ci~lines to be studied with
reasonable sensitivity from one of the best sites on the planet, and ALMA\footnote{http://www.almaobservatory.org/} 
will allow these lines to be studied with much greater sensitivity and angular resolution once Bands 8 and 10 
are commissioned. Moreover, in the longer term, CCAT\footnote{http://www.ccatobservatory.org} will be able to 
rapidly map \ci~emission over large areas of the sky with good sensitivity. The time is therefore ripe for an in-depth 
look at what \ci~emission can tell us about molecular clouds,  using state-of-the-art numerical simulations.

In this paper, we take the first steps towards this goal. We present results from a simulation of the
chemical, thermal and dynamical evolution of a $10^{4} \: {\rm M_{\odot}}$ turbulent molecular cloud
illuminated by the standard interstellar radiation field, 
and examine in detail how the neutral atomic carbon is distributed in this cloud. We also make synthetic
emission maps of the resulting \ci~lines, and investigate how much we can learn about the cloud by looking
at this emission. Although our focus in this paper is the detailed analysis of a single simulation, we plan to
follow this up in future work with a much broader parameter study.

The plan of the paper is as follows. In Section~\ref{method}, we present the numerical approach used to
simulate the cloud and generate the synthetic emission maps. We also outline the initial conditions used
for our simulation. Our main results are presented in Section~\ref{res}, where we examine the chemical
and thermal state of the gas in the cloud, and investigate what we can learn about the properties of the
cloud from the \ci~emission. Finally, in Section~\ref{conc} we present our conclusions.
  
\section{Method}
\label{method}
\subsection{Hydrodynamical model}
\label{hydro}
The chemical and thermal evolution of the gas in our model cloud is simulated using a modified
version of the Gadget 2 SPH code \citep{springel05}. Our modifications include a simplified treatment
of the gas chemistry, based on the work of \citet{gm07a,gm07b} and \citet{nl99}, and described in
more detail in \citet{gc12a}, a detailed atomic and 
molecular cooling function \citep{glo10}, 
and a treatment of the attenuation of the interstellar radiation field within the cloud using the {\sc TreeCol} algorithm
\citep{cgk12}. The same code has previously been used to model star formation in molecular clouds
with a variety of metallicities \citep{gc12b}, the formation of GMCs from converging flows of atomic 
gas \citep{clark12}, and the thermal state of the dense Galactic Centre molecular cloud known as the 
``Brick'' \citep{clark13}. More extensive discussions of the numerical details of the code can be found
in these references. Here, in the interests of brevity, we focus only on a small number of changes that we
have made to the code compared to the version described in these previous works. 

The most important change that we have made involves our treatment of CO photodissociation. 
Previously, this was based on the work of \citet{lee96}, but we have now updated our treatment
to follow the recent paper by \citet{visser09}. Comparison of runs performed with the old and new
treatments of CO photodissociation demonstrates that the main difference in behaviour is a 
sharpening of the transition between regions dominated by neutral atomic carbon and regions
dominated by CO, owing to the steeper dependence of the CO photodissociation rate on the visual
extinction, $A_{\rm V}$, found in the newer work.

Another important change that we have made to the chemical model is the inclusion of the 
effects of  cosmic-ray ionization of atomic carbon
\begin{equation}
{\rm C} \xrightarrow{{\rm c.r.}} {\rm C^{+}} + {\rm e^{-}}, \label{react1}
\end{equation}
and cosmic-ray induced photodissociation of C and CO
%\begin{eqnarray}
%{\rm C} + \gamma_{\rm cr} & \rightarrow &  {\rm C^{+}} + {\rm e^{-}},  \label{react2} \\
%{\rm CO} + \gamma_{\rm cr} & \rightarrow &  {\rm C + O}. \label{react3}
%\end{eqnarray}
\begin{equation}
{\rm C} + \gamma_{\rm cr}  \rightarrow  {\rm C^{+}} + {\rm e^{-}},  \label{react2} 
\end{equation}
\begin{equation}
{\rm CO} + \gamma_{\rm cr}  \rightarrow  {\rm C + O}. \label{react3}
\end{equation}
These processes were not included in the version of the chemical model described in
\citet{gc12a}. The rates that we have adopted for processes~\ref{react1} and \ref{react2} 
are based on the values given in \citet{umist07}, but have been uniformly rescaled in
order to be consistent with our adopted cosmic ray ionization rate. The rate for 
process~\ref{react3} comes from \citet{gld87}, and has been similarly rescaled.

Finally, we have also added a simple treatment of the chemistry of $^{13}$CO, using
an approach similar to that in \citet{laszlo14}. Specifically, we have extended the
\citet{nl99} chemical network to also follow the abundances of $^{13}$C$^{+}$, $^{13}$C,
$^{13}$CO and H$^{13}$CO$^{+}$. For simplicity, we assume that most of the 
reactions involving these species are not isotope-selective and hence adopt the same
reaction rate coefficients for them as for their counterparts that involve $^{12}$C.
However, we also include the reaction
\begin{equation}
{\rm ^{12}CO + \mbox{}^{13}C^{+} \rightarrow \mbox{}^{13}CO + \mbox{}^{12}C^{+}}
\end{equation}
together with the reverse reaction. The forward reaction is exothermic, and has a reaction
rate coefficient $k_{\rm frac} = 2 \times 10^{-10} \: {\rm cm^{3} \: s^{-1}}$, but
the reverse reaction is endothermic and has a rate coefficient $k_{\rm frac, inv} = 
k_{\rm frac} \exp(-35/T)$ \citep{WatsonEtAl1976}. In cold gas, this imbalance between the rates
of the forward and reverse reactions leads to chemical fractionation, increasing the
$^{13}$CO-to-$^{12}$CO ratio significantly above the elemental $^{13}$C-to-$^{12}$C
ratio. In addition, we also account for the fact that in low extinction gas, it is easier
to photodissociate $^{13}$CO than it is to photodissociate $^{12}$CO, owing to the
greater effectiveness of CO self-shielding in the latter case \citep[see e.g.][]{visser09}.
This effect counteracts the effects of fractionation to some extent, but in conditions typical of
local molecular clouds or photodissocation regions, fractionation generally dominates
\citep{roe13,laszlo14}.

\subsection{Post-processing}
\label{post}
We post-process our simulation data to produce synthetic emission maps in
the $^{3}P_{1} \rightarrow \mbox{}^{3}P_{0}$ and $^{3}P_{2} \rightarrow \mbox{}^{3}P_{1}$
fine structure lines of \ci~and the $J = 1 \rightarrow 0$ line of $^{12}$CO and $^{13}$CO
using the {\sc radmc-3d} radiation transfer 
code\footnote{http://www.ita.uni-heidelberg.de/$\sim$dullemond/software/radmc-3d/},
and making use of atomic and molecular data (transition energies, radiative transition rates,
collisional excitation rate coefficients, etc.) from the LAMDA database\footnote{http://home.strw.leidenuniv.nl/$\sim$moldata/}
\citep{sch05}.
We calculate level populations using the large velocity gradient (LVG) approximation, as
described in \citet{shetty11a,shetty11b}. {\sc radmc-3d} cannot presently deal
with SPH data directly, and so before we can use the code to generate our 
synthetic emission maps, we must first interpolate all of the necessary data to a 
Cartesian grid. By default, we use a cubic grid with a side length of 
$5 \times 10^{19} \: {\rm cm}$ (approximately 16~pc) and a grid resolution of 
$256^{3}$ zones. In Appendix~\ref{radmc-resolution}, we demonstrate that this grid 
resolution is sufficient to allow us to fully resolve the \ci~emission produced by our 
model clouds.

We use {\sc radmc-3d} to produce full position-position-velocity (PPV) cubes, with
a velocity channel width of $\Delta v = 0.094 \: {\rm km \: s^{-1}}$, and then produce
maps of velocity-integrated intensity based on these PPV cubes. Our analysis in
this paper focuses on these integrated intensity maps, and we defer any discussion
of the velocity structure of the clouds, as traced by the \ci~emission, to a future
paper. We account for unresolved small-scale velocity structure in the clouds by
including a microturbulent contribution of $0.2 \: {\rm km \: s^{-1}}$ in the emission
and absorption line-widths computed for each grid zone. In practice, this choice
has little influence on our integrated intensity maps. Finally, we note that for 
consistency with \citet{shetty11a,shetty11b} and with observational work in this
field, we express the line intensities in terms of the brightness temperature, 
$T_{\rm B}$, computed using the Rayleigh-Jeans approximation
\begin{equation}
T_{\rm B}(\nu) = \left(\frac{c}{\nu} \right)^{2} \frac{I_{\nu}}{2k_{\rm b}},
\end{equation}
where $I_{\nu}$ is the specific intensity and $k_{\rm b}$ is the Boltzmann constant.
Our velocity-integrated intensity maps therefore have units of ${\rm K \: km \: s^{-1}}$.

%When producing maps of \ci~or $^{12}$CO emission, we can use the C and
%CO abundances directly produced by our SPH simulations. However, the chemical
%model that we use does not track  $^{13}$CO directly, and so to compute our synthetic
%map of the $^{13}$CO emission, we have assumed a fixed ratio $n_{\rm 13CO} / n_{\rm 12CO} = 1 / 60$
%for the number densities of the two isotopologues. In reality, in much of the gas this ratio will vary owing 
%to the competing effects of chemical fractionation \citep{WatsonEtAl1976}, which tends to increase 
%$n_{\rm 13CO} / n_{\rm 12CO}$, and the selective photodissociation of $^{13}$CO 
%\citep[see e.g.][]{visser09}, which works to decrease $n_{\rm 13CO} / n_{\rm 12CO}$. We have 
%examined both of these effects in a separate study \citep{laszlo14} and find that 
%along lines of sight where there is detectable $^{13}$CO emission, the effects of fractionation
%generally dominate. However, the resulting $^{13}$CO emission maps do not differ greatly
%from those that we would obtain simply by assuming a fixed ratio of $^{13}$CO to $^{12}$CO,
%and so for our present purposes,  this approximation is adequate.

\subsection{Initial conditions}
In this paper, we focus on the detailed analysis of the \ci~emission produced 
by a single model cloud. The model cloud we chose to study is very similar to
one of the clouds studied in \citet{gc12b,gc12c}. We adopted a cloud mass of
$10^{4} \: {\rm M_{\odot}}$, which we simulated using two million SPH particles,
giving us a mass resolution of $0.5 \: {\rm M_{\odot}}$.\footnote{We examine the
effects of increasing the numerical resolution in Appendix~\ref{hydro_res}.} The cloud was
initially spherical, with a radius of 6.3~pc, giving it an initial mean hydrogen
nuclei number density of $n_{0} = 276 \: {\rm cm^{-3}}$. The initial gas
temperature was set to $20$~K and the initial dust temperature was set
to 10~K, but as the cloud quickly relaxes to thermal equilibrium following
the start of the simulation, our results are not sensitive to these particular
choices.

We imposed a turbulent velocity field on the clouds, with a power spectrum
$P(k) \propto k^{-4}$, where $k$ is the wavenumber. The RMS velocity of
the turbulence was $v_{\rm rms} = 2.8 \: {\rm km \: s^{-1}}$, chosen so that
the kinetic energy and gravitational energy of the cloud were initially equal.
The turbulence was not driven during the simulation, but was instead allowed 
to freely decay. 

We simulated the cloud using vacuum boundary conditions, but applied an 
external pressure term using the technique described in \citet{benz90} to 
minimize the effects of the pressure gradient that would otherwise exist at the 
edge of the cloud. We stress that the cloud is gravitationally bound and so 
would collapse and form stars even if this external pressure term were absent.

We took the metallicity of the cloud to be solar, with the standard local dust-to-gas ratio, 
and with total abundances of carbon-12, carbon-13 and oxygen relative to hydrogen given by
$x_{\rm ^{12}C} = 1.4 \times 10^{-4}$, $x_{\rm ^{13}C} = x_{\rm ^{12}C} / 60$
and $x_{\rm O} = 3.2 \times 10^{-4}$,  respectively \citep{sem00}. We also assumed that the 
cloud was initially atomic, with all of the carbon initially in the form of C$^{+}$. 

For the strength and spectral shape of the interstellar radiation field, we used the results
of \citet{dr78} in the ultraviolet and \citet{mmp83} at longer wavelengths. The cosmic
ray ionization rate for atomic hydrogen was set to $\zeta_{\rm H} = 10^{-17} \: {\rm s^{-1}}$,
and the corresponding ionization rates for the other chemical species were scaled 
accordingly, as described in Section~\ref{hydro} above. 

We halted the simulation as soon as the first gas reached a density above our maximum
density threshold  $n_{\rm th} \sim 10^{7} \: {\rm cm^{-3}}$. This density is too high to be 
reached purely by turbulent compression given our choice of RMS turbulent velocity, and
finding gas at this density therefore indicates that one or more cores have entered
runaway gravitational collapse and will shortly begin forming stars. We found that this
occurred at a time $t = 1.98 \: {\rm Myr}$ after the beginning of the simulation.

\section{Results}
\label{res}
In this section, we analyze the physical state of the cloud and the \ci~emission that
it produces. Our analysis is carried out at $t \simeq 2\: {\rm Myr}$, immediately
before the onset of star formation in the cloud. We start by briefly examining the
chemical and thermal state of the cloud (section~\ref{chemtherm}), before going
on to examine how much of the cloud is traced by \ci~emission (section~\ref{how}).
We next examine how well we can estimate the \ci~excitation temperature 
based on our synthetic emission maps (section~\ref{calc_tex}). Finally, we explore
the usefulness of \ci~emission as a column density tracer in section~\ref{column},
and compare it with another commonly-used column density tracer, $^{13}$CO,
in section~\ref{13co}.

\subsection{Chemical and thermal state of the cloud}
\label{chemtherm}
\begin{figure*}
\includegraphics[width=0.45\textwidth]{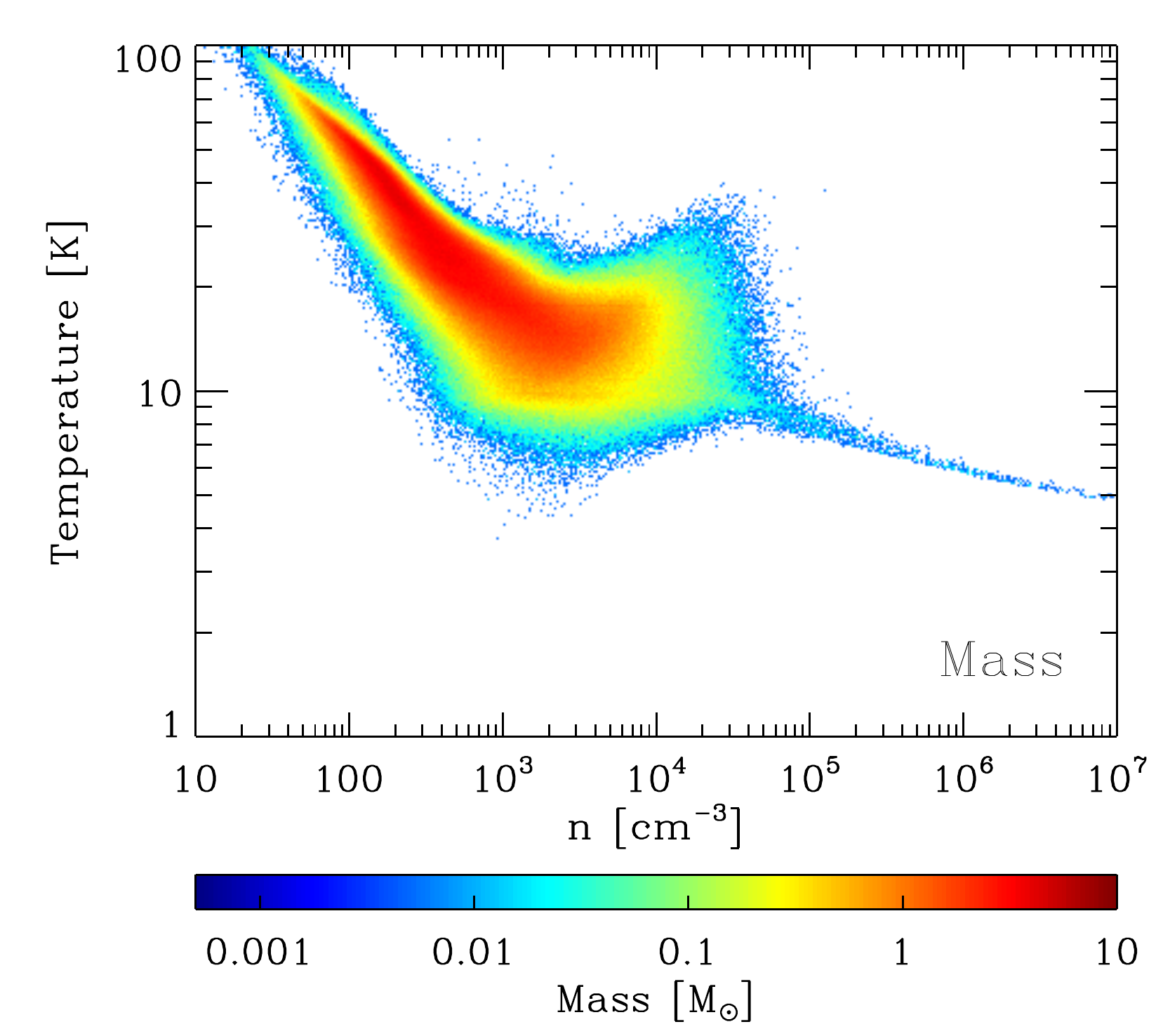}
\includegraphics[width=0.45\textwidth]{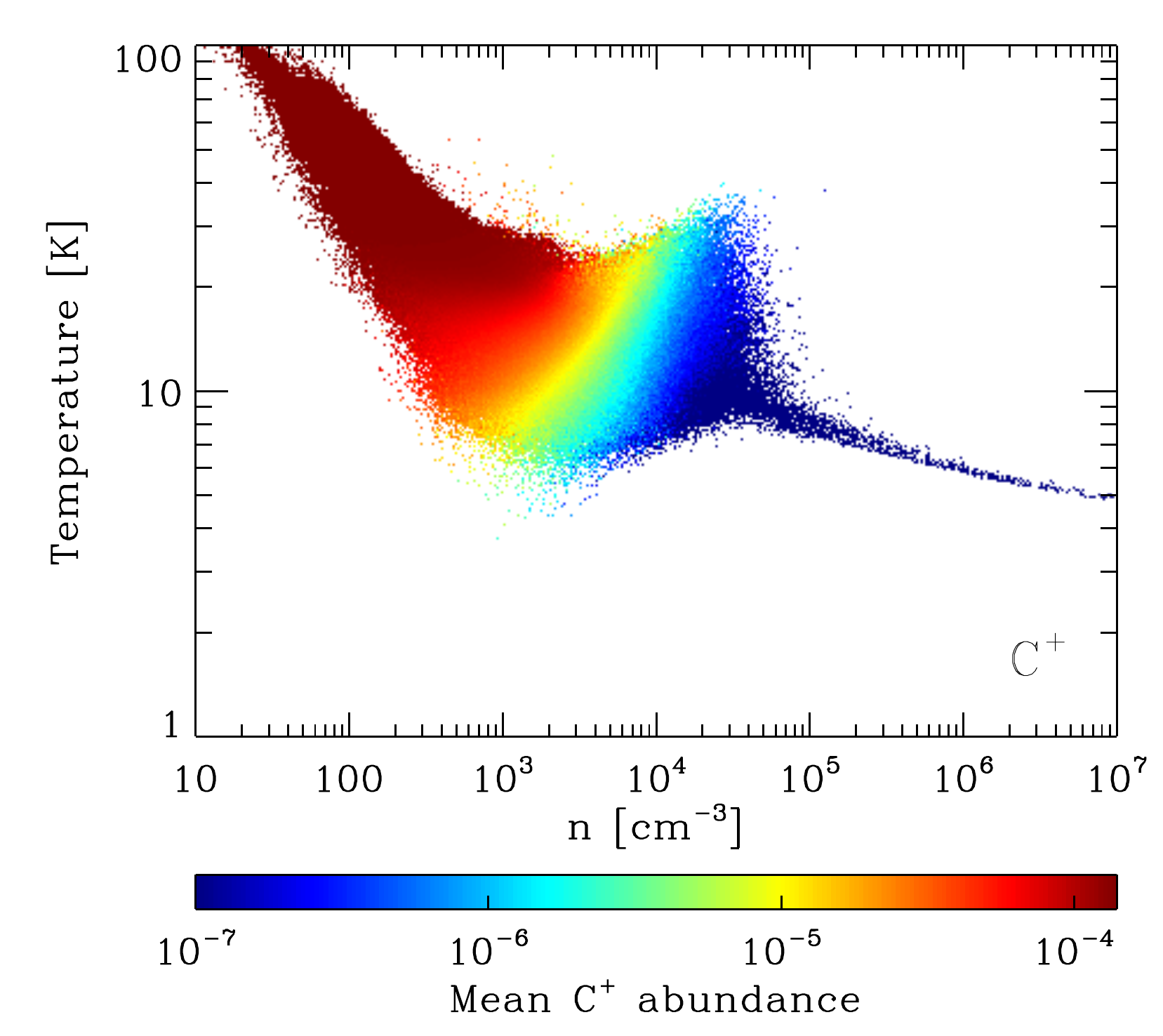}
\includegraphics[width=0.45\textwidth]{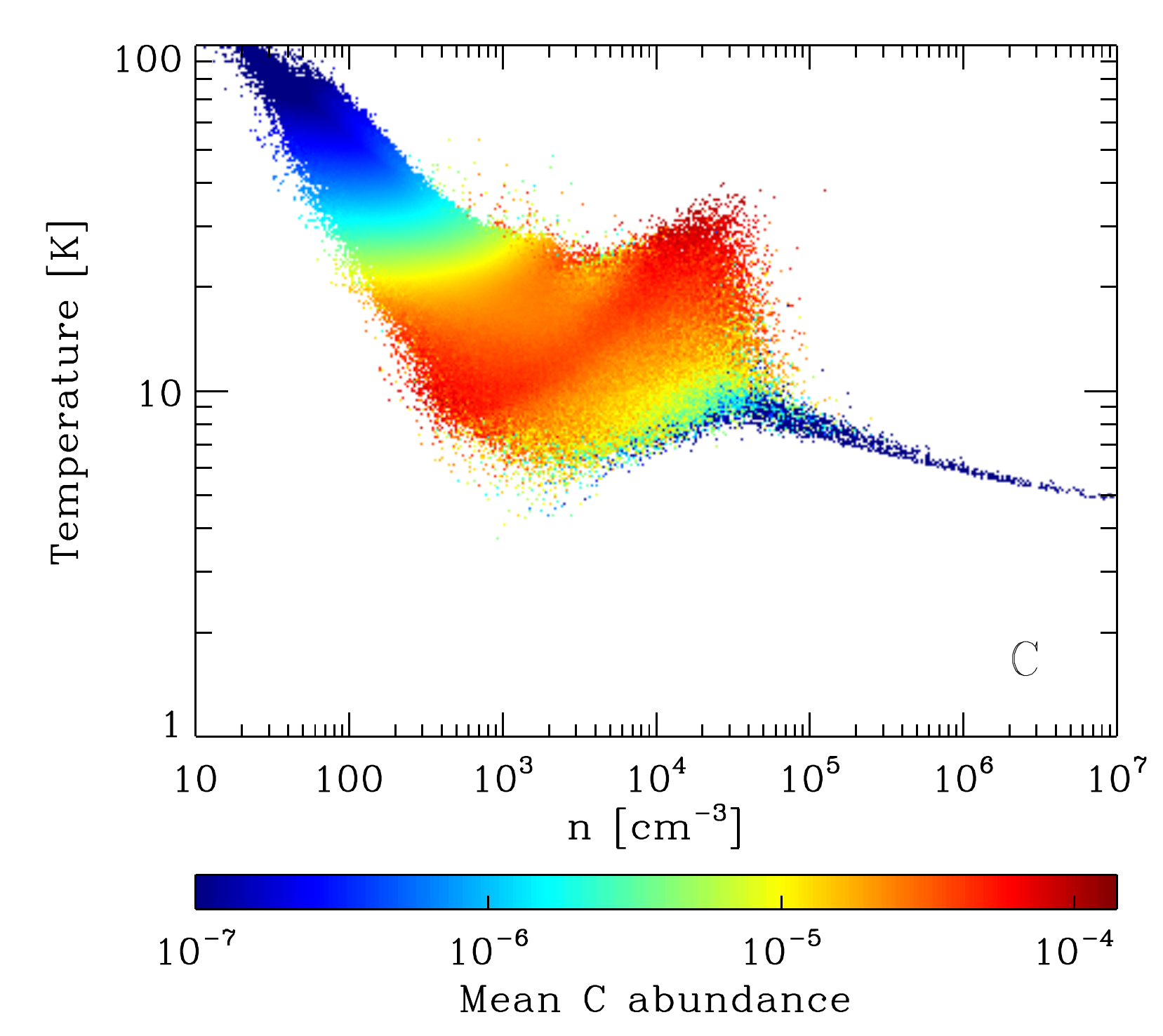}
\includegraphics[width=0.45\textwidth]{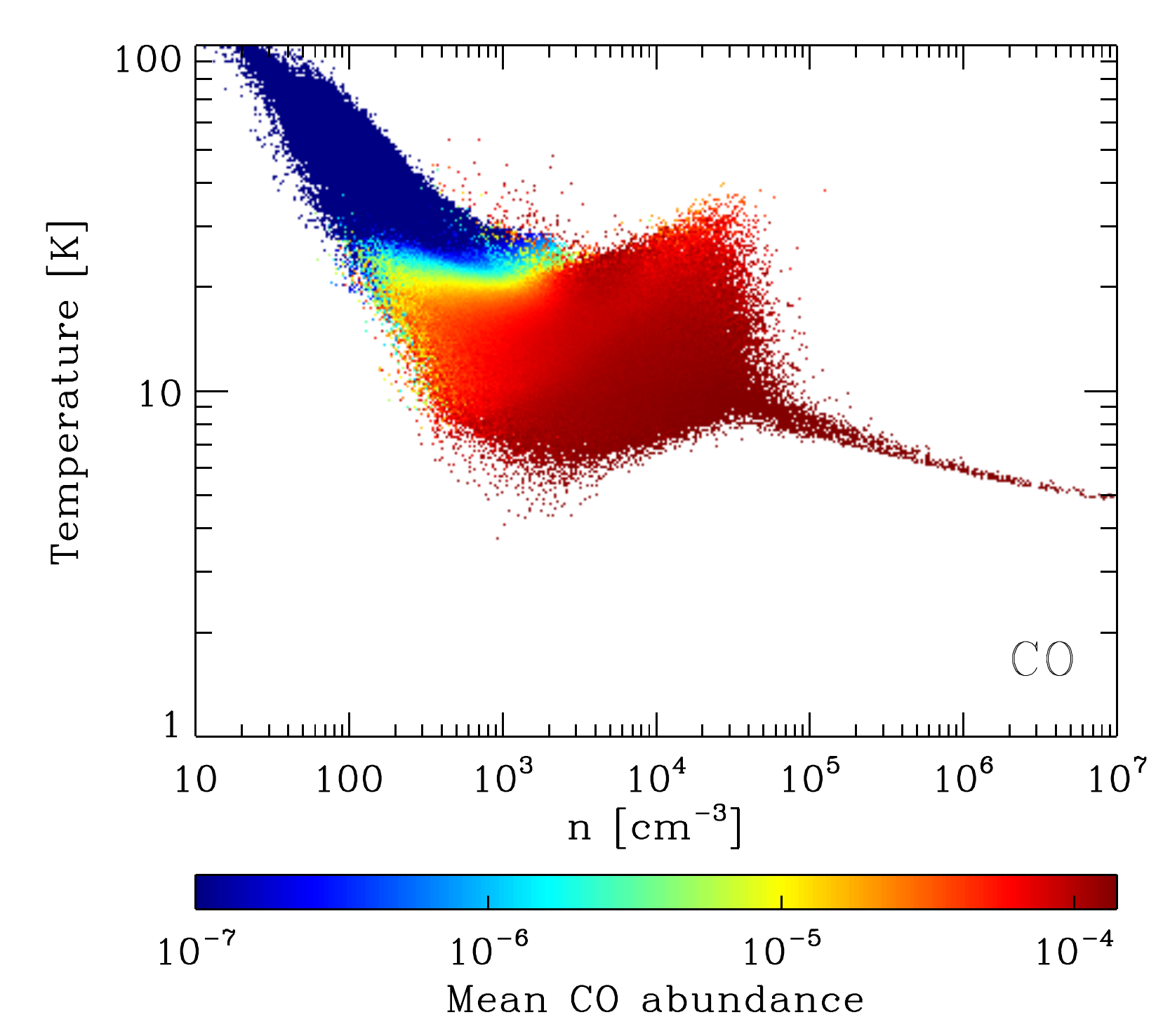}
\caption{Physical state of the cloud just before the onset of star formation. The top left panel
shows the two-dimensional probability density function (PDF) of gas temperatures and densities
within the cloud, weighted by the amount of mass present in each region of density-temperature
space. The remaining panels show the same plot of densities and temperatures, but in this case,
the shading indicates the mean fractional abundance of C$^{+}$, C  or 
CO present at each point.
\label{rhotemp}}
\end{figure*}

In the upper left panel of Figure~\ref{rhotemp}, we show the two-dimensional probability 
density function (PDF) of the gas temperature $T$ and the hydrogen nuclei number 
density $n$ in our model cloud at a point just before the onset of star formation.
Several distinct features are apparent. At low densities, there is a strong inverse
relationship between density and temperature. This corresponds to the part of the
cloud that is optically thin to the ultraviolet portion of the ISRF.
Cooling in this region is dominated by C$^{+}$ fine structure line emission,
while heating comes primarily from the photoelectric effect. As the density increases,
the balance between these two processes shifts in favour of cooling, leading to a drop
in the temperature, as noted in a number of previous studies \citep[see e.g.][]{larson05,
gm07b,gc12c}. 

\begin{figure*}
\includegraphics[width=0.45\textwidth]{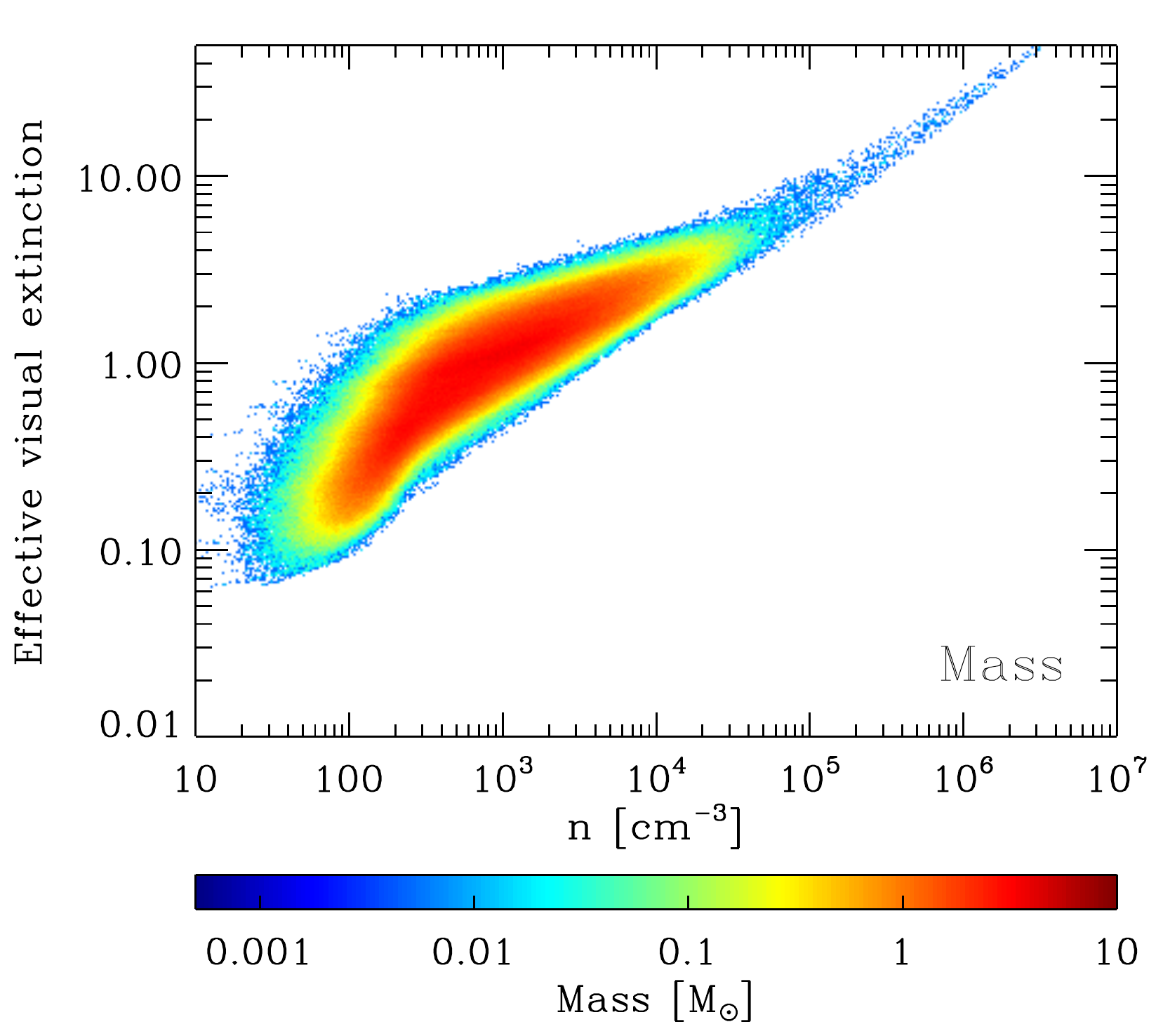}
\includegraphics[width=0.45\textwidth]{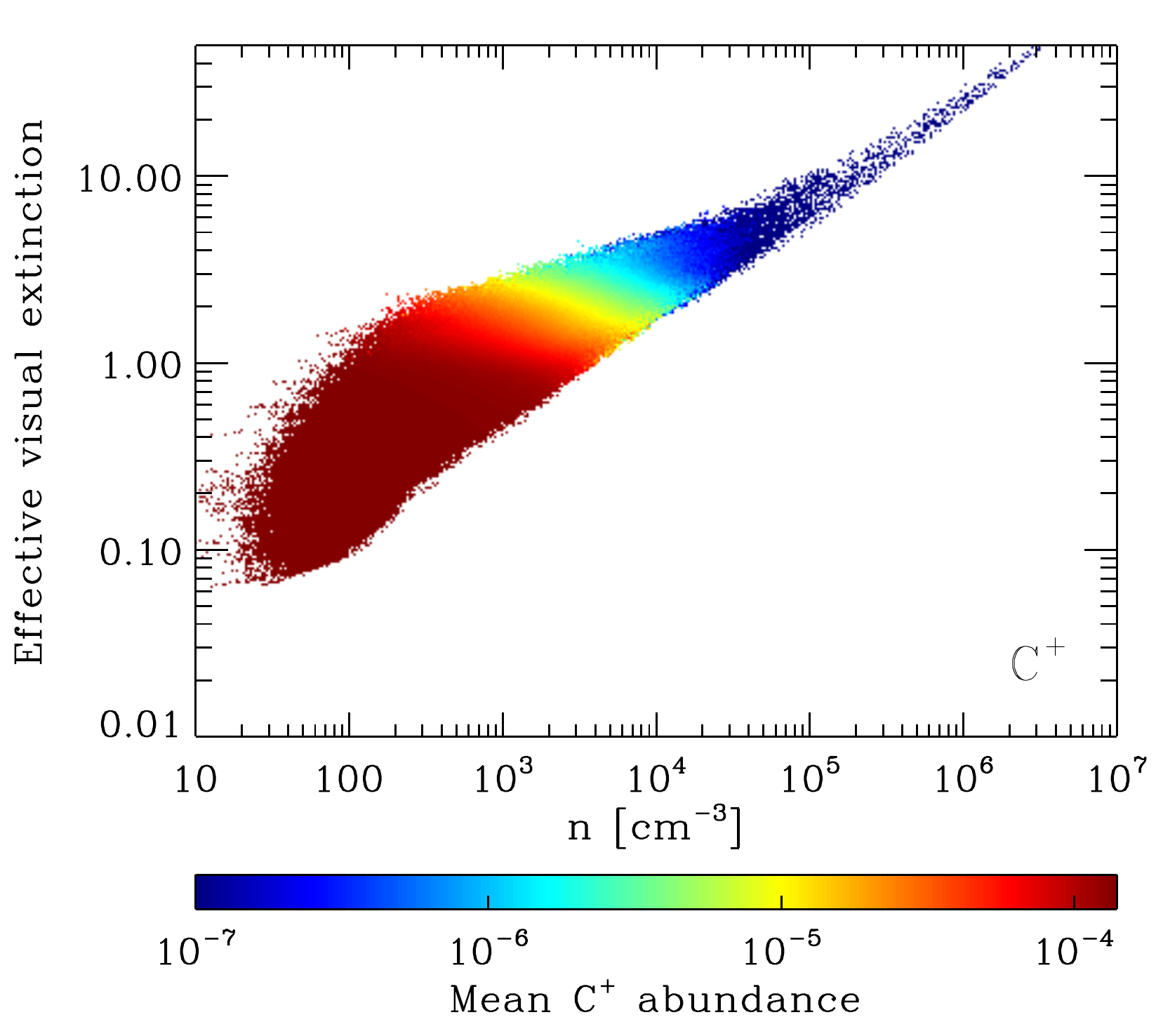}
\includegraphics[width=0.45\textwidth]{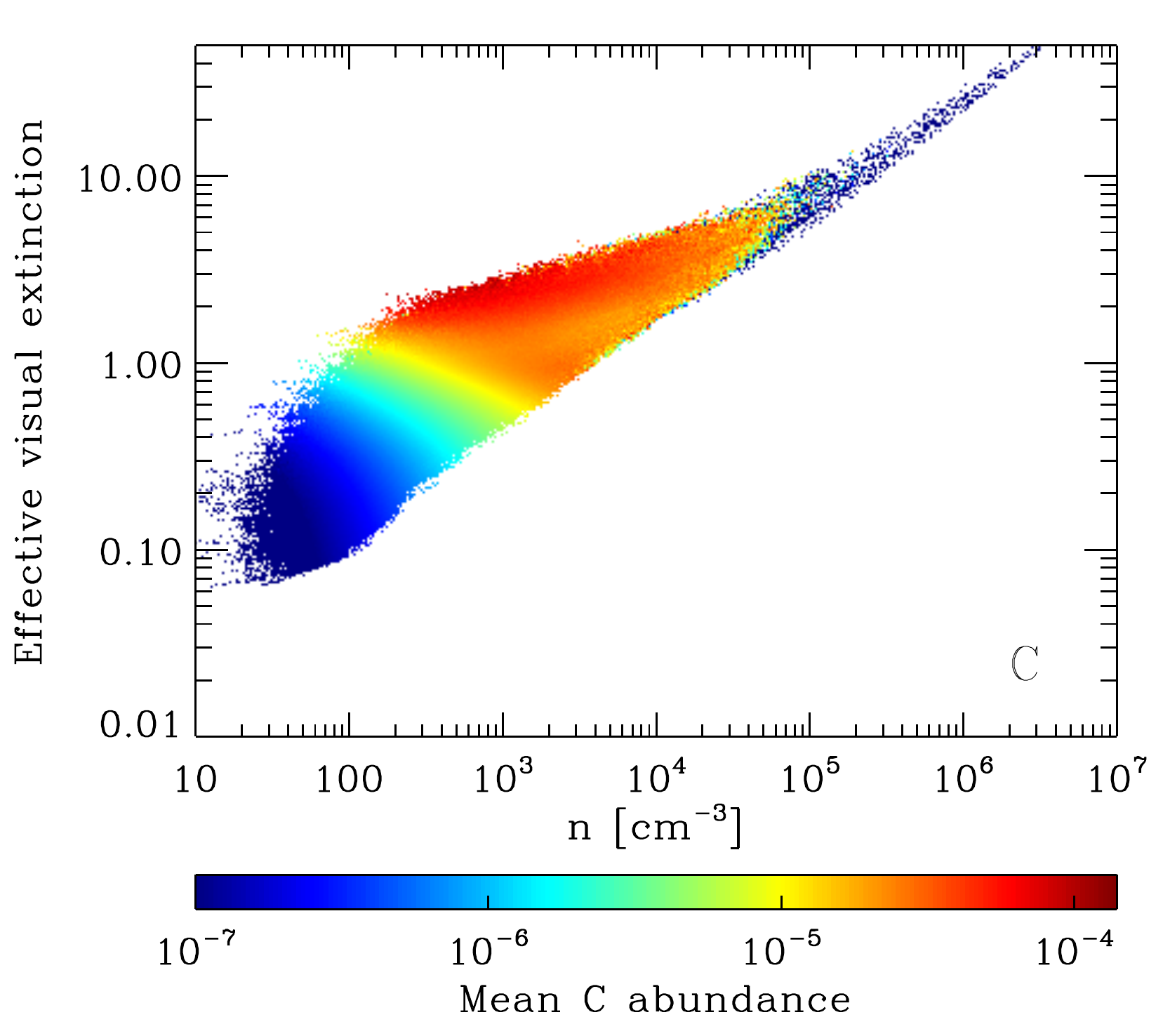}
\includegraphics[width=0.45\textwidth]{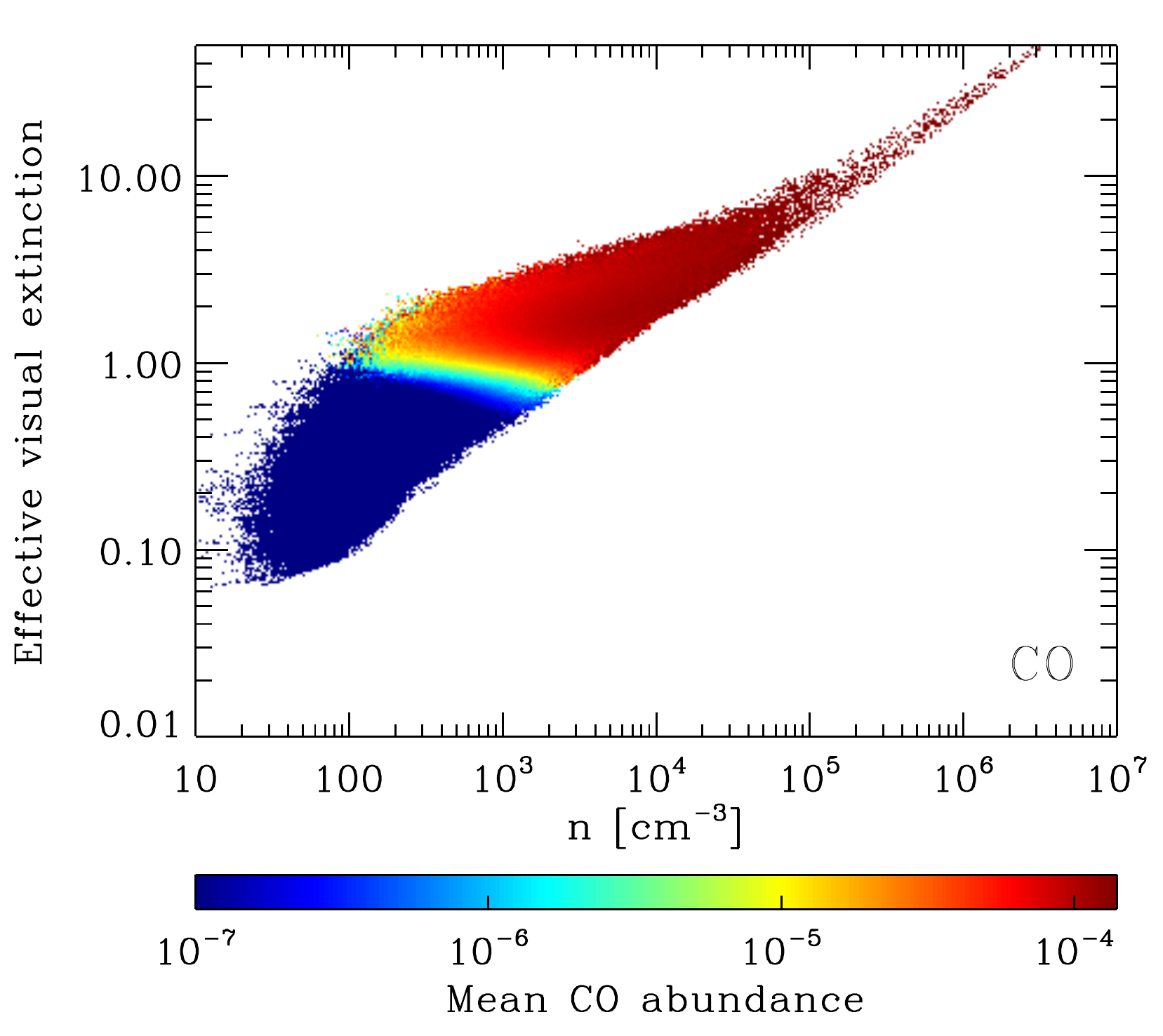}
\caption{As Figure~\ref{rhotemp}, but for the density and the effective visual extinction, $A_{\rm V, eff}$
(see Section~\ref{chemtherm}). \label{rhoAV}}
\end{figure*}

Above a number density $n \sim 1000 \: {\rm cm^{-3}}$, the cloud temperature
becomes largely independent of the density, although since there is considerable scatter
in the temperature in this regime, it is somewhat misleading to describe the gas as isothermal.
This scatter is related to the large scatter in the effective extinction seen by different 
regions of the gas that have similar densities (see Figure~\ref{rhoAV}), which leads to
substantial variations in the efficiency of photoelectric heating, as well as in the chemical
state of the gas. One notable feature in this regime is the distinct bump in the temperature 
distribution at $n \sim 2 \times 10^{4} \: {\rm cm^{-3}}$ and $T \sim 20$~K, caused by 
H$_{2}$ formation heating \citep{gc12c}. 

Finally, at $n > 10^{5} \: {\rm cm^{-3}}$, the scatter in the gas temperature largely disappears,
and the temperature once again begins to decrease with increasing density. This feature
corresponds to the point at which the gas and dust temperatures become strongly coupled,
owing to efficient thermal energy transfer between the gas and the dust. Above this density, the
gas temperature closely tracks the dust temperature. Since the dust temperature is unaffected by 
velocity gradients within the gas and varies only weakly with $A_{\rm V}$ owing to the steep
temperature dependence of the dust cooling rate ($\Lambda_{\rm dust} \propto T_{\rm dust}^{6}$),
most of the scatter in the temperature distribution disappears once we enter this regime.
 
If we now look at the remainder of the panels in Figure 1, which illustrate the chemical state
of the carbon in the gas as a function of temperature and density, we see that as we move to
higher densities, we move from a regime dominated by C$^{+}$ to one in which neutral atomic carbon
dominates, and then finally to a CO-dominated regime. Neutral atomic carbon is present at a variety of
different densities in the range $100 < n < 5 \times 10^{4} \: {\rm cm^{-3}}$, but becomes very rare at higher
densities, where almost all of the available carbon is locked up in CO.
 
In addition, it is clear that there is little neutral atomic carbon present in the warmer gas with
$T > 30$~K. This is easy to understand, as temperatures $T > 30$~K are found
only in regions where the extinction is low, so that photoelectric heating is effective, and
in these regions, photoionization of C to C$^{+}$ occurs rapidly. 
 
It is also interesting to look at how the chemical state of the gas varies as a function of
extinction. However, in this case we need to be careful about how we quantify the
extinction. A common choice is to work in terms of the line-of-sight value, which we
will denote in this paper simply as $A_{\rm V}$. Although this choice is simple, and
is the same as the definition that we would use observationally, it is less than ideal
when one is considering a turbulent cloud, as there is no guarantee that the extinction
in other directions has much to do with the extinction along our particular line-of-sight.
As an extreme example, consider the case where we happen to be looking along the
long axis of a filament: we would see a high extinction, even though the extinction in
directions perpendicular to the filament may be small.

A better way to quantify the amount of extinction seen by a particular SPH particle is by means
of a weighted sum over many different lines of sight from that particle to the edges of the 
cloud.\footnote{Note that by treating the extinction in this fashion, we are implicitly assuming
that the effects of scattering (as opposed to true absorption) are relatively unimportant. This
simplification is a matter of computational expedience, but is unlikely to have a large effect on
our results.} Fortunately, this is precisely the information that our {\sc TreeCol} algorithm provides. As explained
in more detail in Appendix~\ref{treecol_res} and in \citet{cgk12}, the {\sc TreeCol} algorithm provides us 
with a $4\pi$ steradian map of the column density distribution surrounding each of our SPH particles.
Each of these column density distributions is discretized using the {\sc Healpix} pixelation scheme,
so that every SPH particle has a column density map associated with it consisting of $N_{\rm pix}$ 
equal-area ``sky'' pixels. We can use these column density maps to define an effective visual extinction, 
$A_{\rm V, eff}$, for each SPH particle with the help of the following expression:
\begin{equation}
A_{\rm V, eff} = - \frac{1}{3.5} \ln \left[ \frac{1}{N_{\rm pix}} \sum_{i = 1}^{N_{\rm pix}} 
\exp \left(-3.5 \left \{ A_{{\rm V}, i} + A_{\rm V, local} \right \} \right) \right].
\end{equation}
Here, $A_{{\rm V}, i}$ is the extinction corresponding to the column density in the direction defined
by ``sky'' pixel number $i$, and for each SPH particle, we sum over its $N_{\rm pix}$ associated ``sky''
pixels. Our choice of weighting factor is motivated by the fact that the CO photodissociation rate 
scales  with the visual extinction as $R_{\rm pd} \propto e^{-3.5 A_{\rm V}}$ \citep{visser09}. However,
we have also explored the effects of using a smaller weighting factor of 2.5 (motivated by the
fact that the photoelectric heating rate scales with $A_{\rm V}$ as $e^{-2.5 A_{\rm V}}$; see e.g.\
\citealt{bergin04}) but find that in practice our results are not particularly sensitive to this choice.
Finally, we also include in our estimate a term $A_{\rm V, local}$, corresponding to extinction
arising within the gas represented by the SPH particle itself (which is not included in our
{\sc TreeCol} calculation). We assume that this local contribution is proportional to the product 
of the local density and the smoothing length of the SPH particle.

In Figure~\ref{rhoAV}, we show how the chemical state of the gas varies as a function of
density and $A_{\rm V, eff}$. We see that although density and effective extinction are
correlated, there is significant scatter in the relationship, particularly at low densities. 
This scatter can amount to as much as an order of magnitude variation in $A_{\rm V, eff}$
at a given density, and since photochemical rates typically depend exponentially on
$A_{\rm V, eff}$, the scatter in their values at a given density can be far larger
(c.f.\ \citealt{glo10}, who find a similar result for turbulent clouds without self-gravity).

Figure~\ref{rhoAV} also shows us that C$^{+}$ dominates in the region with 
$A_{\rm V, eff} < 1$, as we would expect given the effectiveness of carbon 
photoionization at low $A_{\rm V, eff}$. There is also some neutral atomic carbon 
in this region, but very little CO. Neutral atomic carbon begins to take over 
from C$^{+}$ above $A_{\rm V, eff} = 1$, peaking  at $A_{\rm V, eff} \sim 2$--3, 
and then declining at higher extinctions as CO starts to dominate.

We see, therefore, that the behaviour of C$^{+}$, C and CO in our turbulent clouds
is very similar to that seen in 1D PDR cloud models, which find a similar layering 
as one moves from low to high extinctions.
The main differences here are the much larger scatter that
we have in the turbulent case, owing to the scatter in the $n$-$A_{\rm V, eff}$ relationship,
and the fact that the regions with $A_{\rm V, eff} \sim 2$--3 where C dominates are
not found only at the edges of the cloud, but instead are distributed throughout the
cloud.  

\begin{figure}
\includegraphics[width=3.3in]{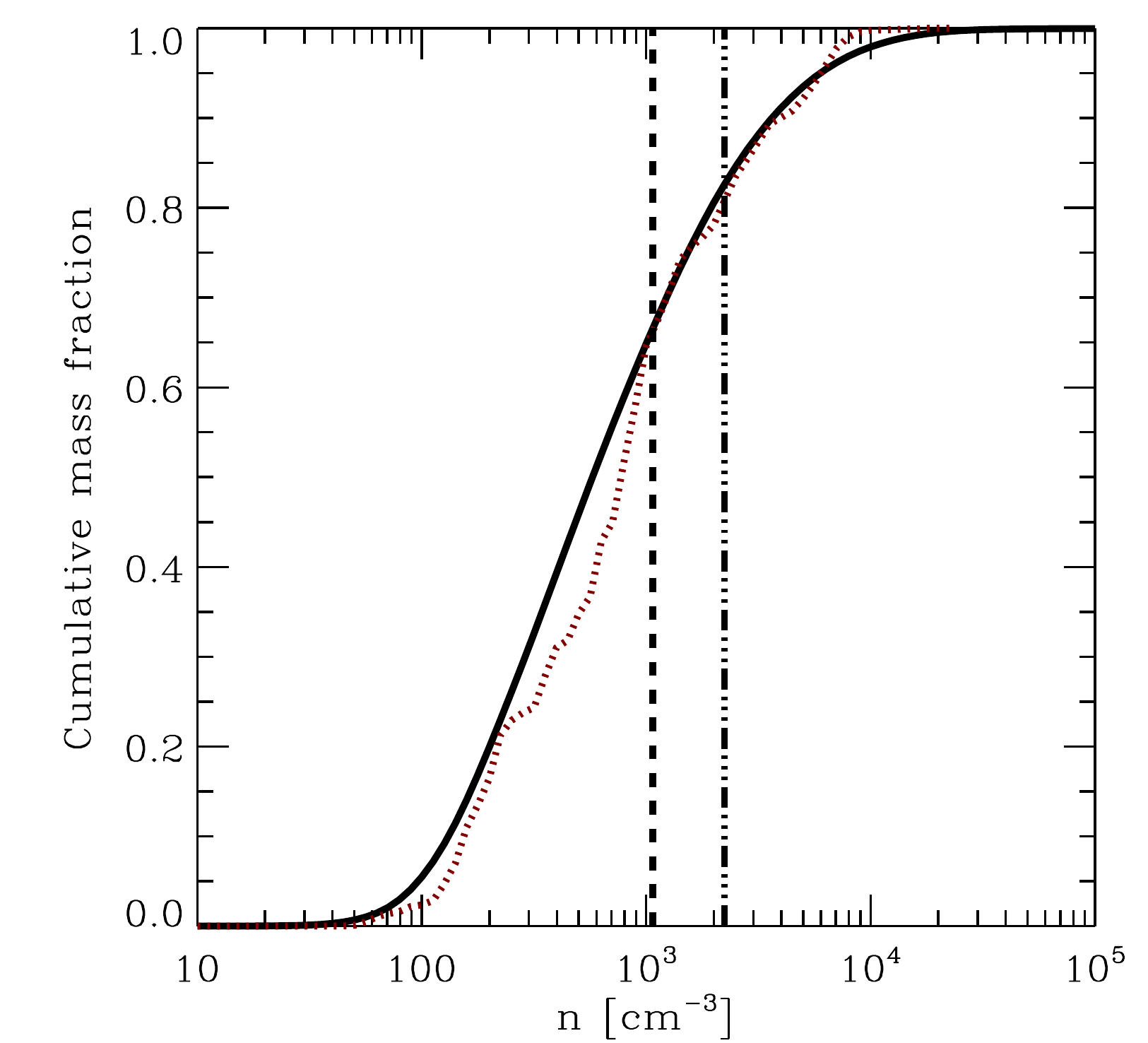}
\caption{Cumulative fraction of the total cloud mass located at or below 
number density $n$ (solid line), plotted as a function of $n$. We also 
show the cumulative atomic carbon mass fraction (dotted line), i.e.\
the fraction of the total amount of atomic carbon located at or below
$n$, plotted as a function of $n$. The vertical dashed line indicates
the critical density of the \ci~$1 \rightarrow 0$ transition, computed
assuming that all of the hydrogen is in molecular form, and that the
gas temperature is 20~K. The vertical dot-dot-dot-dashed line indicates
the critical density for the $2 \rightarrow 1$ transition, computed with
the same assumptions. \label{cumul_dense}}
\end{figure}

It is also useful to quantify how much of the total mass of neutral atomic carbon in
the cloud is located at each density and temperature. In Figure~\ref{cumul_dense},
we show how the mass of the cloud below a density $n$ varies as we vary $n$. 
We see that the majority of the gas is located in the density range $100 < n <
10^{4} \: {\rm cm^{-3}}$, with less than 10\% to be found at higher or lower densities.
We also show in this Figure the same quantity for the neutral atomic carbon, i.e.\
how much of the total mass of C is located below a density $n$ for a range of different
values of $n$. This follows the line for the total gas mass quite closely, although the
fraction of the mass of neutral carbon at $n < 1000 \: {\rm cm^{-3}}$ is a little smaller
than the fraction of the total mass at these densities, since this lower density material
contains significant amounts of C$^{+}$. If we compare the distribution of the neutral
carbon with the critical densities for the $1 \rightarrow 0$ and $2 \rightarrow 1$
transitions (illustrated by the vertical lines), we see that roughly 60\% of the carbon
is located below the critical density for the $1 \rightarrow 0$ transition and 
roughly 80\% is located below the critical density for the $2 \rightarrow 1$
transitions. The values for the critical densities plotted here were computed assuming
a gas temperature of 20~K, but in practice have little dependence on the temperature.
We can therefore conclude that if the \ci~emission is optically thin, much of it will be 
sub-thermal, meaning that analyses that assume local thermodynamic equilibrium
(LTE) level populations for the carbon may give misleading results. We will return to
this point in Section~\ref{calc_tex} below.

\begin{figure}
\includegraphics[width=3.3in]{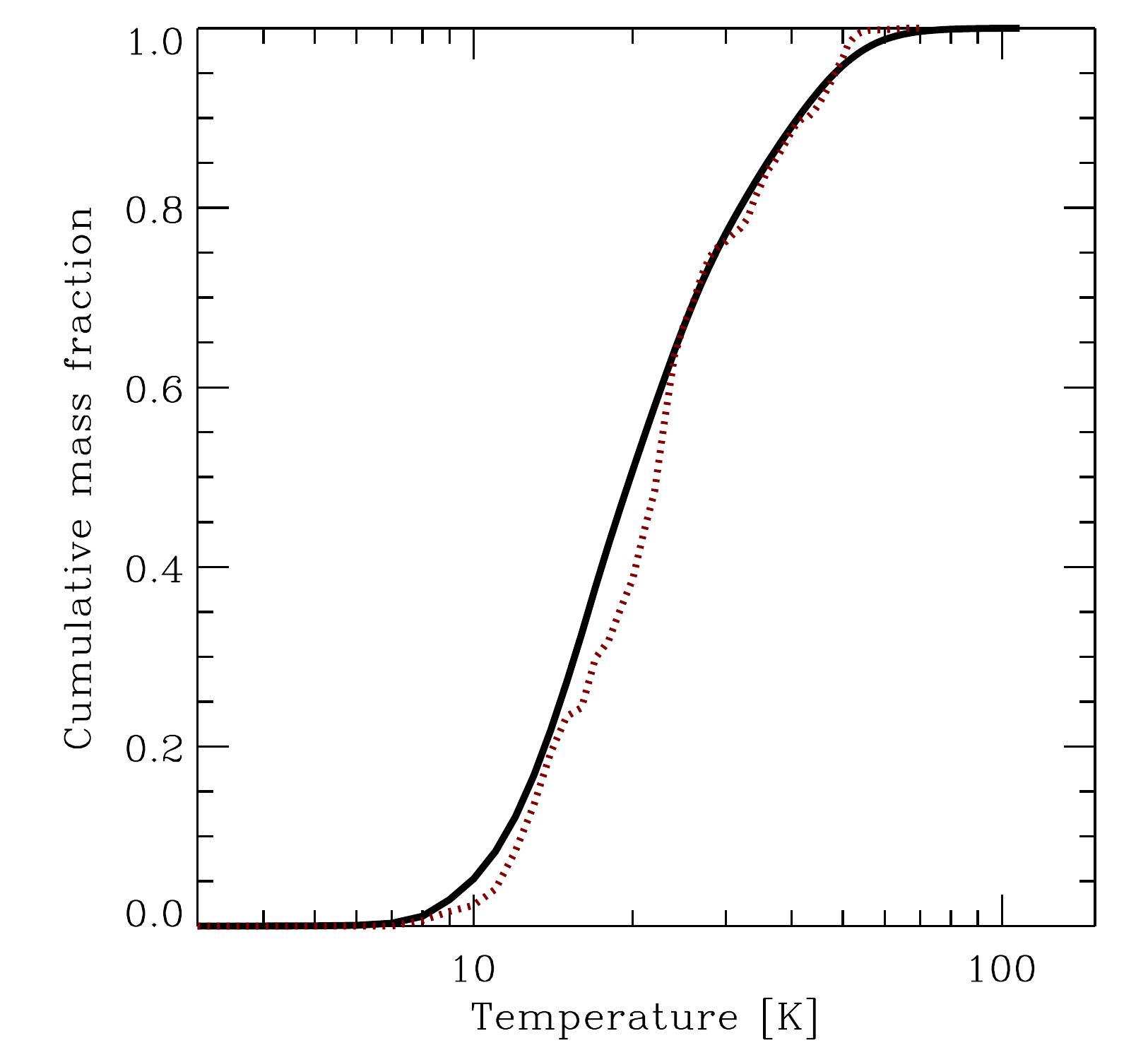}
\caption{Cumulative fraction of the total cloud mass located at or below 
gas temperature $T$ (solid line), plotted as a function of $T$. We also 
show the cumulative atomic carbon mass fraction (dotted line) as
a function of $T$. \label{cumul_temp}}
\end{figure}

Finally, it is also interesting to perform a similar analysis for the cumulative mass 
fractions of total gas and neutral atomic carbon as a function of temperature. This
is illustrated in Figure~\ref{cumul_temp}. We see that almost half of the total gas
mass is located in the temperature range $10 < T < 20$~K, and that there is almost
no gas with $T < 10 \: {\rm K}$ or $T > 100 \: {\rm K}$. We also see that, once again,
the result for the neutral carbon closely tracks that for the total gas mass.

\subsection{How much of the cloud is traced by \ci~emission?}
\label{how}
We saw in the previous section that the range of densities containing
most of the neutral atomic carbon is very similar to the range of densities
containing the bulk of the cloud's mass. Moreover, although most of the 
atomic carbon is found in gas with densities below the critical densities 
for the $1 \rightarrow 0$ and $2 \rightarrow 1$ transitions, it does not
lie very far below $n_{\rm crit}$. Therefore, although we expect the
emission to be sub-thermal, it will not be strongly sub-thermal.
Furthermore, the neutral atomic
carbon is found primarily in gas with temperatures in the range
$10 < T < 50$~K, and these temperatures are high enough to
excite the $1 \rightarrow 0$ transition relatively easily.

\begin{figure*}
\includegraphics[width=0.49\textwidth]{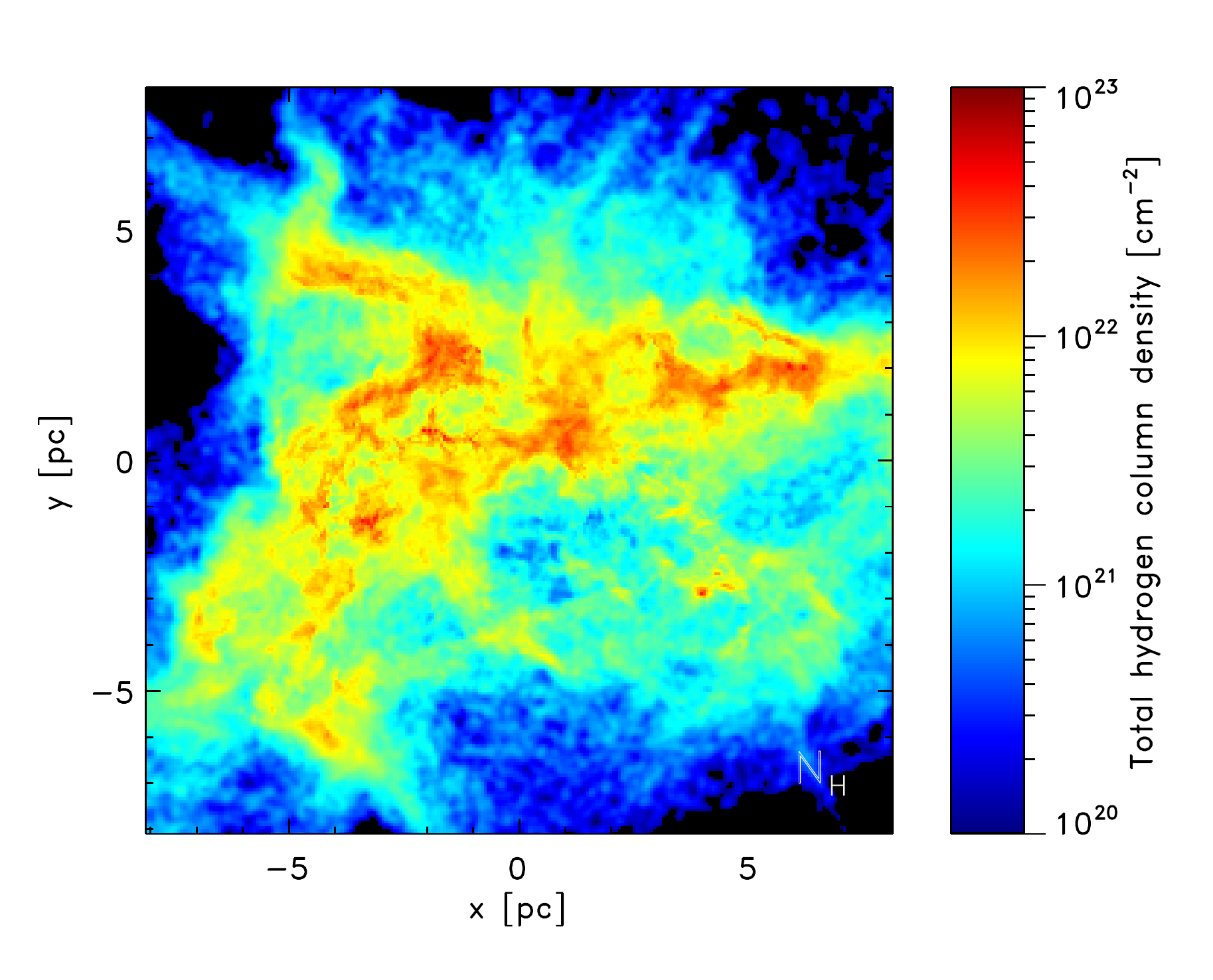}
\includegraphics[width=0.49\textwidth]{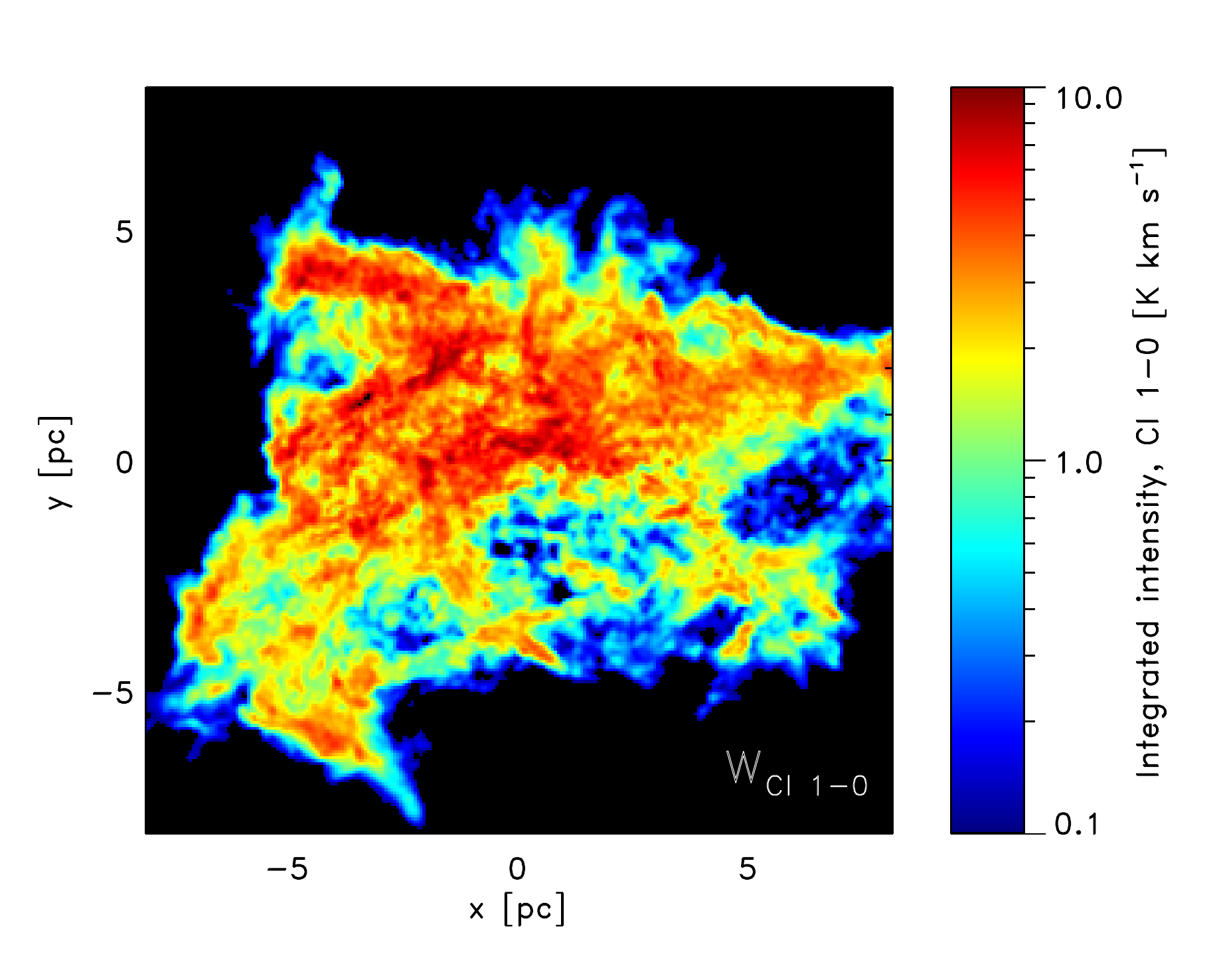}
\includegraphics[width=0.49\textwidth]{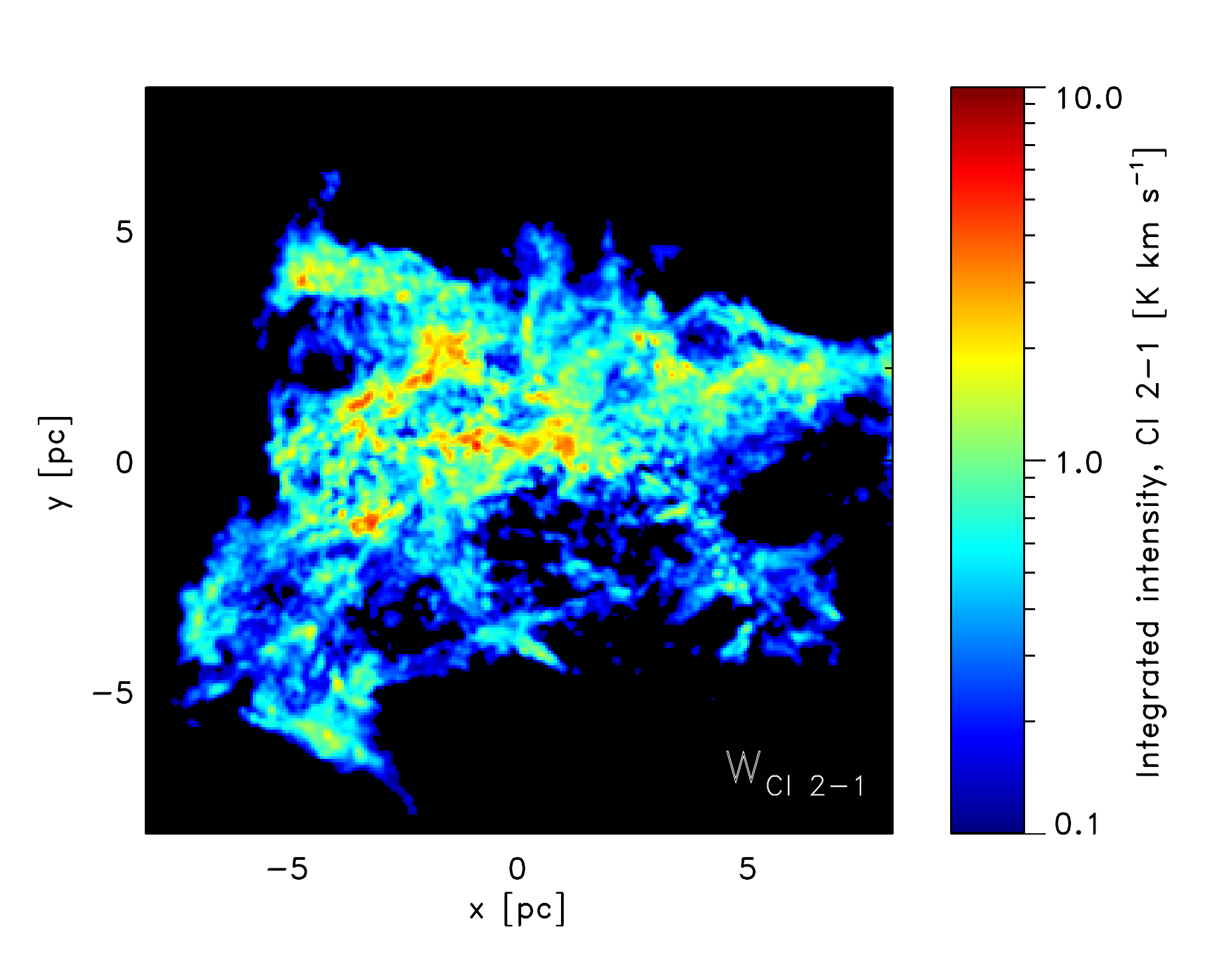}
\includegraphics[width=0.49\textwidth]{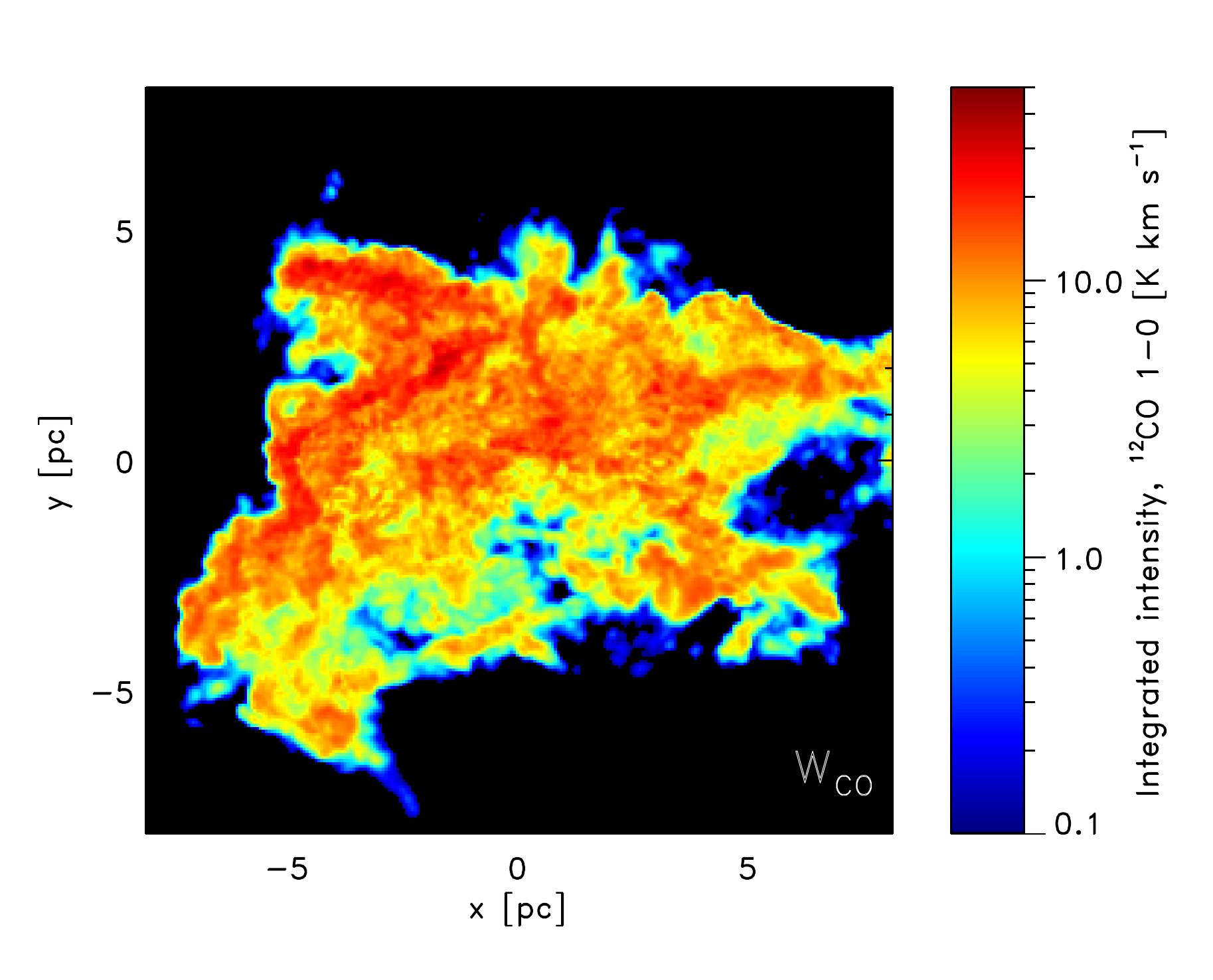}
\caption{{\it Top left}: Map of the total column density of hydrogen nuclei, $N_{\rm H}$. 
The area shown is approximately 16~pc by 16~pc and contains most of the mass of the cloud. 
{\it Top right}: Map of velocity-integrated intensity in the \ci~$1 \rightarrow 0$ transition, computed using 
{\sc radmc-3d} with the large velocity gradient approximation. 
{\it Bottom left}: The same as in the top right panel, but for the \ci~$2 \rightarrow 1$ transition.
{\it Bottom right}: The same as in the top right panel, but for the $J = 1 \rightarrow 0$ transition of 
$^{12}$CO.
\label{cloud_maps}}
\end{figure*}

Taken together, these factors suggest that \ci~emission should be a good
tracer of the gas in the cloud. To investigate whether this is the case, we
have used {\sc radmc-3d} to make maps of the velocity-integrated intensity
of the  \ci~$1 \rightarrow 0$ and  $2 \rightarrow 1$ lines produced by the
cloud. These are shown in Figure~\ref{cloud_maps}, together with the cloud
column density and the integrated intensity in the $^{12}$CO $J = 1 \rightarrow
0$ transition. We see that there is a good correlation between the structure that
we see in the column density map and in the \ci~$1 \rightarrow 0$ emission
map. The \ci~emission traces most of the structure down to a total hydrogen
column density of approximately $N_{\rm H} = 10^{21} \: {\rm cm^{-2}}$, 
but does not pick out the low density gas at the edges of the cloud. Comparing
this map with the CO emission map, we see that both tracers pick out the same
basic cloud structure, but that \ci~appears to do a slightly better job close to the 
edges of the cloud. The \ci~$2 \rightarrow 1$ emission also seems to be a good
tracer of cloud structure in the denser regions, but becomes extremely faint
towards the edges of the cloud.

The widespread \ci~emission that we see in our map is very different from the limb-brightened
emission predicted by simple 1D PDR models, but is qualitatively consistent with the predictions
of clumpy PDR models and with observations of \ci~in real GMCs, which typically find a
good correlation between the \ci~and CO emission \citep[see e.g.][]{svd94,kra08}.
This is a testament
to the central role that turbulence plays in structuring the cloud. As we have already seen, the
neutral atomic carbon in our model cloud is found at a similar range of effective visual
extinctions to those that we would expect based on PDR model results. The key difference is
that gas with this effective extinction is not located in a single coherent structure at the edge of
the cloud. Because of the turbulent motions, the cloud is highly structured, allowing radiation
to penetrate deep into the interior \citep[see also][]{beth04,beth07}. The \ci~emitting surfaces 
that we see are therefore
found throughout the cloud and are present along most lines-of-sight. The crucial role that 
density substructure within GMCs plays in explaining the observed \ci~emission has been 
understood for some time \citep[see e.g.][]{stut88,genz88,pjk94,kram04}, and our results strongly 
support this picture.

\begin{figure}
\includegraphics[width=3.3in]{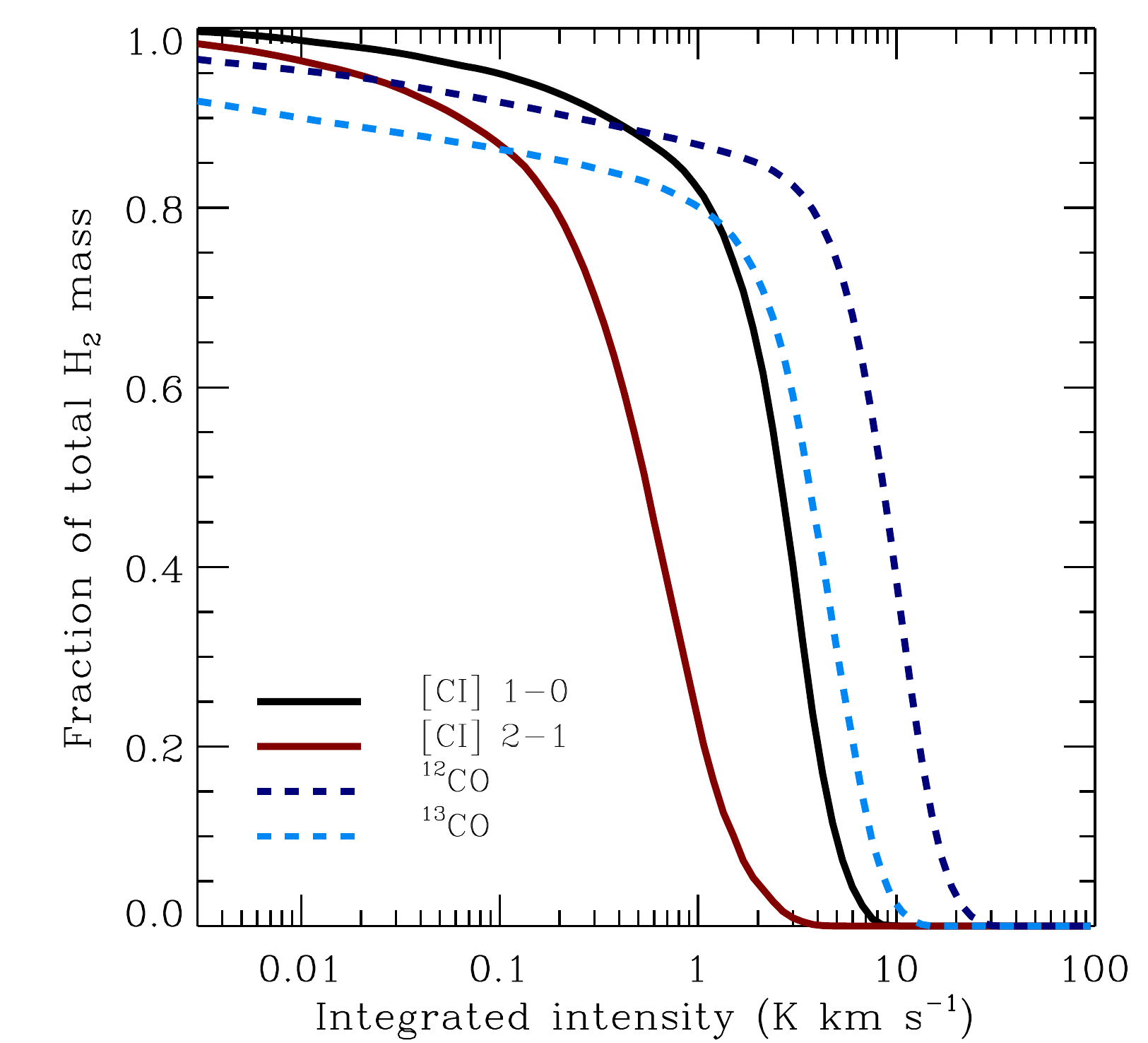}
\caption{Cumulative fraction of the H$_{2}$ mass in the cloud traced by the \ci~$1 \rightarrow 0$ 
transition (black solid line) and the $2 \rightarrow 1$ transition (red solid line), plotted as a function
of the minimum integrated intensity in the line. The same quantity is also shown for the
$^{12}$CO $J = 1 \rightarrow 0$ transition (upper blue dashed line) and the 
$^{13}$CO $J = 1 \rightarrow 0$ transition (lower blue dashed line).
\label{cumul_emission}}
\end{figure}

Having established that the \ci~$1 \rightarrow 0$ line appears to be a promising gas tracer,
the next step is to quantify what fraction of the molecular gas in the cloud it allows us to observe.
A simple starting point is to compute what fraction of the total H$_{2}$ mass is found along
lines-of-sight with \ci~integrated intensities greater than some minimum value 
$W_{\rm CI, 1-0, min}$, and then to look at how this value changes as we vary
$W_{\rm CI, 1-0, min}$. The results of this analysis are shown in 
Figure~\ref{cumul_emission}, together with similar results for the  \ci~$2 \rightarrow 1$ line, 
the $^{12}$CO $J = 1 \rightarrow 0$ line and the $^{13}$CO $J = 1 \rightarrow 0$ line.

We see that for all of the lines, the fraction of the cloud that we detect is
a strong function of the minimum integrated intensity. In the case of 
\ci~$1 \rightarrow 0$, almost 80\% of the molecular mass of the cloud is
located along lines of sight with integrated intensities in the narrow range
$1 < W_{\rm CI, 1-0, min} < 5 \: {\rm K \: km \: s^{-1}}$. In the case of the \ci~$2 \rightarrow 1$ 
line, we see similar behaviour, but the minimum integrated intensity required in order 
to detect any given amount of the molecular gas is typically at least a factor of a few 
smaller. 

If we compare this with the behaviour of $^{12}$CO, we see that if our minimum integrated intensity
for both lines is greater than $0.3  \: {\rm K \: km \: s^{-1}}$, then $^{12}$CO is a superior tracer of the
molecular mass. Only once we start to look along very faint sight-lines does the \ci~emission
start to trace regions that would be unobserved in $^{12}$CO. For $^{13}$CO, the story is rather different. 
For integrated intensities greater than around $1 \: {\rm K \: km \: s^{-1}}$, the behaviour of the 
$^{13}$CO $J = 1 \rightarrow 0$ line and the \ci~$1 \rightarrow 0$ line are very similar. At lower integrated 
intensities, the \ci~line becomes the better tracer, allowing one to study the 15--20\% of the molecular
mass of the cloud that is not well traced by $^{13}$CO.

\subsection{Determining the \ci~excitation temperature}
\label{calc_tex}
As a first step towards quantifying the properties of the \ci~emission produced by our model
cloud, we investigate in this section how accurately one can infer the excitation temperature
of the carbon atoms based on the synthetic emission maps. For an atomic or molecular
system with a lower energy level $l$ and an upper energy level $u$, we can define an
associated excitation temperature
\begin{equation}
T_{\rm ex} \equiv -\frac{E_{ul}}{k} \left[ \ln \left( \frac{g_{l}}{g_{u}} \frac{f_{u}}{f_{l}} \right) \right]^{-1},
\end{equation}
where $f_{l}$ and $f_{u}$ are the fractional level populations of the lower and upper levels,
$g_{l}$ and $g_{u}$ are the associated statistical weights of these levels, and $E_{ul}$ is
the energy separation between the two levels. In the case of neutral atomic carbon, we
have three fine structure levels, with total angular momentum $J = 0, 1, 2$, 
and hence three different excitation temperatures, corresponding to the three different 
ways in which we can pair up two out of these three levels. In local thermodynamic
equilibrium, all three excitation temperatures are equal to each other and to the kinetic
temperature of the gas, $T_{\rm K}$. If the carbon is not in LTE, however, then the
excitation temperatures may differ from each other. Because the critical densities of the 
$1 \rightarrow 0$ and $2 \rightarrow 1$ transitions are quite similar, it is often assumed
that the excitation temperatures corresponding to these transitions are equal, but as we
will see below, this is not always a safe assumption.

One common technique for estimating $T_{\rm ex}$ from maps of \ci~emission involves the
assumption that the \ci~lines are optically thin and that the contribution of the ISRF (including
the CMB) at the frequencies of the \ci~lines can be neglected \citep[see e.g.][]{frerking89,sch03}. 
In this case, the column densities of  carbon atoms in the $J = 1$ and $J = 2$ states are directly 
proportional to the integrated intensities of the $1 \rightarrow 0$ and $2 \rightarrow 1$ lines, and 
we can estimate $T_{\rm ex}$ using the expression  \citep{sch03}
\begin{eqnarray}
 T_{\rm ex, est} & = & \frac{h\nu_{21}}{k} \left[ \ln \left(\frac{N_{1}}{N_{2}} \frac{g_{2}}{g_{1}} \right) \right]^{-1}, \\
 & = & 38.8 \left[ \ln \left(\frac{2.11}{R}\right) \right]^{-1} \: {\rm K},  \label{tx_sch}
\end{eqnarray}
where $R = \int T_{\rm B, 2-1} {\rm d}v / \int T_{\rm B, 1-0} {\rm d}v$ is the ratio of the integrated intensity
of the two lines.

%\begin{eqnarray}
%N_{1} & = & \frac{8\pi k \nu_{10}^{2}}{h c^{3} A_{10}} \int T_{\rm B, 1-0} {\rm d}v,  \label{N1} \\
%N_{2} & = & \frac{8\pi k \nu_{21}^{2}}{h c^{3} A_{21}} \int T_{\rm B, 2-1} {\rm d}v,  \label{N2}
%\end{eqnarray} 
%where $T_{\rm B, 1-0}$ is the brightness temperature of the $1 \rightarrow 0$ transition,
%$T_{\rm B, 2-1}$ is the same for the $2 \rightarrow 1$ transition, $\nu_{10}$, $\nu_{21}$
%are the frequencies of the two transitions and $A_{10}$, $A_{21}$ are the corresponding
%Einstein coefficients. Given $N_{1}$ and $N_{2}$, we can then estimate the mean excitation
%temperature along a line of sight via the following expression:
%\begin{equation}
%\frac{N_{2}}{N_{1}} = \frac{g_{2}}{g_{1}} \exp\left(-\frac{h \nu_{21}}{kT_{\rm ex, est}} \right),
%\end{equation}
%where $g_{1} = 3$ and $g_{2} = 5$ are the statistical weights of the $J=1$ and $J=2$
%levels. Rearranging this yields
%\begin{eqnarray}
% T_{\rm ex, est} & = & \frac{h\nu_{21}}{k} \left[ \ln \left(\frac{N_{1}}{N_{2}} \frac{g_{2}}{g_{1}} \right) \right]^{-1}, \\
% & = & 38.8 \left[ \ln \left(\frac{2.11}{R}\right) \right]^{-1} \: {\rm K},  \label{tx_sch}
%\end{eqnarray}
%where $R = \int T_{\rm B, 2-1} {\rm d}v / \int T_{\rm B, 1-0} {\rm d}v$ is the ratio of the integrated intensity
%of the two lines.
 
Alternatively, if the \ci~emission is optically thick, we can determine the excitation temperature using the
expression \citep{dick78,rw13} 
%we assume that the \ci~emission is optically thick,  we can estimate the excitation temperature
%using an approach originally introduced by \citet{dick78}. For a static cloud with a constant excitation temperature,
%we can write the brightness temperature of the \ci~$1 \rightarrow 0$  line as 
%\begin{equation}
%\frac{kT_{\rm B, 1-0}}{E_{10}} = \left[\frac{1}{e^{E_{10} / kT_{\rm ex}} - 1} - \frac{1}{e^{E_{10} / kT_{\rm bg}} -1} 
%\right] \left(1 - e^{-\tau} \right)  \label{dic1}
%\end{equation}
%where $E_{10} = h\nu_{10}$ and $T_{\rm bg}$ is the brightness temperature of the background 
%radiation field, which in this case is dominated by the cosmic microwave background. In a real cloud,
%of course, the gas is not at rest and there is no {\it a priori} reason to expect $T_{\rm ex}$ to be constant.
%Therefore, when we apply this expression to observations of real clouds, or to the synthetic observations
%presented here, we need to make a couple of approximations. First, we assume that the appropriate value to
%use for $T_{\rm B, 1-0}$ is the maximum value that we measure along the line of sight. Second, we interpret
%$T_{\rm ex}$ as an average value for the line of sight. Having made these approximations, it is then easy
%to show that in the limit $\tau \gg 1$, this expression yields the following estimate for the average
%excitation temperature: 
\begin{equation}
T_{\rm ex, est, 2} = \frac{23.6}{\ln \left[ 1 + 23.6 / (T_{\rm B, 1-0, max}) \right]}, \label{tx_dick}
\end{equation}
where $T_{\rm B, 1-0, max}$ is the maximum value of the brightness temperature along the
line of sight, and where we have again assumed that the energy density of any background
radiation field (the ISRF, the CMB, etc.) at the frequency of the line is negligible in comparison
to the emission produced within the cloud itself.

%In \citeauthor{dick78}'s original analysis he also assumed that the gas was in LTE, in which
%case this estimate for $T_{\rm ex}$ also serves as an estimate of the kinetic temperature of the 
%gas. However, Equation~\ref{dic1} does not rely on this assumption, and holds even when the
%gas is not in LTE.

\begin{figure*}
\includegraphics[width=0.49\textwidth]{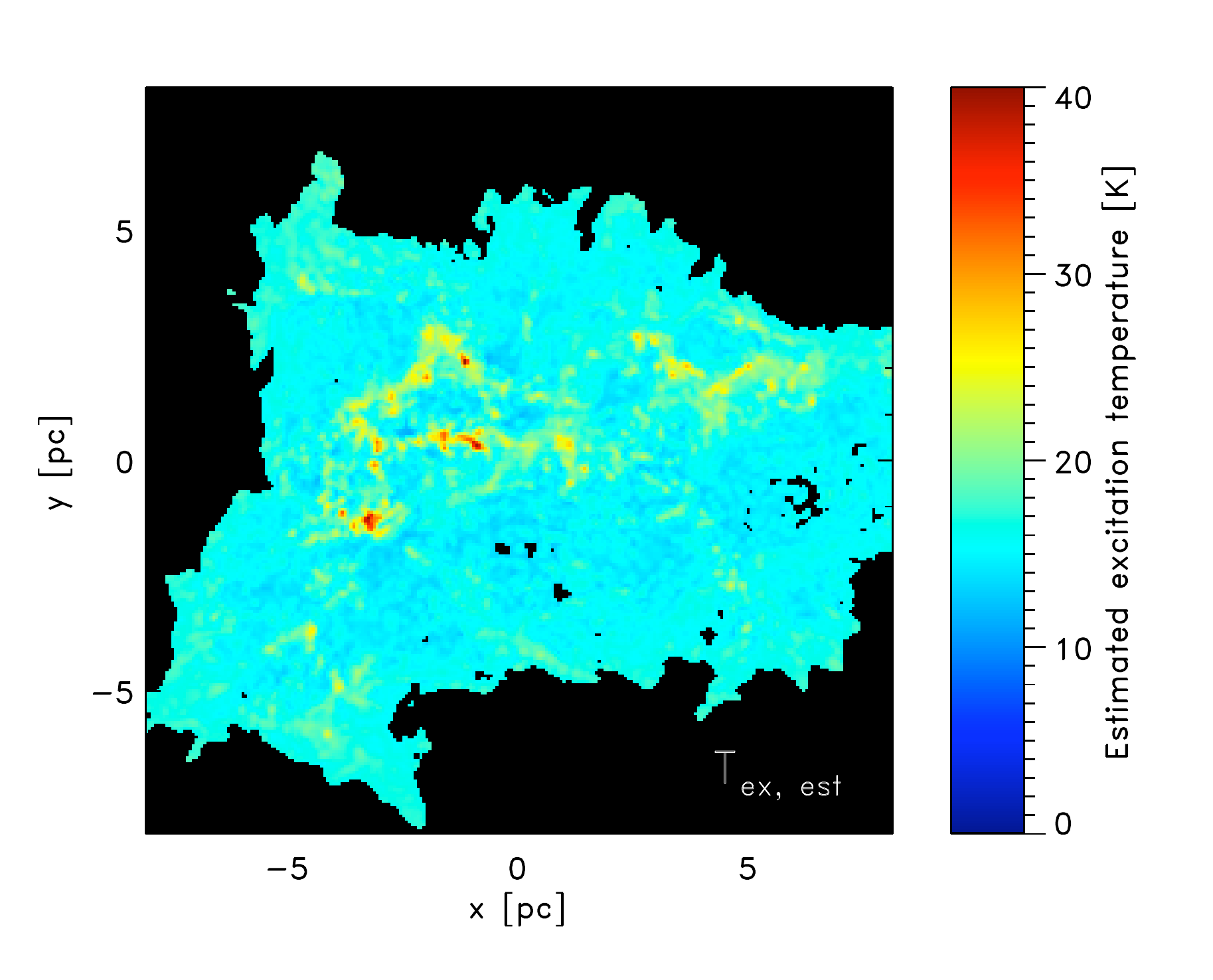}
\includegraphics[width=0.49\textwidth]{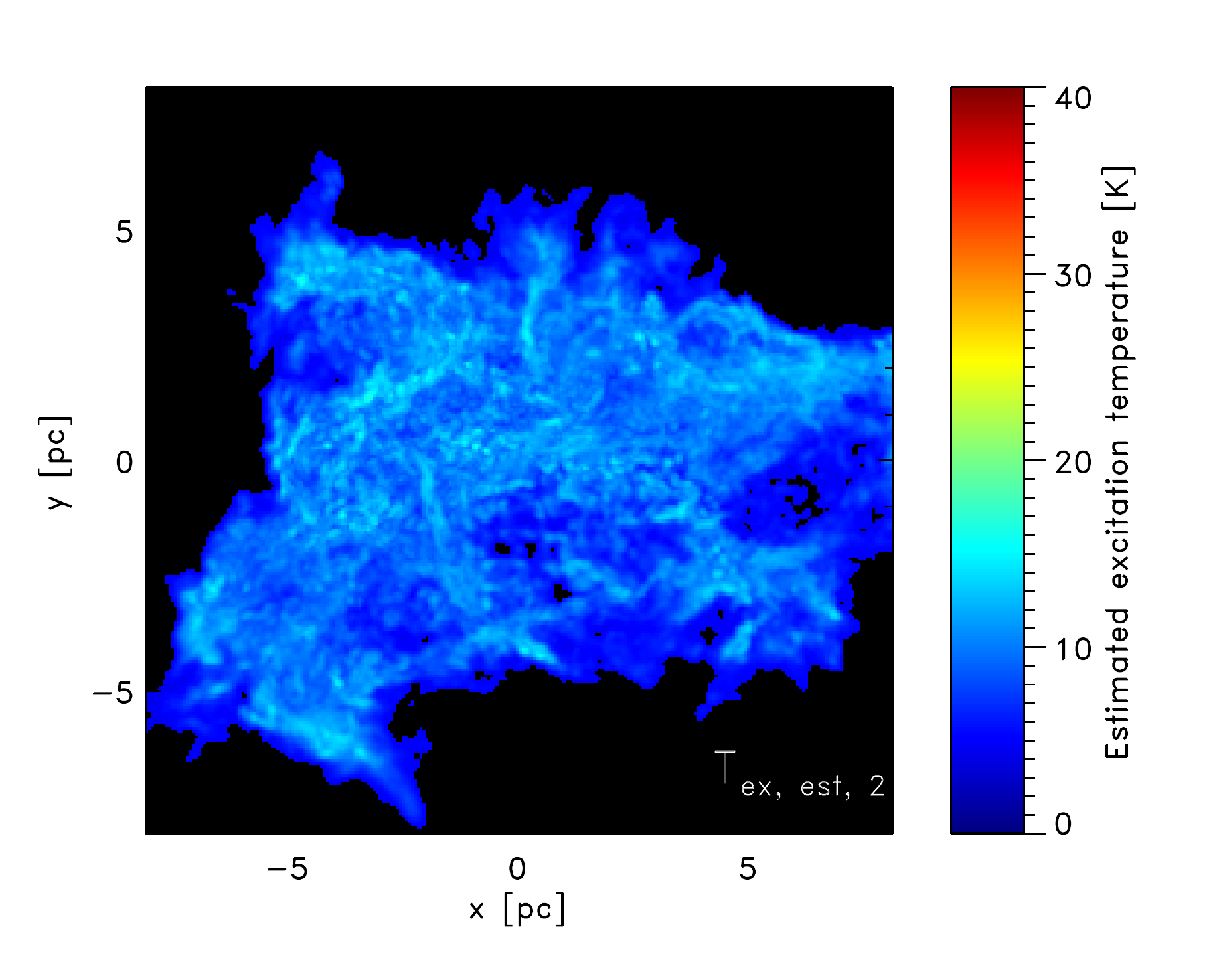}
\includegraphics[width=0.49\textwidth]{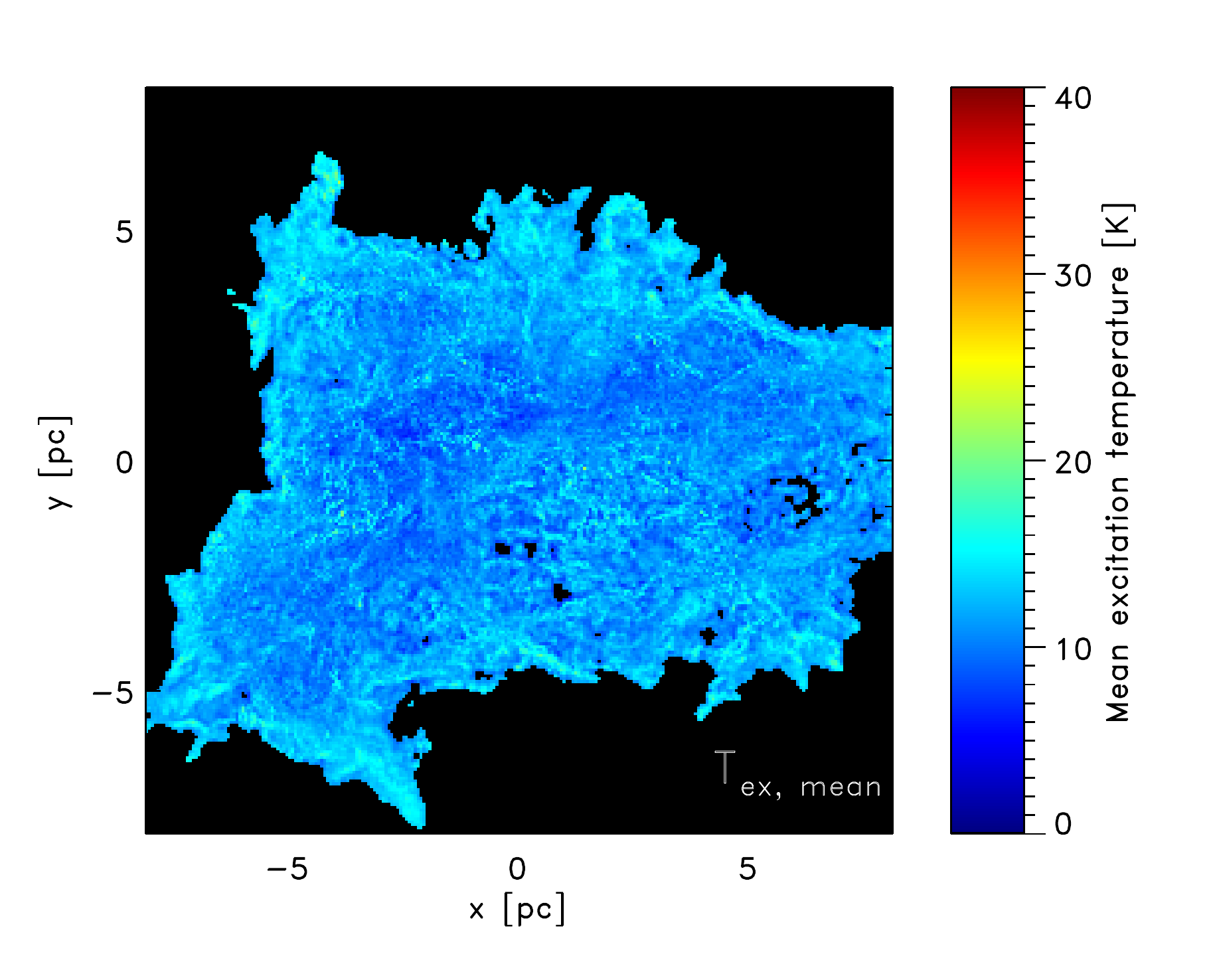}
\includegraphics[width=0.49\textwidth]{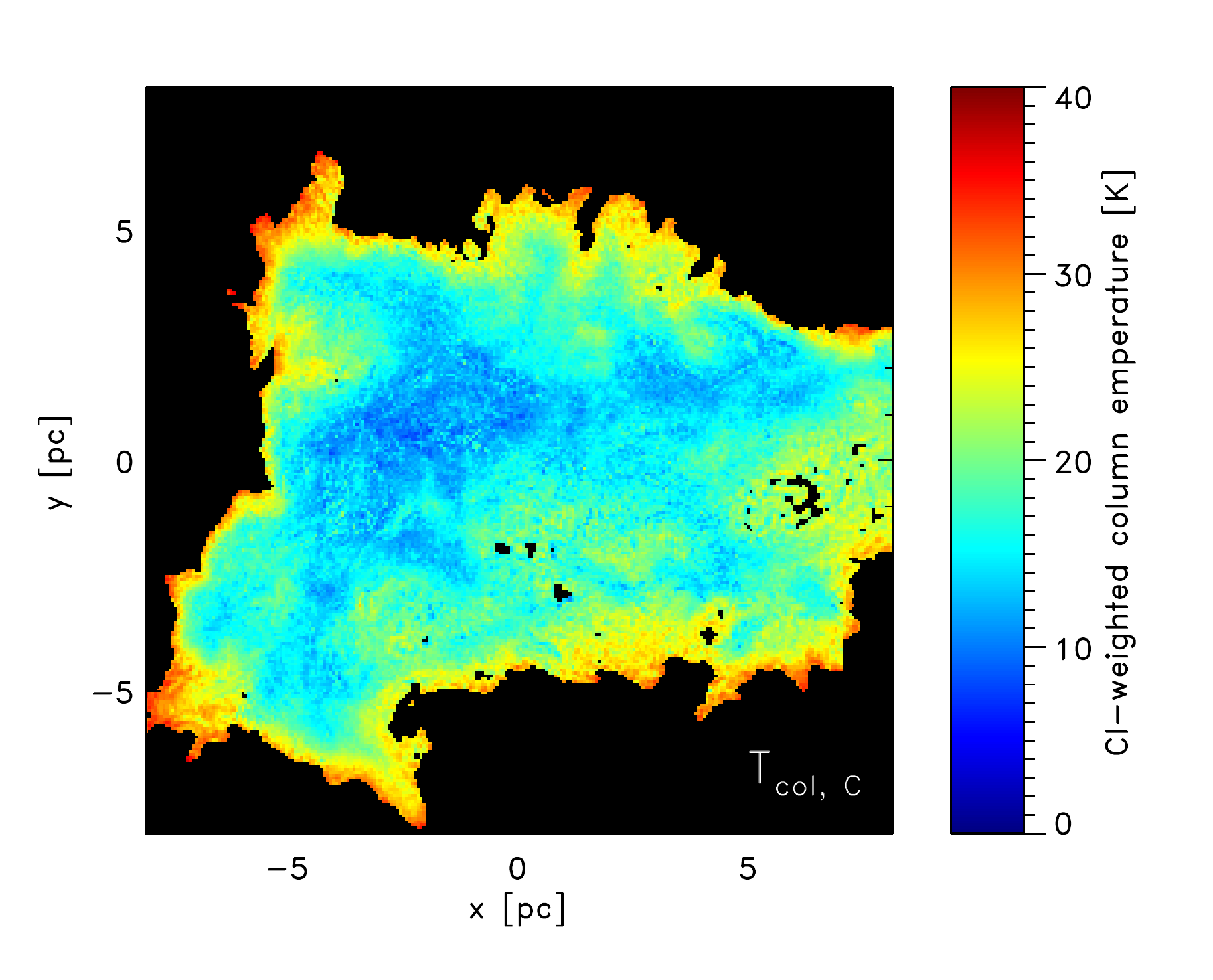}
\caption{{\it Top left}: Estimate of the mean excitation temperature of the neutral
carbon atoms along each line of sight, computed from the ratio of the integrated 
intensities of the \ci~$2 \rightarrow 1$ and $1 \rightarrow 0$ transitions
(Equation~\ref{tx_sch}). This estimate assumes that the gas is optically thin
in both transitions, and hence yields artificially high values towards optically
thick regions.
{\it Top right}: An alternative estimate of $T_{\rm ex}$, computed using Equation~\ref{tx_dick},
based on a procedure introduced
by \citet{dick78}. This estimate assumes that the \ci~$1 \rightarrow 0$ line is optically
thick and so underestimates $T_{\rm ex}$ along optically thin sightlines.
{\it Bottom left}: Mean value of $T_{\rm ex}$ for the $1 \rightarrow 0$ transition along each sight-line, 
computed as a weighted average of the value in each grid cell, with the number density of atomic 
carbon atoms used as the weighting function (Equation~\ref{tex_true_av}). 
{\it Bottom right}: Column temperature of the carbon atoms along each line of sight, computed
using Equation~\ref{tex_col_C}. 
\label{Tex_maps}}
\end{figure*}

In Figure~\ref{Tex_maps}, we show the results obtained using these two approaches. In both cases, we
restrict the region plotted to the set of sight-lines with $W_{\rm CI, 1-0} > 0.1 \: {\rm K} \: {\rm km \: s^{-1}}$.
By enforcing this restriction, we avoid the poorly resolved regions at the edge of the cloud that are too
faint in \ci~to be observable with current instruments. We see from this Figure that Equation~\ref{tx_sch} 
predicts values for $T_{\rm ex}$ of around 15--20~K, with significantly higher values of up to 40~K along 
the lines of sight passing through the densest gas.  On the other hand, Equation~\ref{tx_dick} predicts 
values of $T_{\rm ex}$ that are close to 10~K through much of the cloud, dropping to only a few K at the 
edges of the cloud. 

For comparison, we also show in Figure~\ref{Tex_maps} the mean excitation temperature 
along each sight-line for the $1 \rightarrow 0$ transition. These values are derived directly from
the level populations computed during our radiation transfer calculations.
To compute this mean value, we first compute $T_{\rm ex}$ in each cell in our three-dimensional 
grid using the expression
\begin{equation}
T_{\rm ex} = -\frac{h\nu_{10}}{k} \left[ \ln \left(\frac{n_{1}}{3n_{0}} \right) \right]^{-1},
\end{equation}
where $n_{0}$ and $n_{1}$ are the number densities of neutral atomic carbon in the $J = 0$ and
$J = 1$ levels, respectively. We then compute a weighted mean value along each sight-line, using 
the local number density of atomic carbon as our weighting function:
\begin{equation}
T_{\rm ex, mean} = \frac{\sum_{k} T_{\rm ex}(k) n_{\rm C}(k)}{\sum_{k} n_{\rm C}(k)}, \label{tex_true_av}
\end{equation}
where we sum over all cells along a line of sight. 
%An alternative way of computing the mean
%excitation temperature is to first compute the column densities of carbon atoms in the 
%$J = 0$ and $J = 1$ levels (which we denote as $N_{0}$ and $N_{1}$, respectively),
%and then to compute a mean $T_{\rm ex}$ from the expression\footnote{Note that the values resulting from this approach are not shown in Figure~\ref{Tex_maps}.}
%\begin{equation}
%T_{\rm ex} = -\frac{h\nu_{10}}{k} \left[ \ln \left(\frac{N_{1}}{3N_{0}} \right) \right]^{-1}.  \label{tex_true_col}
%\end{equation}
%In practice, these two methods give results that agree to within around 10\% (i.e.\ around 1~K), with 
%Equation~\ref{tex_true_col} typically yielding slightly larger values than Equation~\ref{tex_true_av}.

Comparison of the different maps shows that both of our approximations give biased views of $T_{\rm ex}$.
Equation~\ref{tx_sch} predicts excitation temperatures that are systematically too large, particularly for the
densest lines of sight. At low column densities, this error comes about because Equation~\ref{tx_sch} is
actually a measure of the excitation temperature associated with the $2 \rightarrow 1$ transition.
Although this is often assumed to be the same as the excitation temperature associated with the 
$1 \rightarrow 0$ transition, we find that in practice the two excitation temperatures  differ by several K
in much of the \ci~emitting gas. Towards dense regions, the performance of Equation~\ref{tx_sch}
becomes even worse because of its neglect of the effects of line opacity. If one or both of the
\ci~lines becomes optically thick, then the ratio of integrated brightness temperatures for the
two transitions is no longer proportional to the ratio of the column densities of the $J = 1$ and $J = 2$
level, and hence Equation~\ref{tx_sch} is no longer valid. If the $2 \rightarrow 1$ transition becomes
optically thick while the $1 \rightarrow 0$ transition remains optically thin, then this leads to an
underestimate of $N_{2}/N_{1}$ and hence and underestimate of the excitation temperature. In
the far more likely case that the $1 \rightarrow 0$ transition becomes optically thick while the
$2 \rightarrow 1$ transition remains optically thin (see Appendix~\ref{op_depths}), then we overestimate
$N_{2}/N_{1}$ and hence overestimate $T_{\rm ex}$, potentially by a large amount.

As one would expect, Equation~\ref{tx_dick} performs much better along the optically thick sight-lines,
predicting  values for $T_{\rm ex}$ that are within a few percent of the true ones, providing that we restrict
our attention to lines of sight with $W_{\rm CI, 1-0} > 3 \: {\rm K} \:
{\rm km} \: {\rm s^{-1}}$. However, it systematically underestimates $T_{\rm ex}$ along lines of sight that 
have low $\tau$ and low $W_{\rm CI, 1-0}$, such as those near the edges of the cloud. In practice, 
the fact that the real excitation temperature does not vary very strongly over the cloud means that
we can construct a reasonably accurate estimate of $T_{\rm ex}$ by using Equation~\ref{tx_dick}
for lines of sight with $W_{\rm CI, 1-0} > 3 \: {\rm K} \: {\rm km} \: {\rm s^{-1}}$ and adopting a constant
value for lines of sight with lower $W_{\rm CI, 1-0}$ that is the mean of the value inferred for the
bright sight-lines.

It is also interesting to compare the excitation temperature of the gas with the actual kinetic temperature.
Since we are observing the cloud in projection, our observations inevitably involve some degree of
averaging of temperatures along each line of sight. Therefore, the appropriate quantity to compare to
our derived excitation temperatures is some form of line-of-sight averaged kinetic temperature, or
``column'' temperature \citep{shetty09,mol13}. In the present case, we expect that the largest contribution to the 
observed \ci~emission will come from the gas with the highest number density of neutral carbon. It therefore
makes sense to use as a weighted average
\begin{equation}
T_{\rm col, C}(i,j) = \frac{\sum_{k} T_{\rm kin}(i,j,k) n_{\rm C}(i,j,k)}{\sum_{k} n_{\rm C}(i,j,k)}. \label{tex_col_C}
\end{equation}
Here, $T_{\rm kin}(i,j,k)$ is the kinetic temperature of the gas in the cell with coordinates $(i,j,k)$
and $n_{\rm C}(i,j,k)$ is the number density of neutral carbon atoms in the same cell. 

The resulting column temperature map is shown in panel (d) of Figure~\ref{Tex_maps}. 
If we compare it with the map of mean excitation temperatures, we see that there is relatively good agreement 
between $T_{\rm col, C}$ and $T_{\rm ex, mean}$ along lines-of-sight that pass through the highest density 
parts of the cloud, but that the values diverge significantly as we move towards the edges of the \ci~map and 
look along lines-of-sight probing lower density gas. This behaviour is simple to explain. Along high column 
density lines-of-sight, most of the \ci~emission that we see comes from high density gas that lies close to or 
above the \ci~critical density. The carbon atoms in this gas are in local thermodynamic equilibrium, and hence 
$T_{\rm ex} \simeq T_{\rm kin}$. Along lower column density lines-of-sight, the mean kinetic temperature of the 
gas is higher, since the mean gas density is lower, and the gas is also less well-shielded from the external 
radiation field. However, the lower gas density also means that most of the carbon atoms are in regions with 
$n < n_{\rm crit}$, and hence are sub-thermally excited, with $T_{\rm ex} < T_{\rm kin}$. (Similar results have 
recently been reported for $^{12}$CO $J = 1 \rightarrow 0$ line emission by \citealt{mol13}).
In practice, we find that if we average over all lines of sight with detectable \ci~emission, the resulting mean 
$T_{\rm ex}$ is around 30\% lower than the mean column temperature. We can reduce the discrepancy to 
$\sim 10$\% if we restrict our attention to lines of sight with $W_{\rm CI, 1-0} > 4 \: {\rm K} \: {\rm km \: s^{-1}}$,
and to $\sim 5$\% if we look only at regions with $W_{\rm CI, 1-0} > 6 \: {\rm K} \: {\rm km \: s^{-1}}$.

\begin{figure}
\includegraphics[width=3.3in]{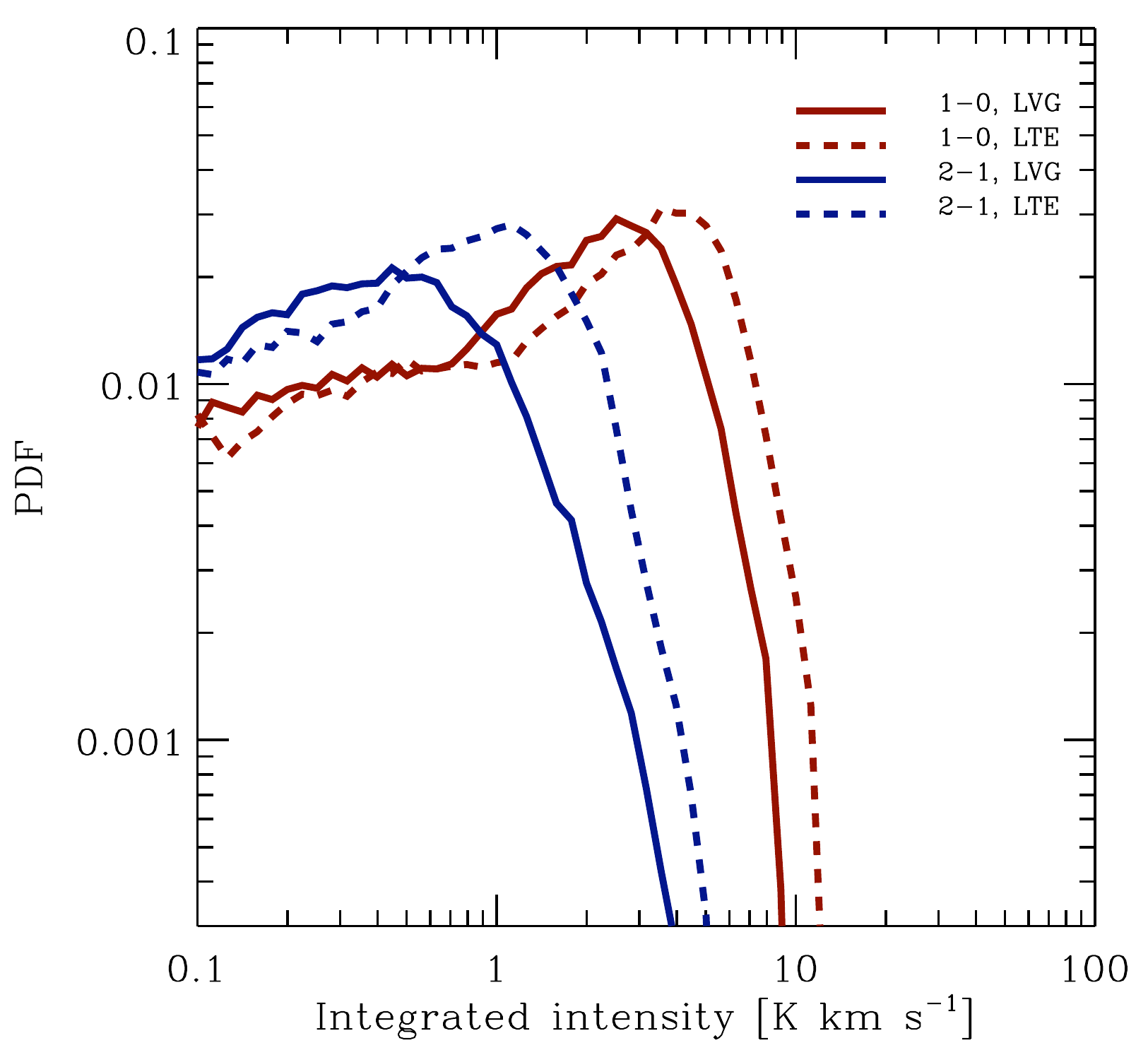}
\caption{PDF of the velocity-integrated emission in the \ci~$1 \rightarrow 0$ (red)
and $2 \rightarrow 1$ (blue) transitions. The solid lines gives the results of our
LVG analysis, while the dashed lines show the distribution of integrated intensities
that we get if we assume LTE level populations. The difference between the LVG
and LTE results demonstrates that a significant fraction of the carbon atoms are
sub-thermally excited.  \label{WCI_LTE}}
\end{figure}

An important lesson that we can draw from these results is that, contrary to what is often assumed, a
significant fraction of the neutral atomic carbon in molecular clouds is not in LTE, and is only sub-thermally
excited. We can confirm this interpretation by comparing the PDF of the integrated intensities of the two 
\ci~transitions that we obtain from our LVG analysis with the corresponding PDFs that we get if we assume
LTE level populations throughout the gas (Figure~\ref{WCI_LTE}). We see that the LVG approach produces
systematically fainter emission than the LTE analysis, consistent with what we expect if  a significant proportion 
of the carbon atoms are subthermally excited.

\subsection{\ci~emission as a probe of column densities}
\label{column}

\subsubsection{Tracing the total column density}
\label{totcol}
\begin{figure}
\includegraphics[width=3.3in]{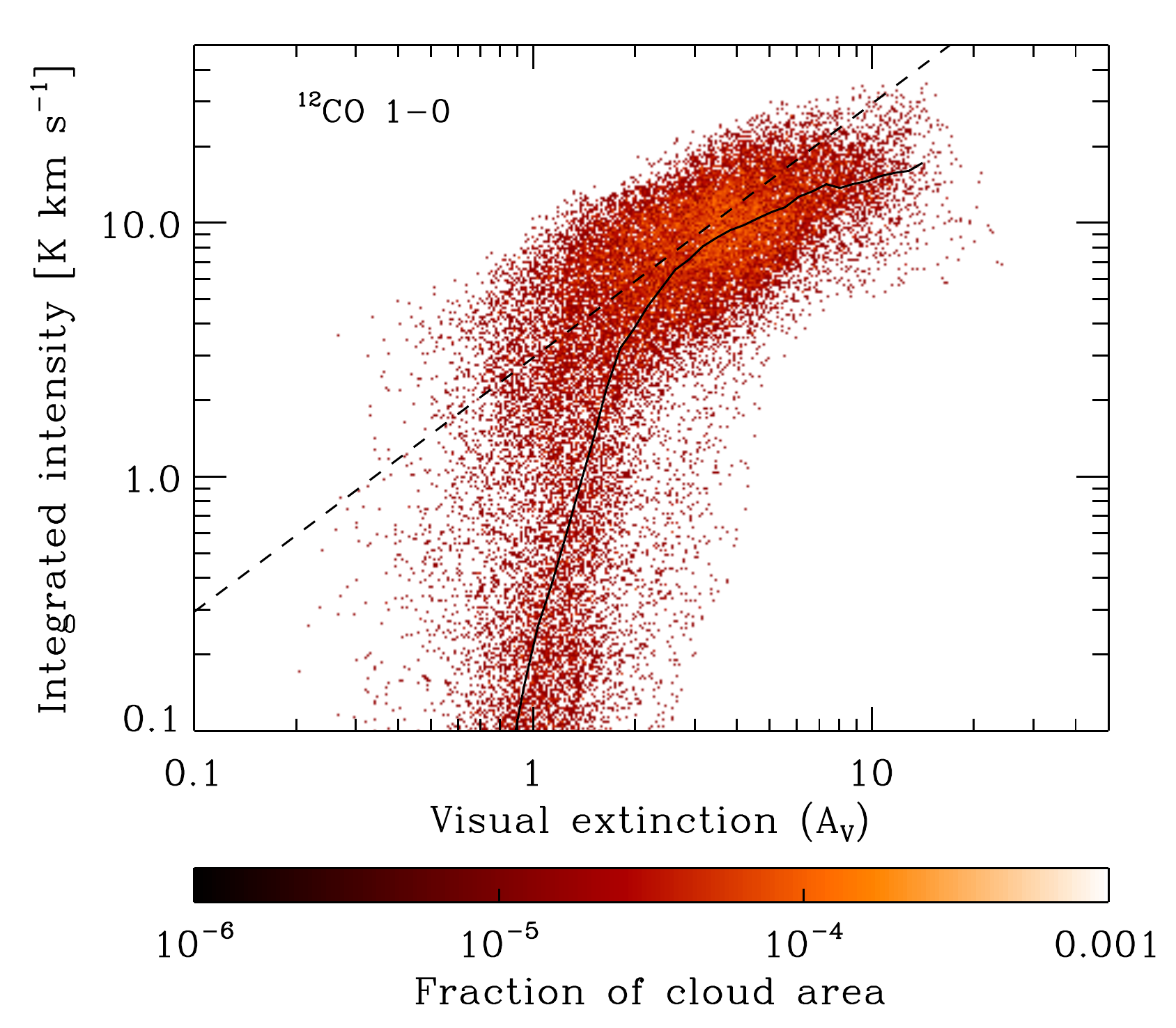}
\caption{Two-dimensional PDF of the integrated intensity in the $^{12}$CO $J = 1\rightarrow 0$ line,
$W_{\rm CO}$, plotted as a function of the visual extinction. The dashed line indicates a linear 
relationship between $W_{\rm CO}$ and $A_{\rm V}$ and is there to guide the eye. 
The solid line shows the geometric mean of $W_{\rm CO}$ as a function of $A_{\rm V}$.
The points are colour-coded according to the fraction of the total cloud area that they 
represent.
\label{WCO_AV}}
\end{figure}

CO emission is often used as a tracer of column densities within molecular clouds. However, its ability to
do this is hampered by a couple of significant problems. It has been understood for a long time that
at high column densities, $^{12}$CO emission 
becomes optically thick, meaning that it is no longer a good tracer of $N_{\rm CO}$ or of the total column
density of the cloud \citep[see e.g.][for an early discussion of this point]{ss74}.
%\citep[see e.g.][for a detailed discussion of this point]{shetty11a}. 
On the other hand,
at low column densities, $W_{\rm CO}$ falls off sharply with decreasing $A_{\rm V}$ owing to the effects
of CO photodissociation \citep{good09,shetty11a}. This is illustrated in Figure~\ref{WCO_AV}, where
we plot the two-dimensional PDF of $W_{\rm CO}$ versus $A_{\rm V}$ that we find in our simulation.
Although the two quantities are clearly correlated, the correlation is close to linear only over a small
range of visual extinctions around $A_{\rm V} \sim 3$. At higher $A_{\rm V}$, the correlation becomes
sub-linear, owing to the effects of the line opacity, while at lower $A_{\rm V}$, there is a rapid fall-off
in $W_{\rm CO}$ owing to the loss of CO from the gas due to photodissociation. Moreover, it is also
clear that even in the regime where $W_{\rm CO}$ and $A_{\rm V}$ have a correlation that is close
to linear, there is substantial scatter around this correlation. 

\begin{figure}
\includegraphics[width=3.3in]{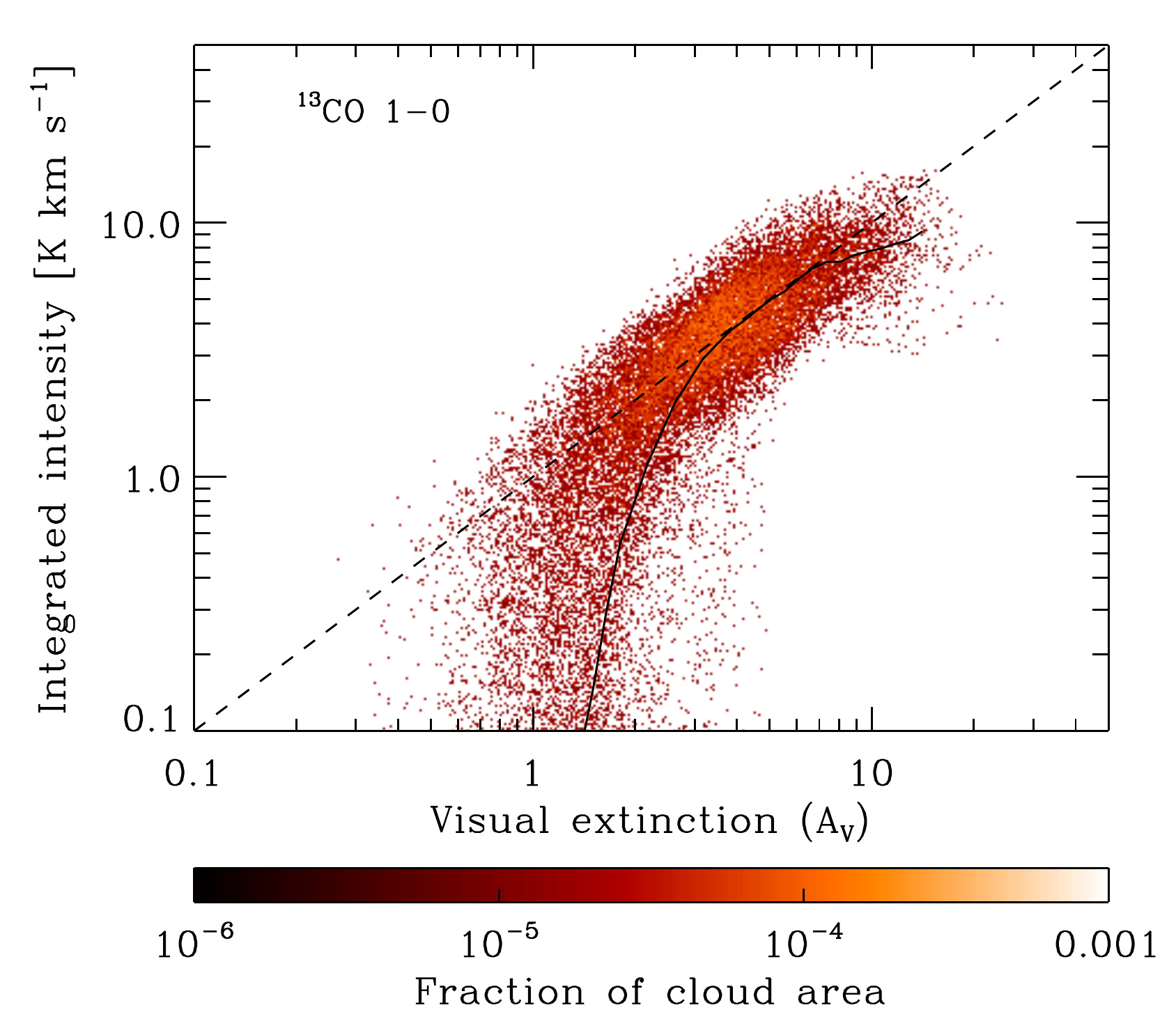}
\caption{As Figure~\ref{WCO_AV}, but for the $J = 1 \rightarrow 0$ transition of $^{13}$CO. 
To compute the $^{13}$CO emission map, we have assumed a constant ratio of 60:1 between
$^{12}$CO and $^{13}$CO. \label{WCO13_AV}}
\end{figure}

We can improve on this situation by looking at emission from $^{13}$CO instead of $^{12}$CO.
The abundance of carbon-13 in the interstellar medium is much smaller than that of carbon-12,
and hence the $^{13}$CO abundance is much smaller than the  $^{12}$CO abundance. 
Consequently, the optical depths of the $^{13}$CO rotational transitions are much smaller than
for $^{12}$CO, making $^{13}$CO a more reliable tracer of the gas column density along high
extinction sight-lines \citep{pin08,good09}. This is illustrated in Figure~\ref{WCO13_AV}, where
we show the 2D PDF of the integrated intensity in the $J = 1 \rightarrow 0$ line of $^{13}$CO,
$W_{\rm 13CO}$, versus the visual extinction. 
%As before, to compute the $W_{\rm 13CO}$ map,
%we assume a fixed ratio $n_{\rm 13CO} / n_{\rm 12CO} = 1 / 60$ between the number densities
%of $^{13}$CO and $^{12}$CO.

Figure~\ref{WCO13_AV} demonstrates that, as expected, $^{13}$CO is a significantly better
tracer of the column density than $^{12}$CO. The relationship between $W_{\rm 13CO}$
and $A_{\rm V}$ is extremely close to linear for visual extinctions in the range
$3 < A_{\rm V} < 10$, although there are hints that opacity effects are beginning to
become important at $A_{\rm V} \sim 10$. However, as in the case of $^{12}$CO, 
the relationship breaks down at low $A_{\rm V}$ owing to the photodissociation of CO.

\begin{figure}
\includegraphics[width=3.3in]{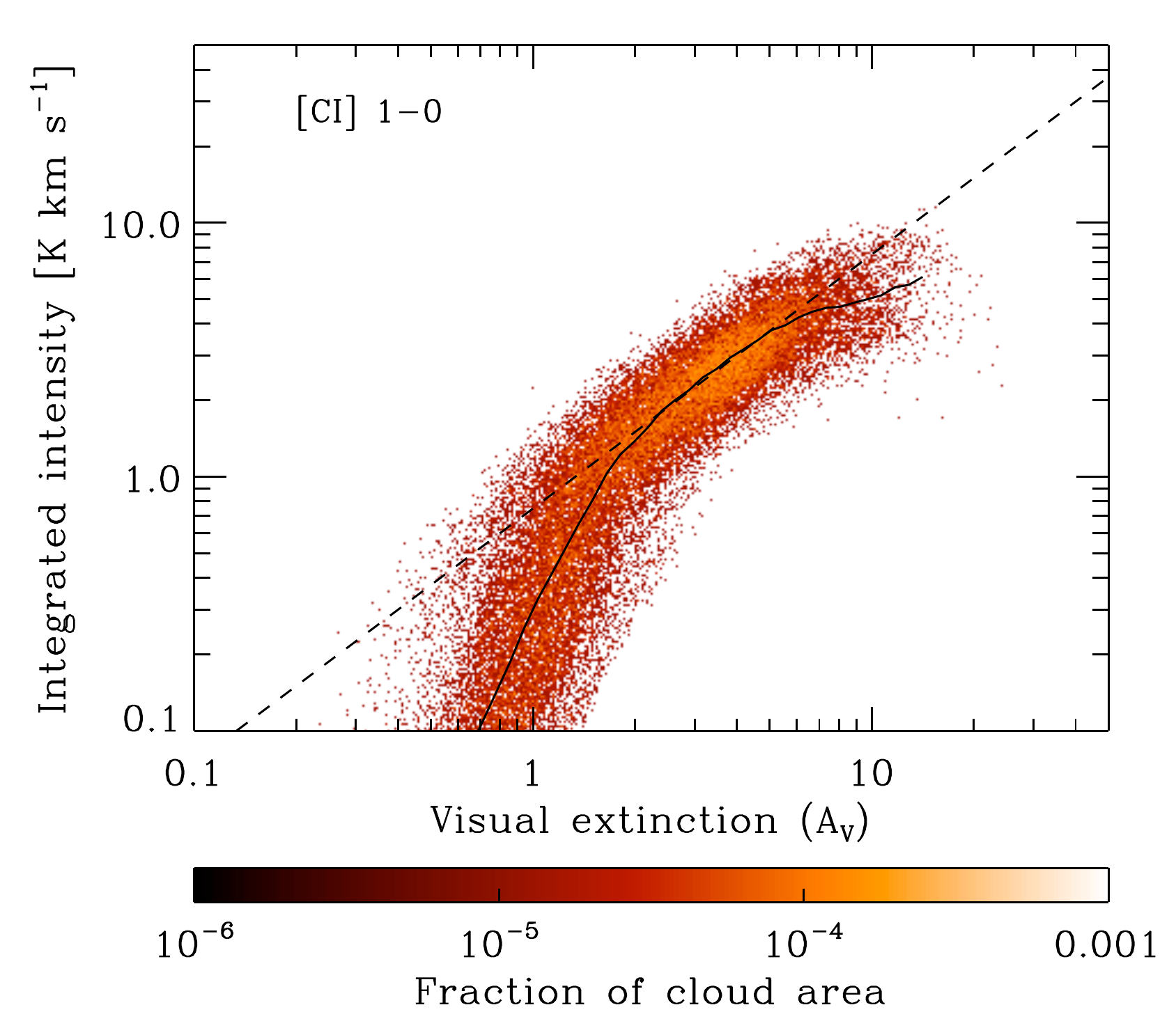}
\caption{As Figure~\ref{WCO_AV}, but for the \ci~$1 \rightarrow 0$ line. \label{WCI_AV}}
\end{figure}

\begin{figure}
\includegraphics[width=3.3in]{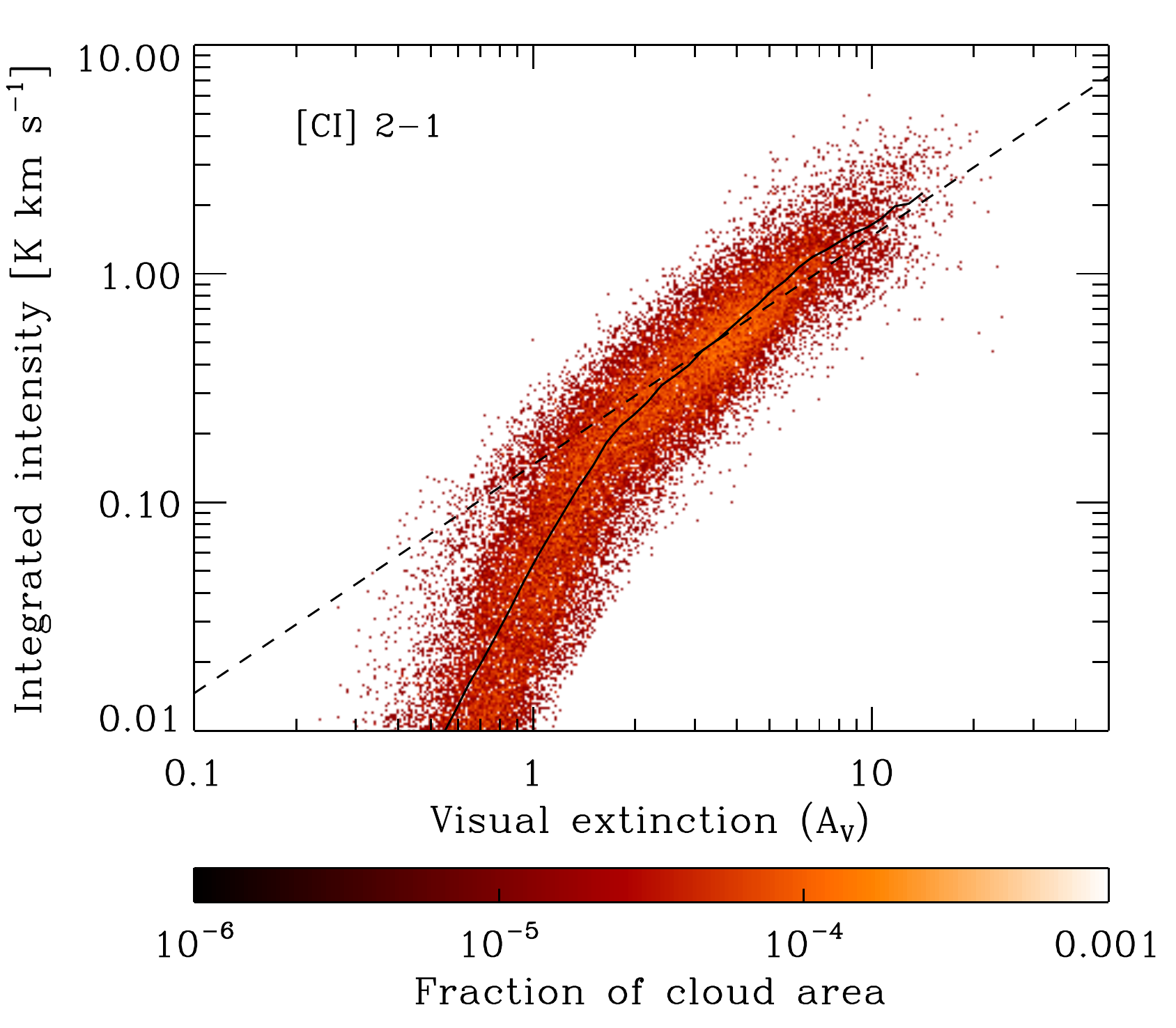}
\caption{As Figure~\ref{WCO_AV}, but for the \ci~$2 \rightarrow 1$ line. 
Note the change in the range of values displayed on the $y$-axis. \label{WCI_21_AV}}
\end{figure}

Since neutral atomic carbon is primarily found at lower gas densities and lower extinctions
than CO, we might reasonably expect \ci~emission emission to be a better tracer of $A_{\rm V}$
at low extinctions than CO. To test this, we plot in Figure~\ref{WCI_AV} the 2D PDF of the integrated 
intensity of the \ci~$1 \rightarrow 0$ line. We see that there is good linear correlation between
$W_{\rm CI, 1-0}$ and $A_{\rm V}$ for visual extinctions in the range $1.5 < A_{\rm V} < 7$. At higher
extinctions, the correlation becomes sub-linear, possibly because an increasing fraction of this
high extinction gas consists of CO rather than C, while at $A_{\rm V} < 1.5$, the correlation breaks
down owing to the photoionization of C. Comparing our results for C and $^{13}$CO, we see that
neutral atomic carbon is a better tracer of low column density material, while $^{13}$CO is to
be preferred at high column densities. We also note that although the difference between
$A_{\rm V} = 1.5$ and $A_{\rm V} = 3$ may seem minor, it actually corresponds to approximately
20\% of the total mass of the cloud, consistent with our results in Section~\ref{how}.

We have also examined how well the \ci~$2 \rightarrow 1$ line traces the total column density
(Figure~\ref{WCI_21_AV}). As with the lower energy transition, we see that there is a good linear correlation between 
the integrated intensity of the line and the visual extinction over a wide range of $A_{\rm V}$.  At low
$A_{\rm V}$, the correlation breaks down at $A_{\rm  V} \sim 2$, a slightly higher value than for 
the $1 \rightarrow 0$ line, but at high $A_{\rm V}$ the $2 \rightarrow 1$ line performs better than the
$1 \rightarrow 0$ line, with the correlation between $W_{\rm CI, 2-1}$ and $A_{\rm V}$ remaining 
approximately linear up to the highest extinctions probed in our study.

However, it is important to note that even at the highest $A_{\rm V}$ examined here, the $2 \rightarrow 1$ 
line has a much lower integrated intensity than the $1 \rightarrow 0$ line. This fact, plus the greater
atmospheric opacity at the frequency of the $2 \rightarrow 1$ line compared to the $1 \rightarrow 0$ line
means that in practice, the small improvement that one gets by observing the $2 \rightarrow 1$ line
is unlikely to justify the additional observing time required to map it to the same level of sensitivity.
Whether this is true for all GMCs or merely those with properties similar to those of the cloud modelled
here is an interesting question, but is beyond the scope of the present study.

In the regime where $W_{\rm CI, 1-0}$ scales linearly with $A_{\rm V}$, the relationship between 
the two can be described by the following empirical fit
\begin{equation}
W_{\rm CI, 1-0} = 0.75 A_{\rm V}  \: {\rm K \: km \: s^{-1}}.  \label{emp_fit}
\end{equation}
If we apply this fit to each pixel in our synthetic \ci~emission map, we can generate an estimate
of $A_{\rm V}$ and from that can estimate the cloud mass. The estimate that we obtain in this 
way is approximately 80\% of the true mass of the cloud; the ``missing'' 20\% is the low 
$A_{\rm V}$ material that is not traced well by \ci.

Finally, we note that our finding here that \ci~emission is a relatively good tracer of the column
density over a significant range of $A_{\rm V}$ was already anticipated to some extent 
observationally. For example, in their study of NGC~891, \citet{gp00} found a good spatial
correlation between \ci~$1 \rightarrow 0$ emission and 1.3~mm dust continuum emission,
which they interpreted as evidence that \ci~emission is a good tracer of the gas column
density.

\subsubsection{Tracing the H$_2$ column density}
In the previous section, we saw that one of the main drawbacks with using \ci~to trace the
structure of the cloud is that it becomes very faint along low $A_{\rm V}$ lines-of-sight, and
hence misses material at the edges of the cloud. However, this low density, low extinction
gas will largely be composed of H{\sc i}, rather than H$_{2}$, and so it is  worthwhile
to examine whether there is a better correlation between the \ci~integrated intensity and the 
H$_{2}$ column density than there is between the \ci~integrated intensity and the total column
density. 

\begin{figure}
\includegraphics[width=3.3in]{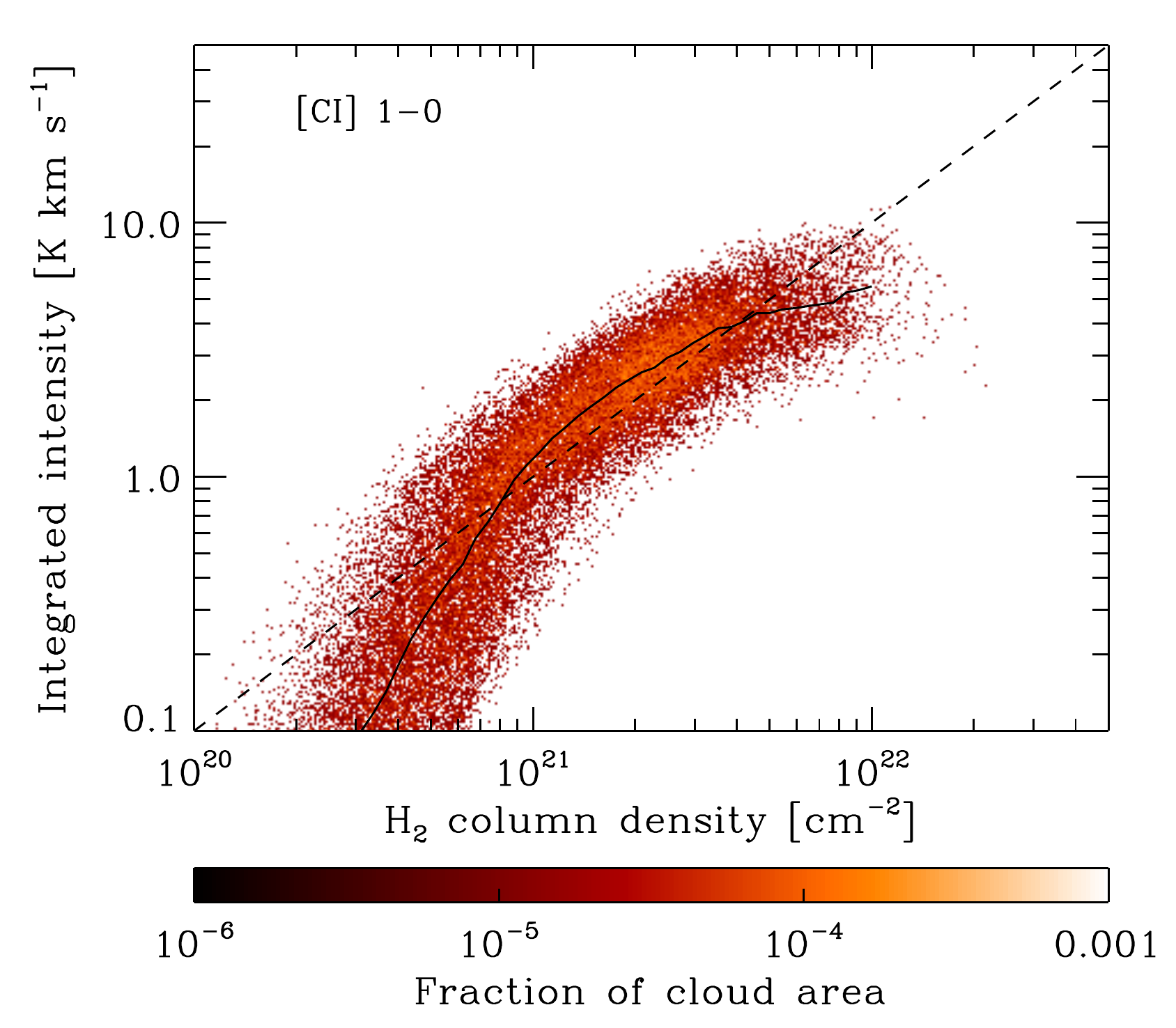}
\caption{Two-dimensional PDF of the integrated intensity in the \ci~$1 \rightarrow 0$ line,
plotted as a function of the H$_{2}$ column density, $N_{\rm H_{2}}$. The dashed line indicates 
a linear relationship between $W_{\rm CI, 1-0}$ and $N_{\rm H_{2}}$ and is there to guide the eye. 
The solid line shows the geometric mean of $W_{\rm CI, 1-0}$ as a function of $N_{\rm H_{2}}$.
The points are colour-coded according to the fraction of the total cloud area that they 
represent. \label{WCI_NH2}}
\end{figure}

In Figure~\ref{WCI_NH2}, we show a 2D PDF of  $W_{\rm CI, 1-0}$ as a function of the H$_{2}$ column
density, $N_{\rm H_{2}}$. The behaviour we see in this plot is qualitatively similar to the behaviour
of the $W_{\rm CI, 1-0}$--$A_{\rm V}$ relationship. \ci~emission traces H$_{2}$ well at intermediate
H$_{2}$ column densities, $5 \times 10^{20} < N_{\rm H_{2}} < 10^{22} \: {\rm cm^{-2}}$, but underestimates
the molecular gas column density at high column densities, owing to the optical depth of the \ci~line.
The correlation between \ci~emission and H$_{2}$ column density also breaks down at low column 
densities, but it is clear from comparing Figures~\ref{WCI_AV} and \ref{WCI_NH2} that \ci~is a better
tracer of the H$_{2}$ column density than of the total column density in this regime.

The results discussed here agree well with those from the recent study of \citet{offner14}, which also
examined how well \ci~emission traces H$_{2}$ column density in turbulent clouds. \citet{offner14}
studied how a quantity they term $X_{\rm CI}$, namely
\begin{equation}
X_{\rm CI} = \frac{N_{\rm H_{2}}}{W_{\rm CI, 1-0}},
\end{equation}
varies as a function of $N_{\rm H_{2}}$ within their simulated clouds. They found that for H$_{2}$
column densities between a few times $10^{20} \: {\rm cm^{-2}}$ and $10^{22} \: {\rm cm^{-2}}$,
the value of $X_{\rm CI}$ has little dependence on the H$_{2}$ column density, implying that in
this regime, $W_{\rm CI, 1-0} \propto N_{\rm H_{2}}$, consistent with the results that we present
here. 

\citet{offner14} also computed the mean value of $X_{\rm CI}$ for each of their simulations.
For their run with $G_{0} = 1.0$, which is probably the closest analogue to our simulated cloud,
they obtained a value $X_{\rm CI} = 1.1 \times 10^{21} \: {\rm cm^{-2} \: K^{-1} \: km^{-1} \: s}$.
For comparison, we find a value $X_{\rm CI} = 1.01 \times 10^{21} \: {\rm cm^{-2} \: K^{-1} \: km^{-1} \: s}$,
within 10\% of their result.

\subsubsection{Measuring the column density of C}
\label{ccol} 
In Section~\ref{totcol}, we saw that \ci~emission is a good tracer of the total column density of 
the cloud over a wide range of visual extinctions. However, the relationship we derived between
$W_{\rm CI, 1-0}$ and $A_{\rm V}$ (Equation~\ref{emp_fit}) was completely empirical, and hence
we cannot be confident that the same relationship will hold for other molecular clouds, whether
simulated or real. Ideally, what we would like to be able to do is to infer the column density of
neutral atomic carbon, $N_{\rm C}$, from our observations and then to infer $A_{\rm V}$ from
$N_{\rm C}$. However, this is not straightforward, as we demonstrate below.

To begin with, we face the issue of how to infer $N_{\rm C}$ given the intensities of the two
\ci~transitions. One approach is to assume that both lines are optically thin. In this case, 
the column densities of neutral atomic carbon atoms in the two excited levels follow directly
from the integrated intensities of the $1 \rightarrow 0$ and $2 \rightarrow 1$ transitions. To
determine the column density of carbon in the $J = 0$ level, we need one additional piece
of information: the excitation temperature of the $J = 1$ level. Given this, the fact that,
by definition,
\begin{equation}
\frac{N_{1}}{N_{0}} = \frac{g_{1}}{g_{0}} \exp \left(-\frac{E_{10}}{kT_{\rm ex}} \right)  \label{tex_defn}
\end{equation}
allows us to solve for $N_{0}$ given $N_{1}$ and $T_{\rm ex}$. The total column density
of carbon then follows trivially if we sum the column densities of all three levels.

Although we have seen already that the \ci~$1 \rightarrow 0$ line does not remain
optically thin over the entire range of column densities probed in our model cloud, it is
nevertheless interesting to examine how well we can recover $N_{\rm C}$ in the case
where we ignore any optical depth effects, and how large an error we make in the total
neutral atomic carbon abundance we derive for the cloud when using this simplified 
approach.
% since this procedure is often used to analyze real observations 

\begin{figure*}
\includegraphics[width=0.49\textwidth]{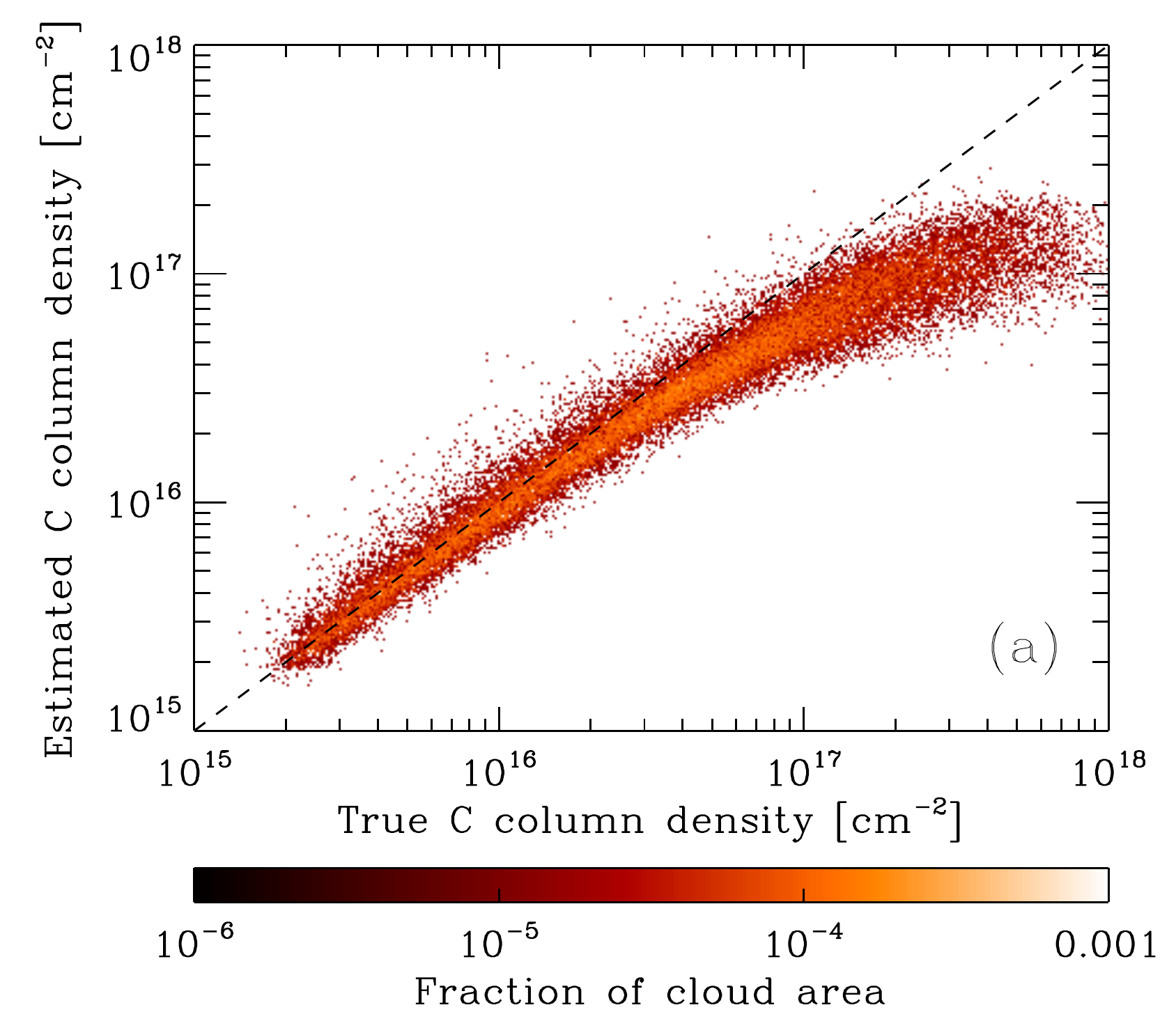}
\includegraphics[width=0.49\textwidth]{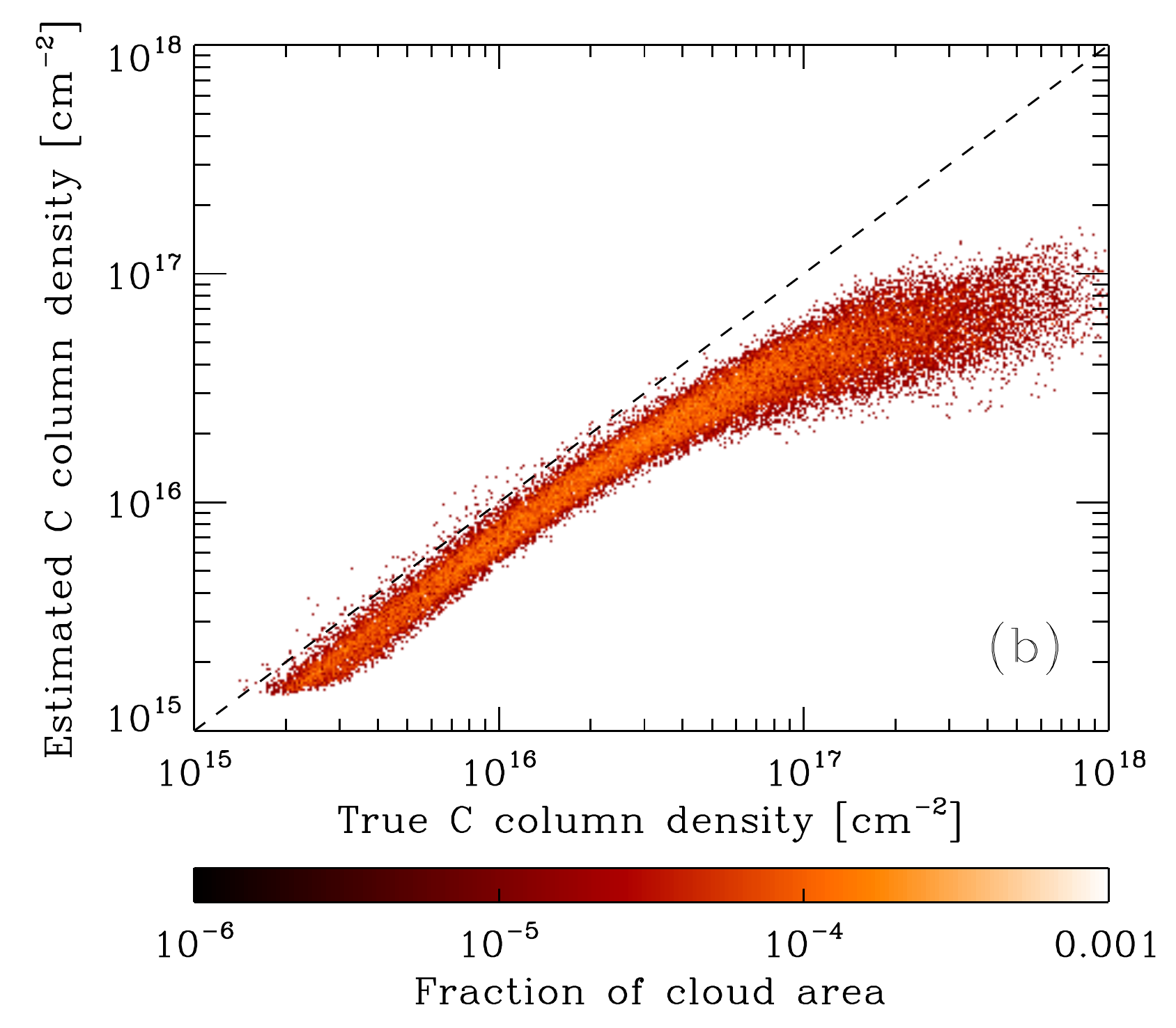}
\includegraphics[width=0.49\textwidth]{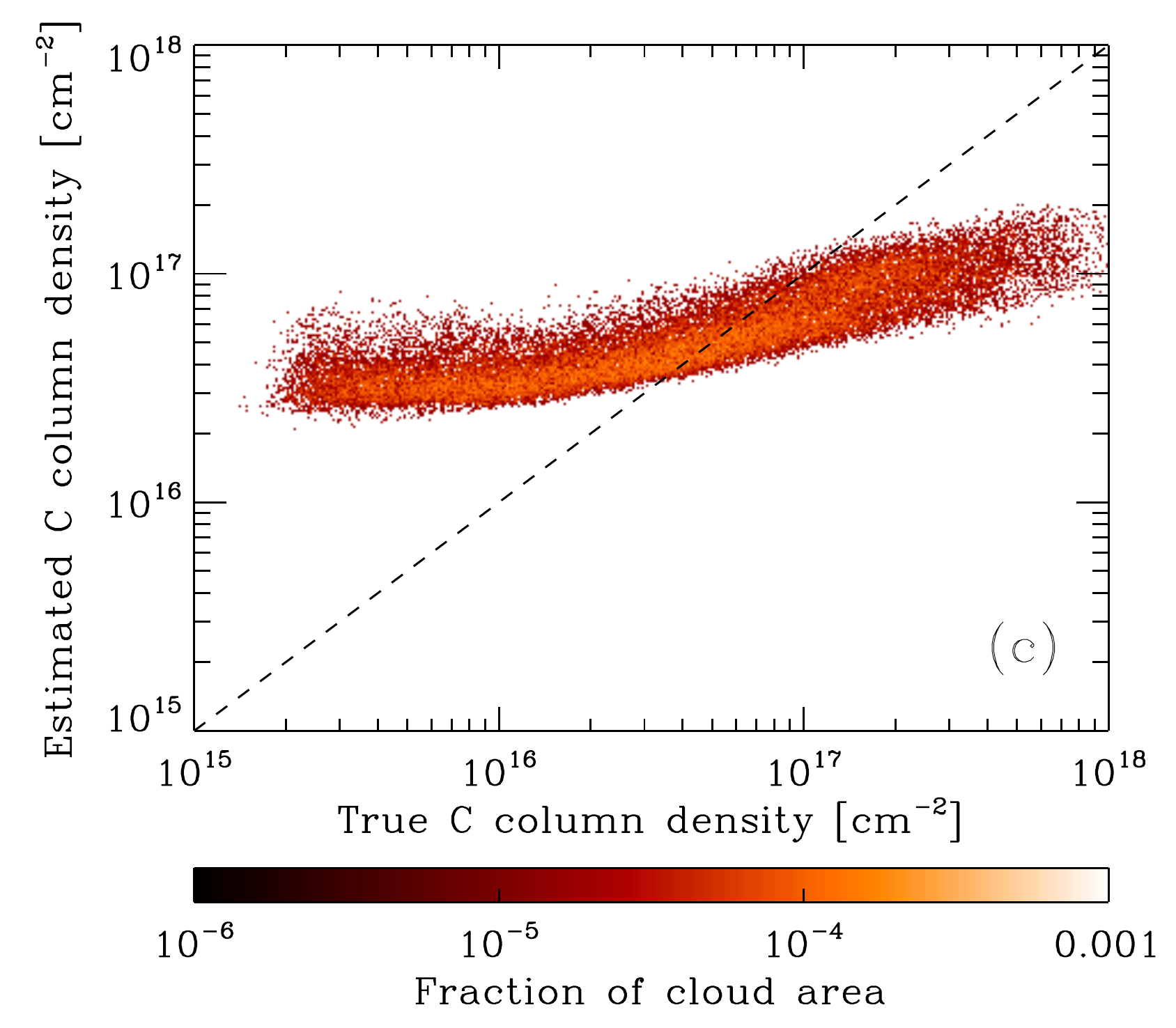}
\includegraphics[width=0.49\textwidth]{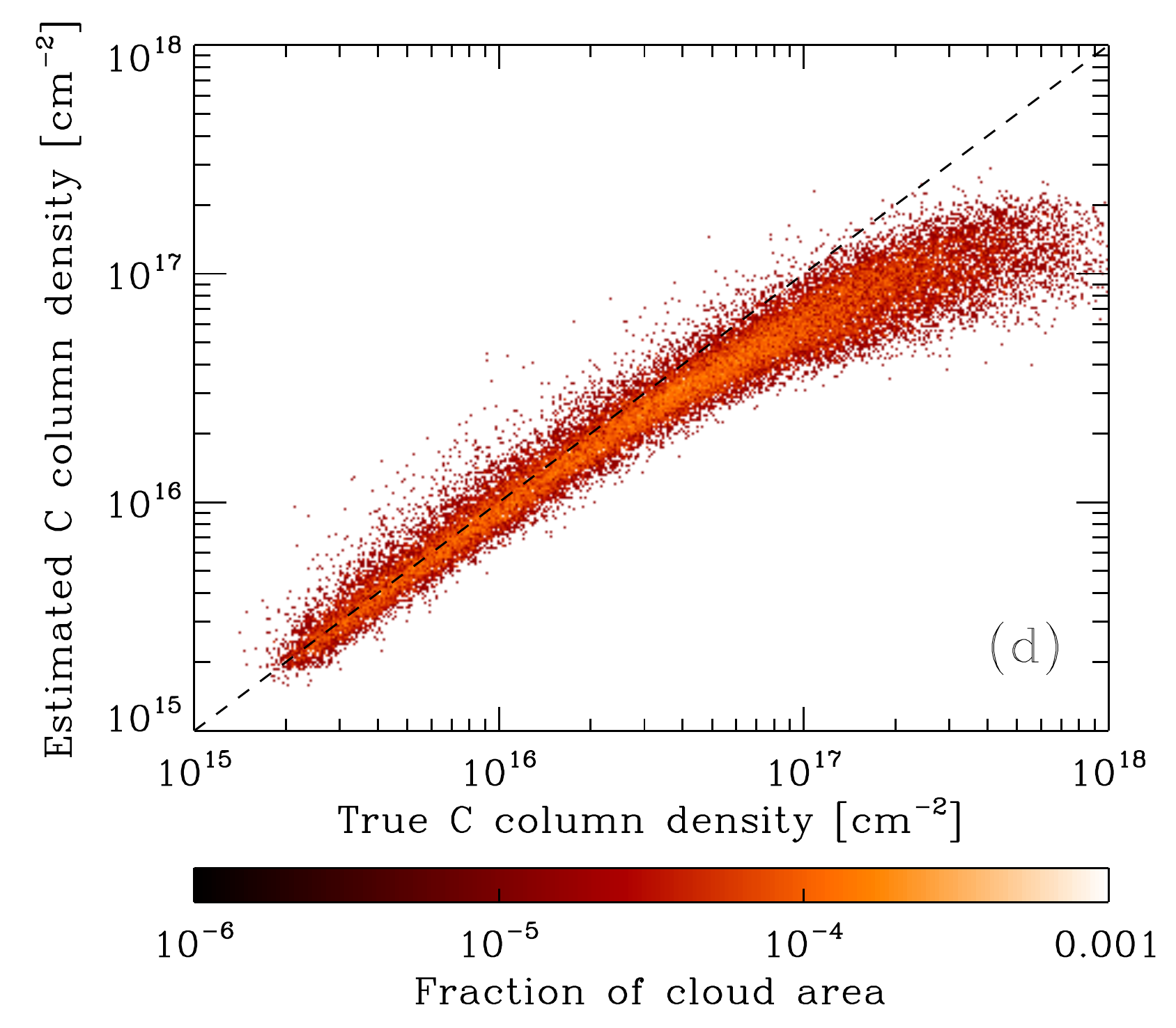}
\caption{(a) 2D PDF showing the estimated column density of neutral atomic carbon as 
a function of the true column density. The estimated column density was computed using the
procedure described in Section~\ref{ccol}, with the excitation 
temperature for each line of sight taken to be a weighted mean of the actual excitation temperature. 
At carbon column densities greater than a few times $10^{16} \: {\rm cm^{-2}}$, we underestimate
$N_{\rm C}$ because we do not account for the effects  of line opacity.
As in Figures~\ref{WCO_AV}--\ref{WCI_AV}, the colour-coding corresponds to the fraction of the total
area of the cloud represented by each point.
(b) As panel (a), but using the estimate of $T_{\rm ex}$ given by Equation~\ref{tx_sch}.
We see that because Equation~\ref{tx_sch} overestimates the excitation temperature 
associated with the $1 \rightarrow 0$ transition, our derived values of $N_{\rm C}$ systematically
underestimate the true values.
(c) As panel (a), but using the estimate of $T_{\rm ex}$ given by Equation~\ref{tx_dick}.
This estimate performs very badly: at low column densities, we underestimate $T_{\rm ex}$ and 
hence overestimate $N_{\rm C}$, while at high column densities, our values for $T_{\rm ex}$
are much more accurate but the neglect of optical depth effects means that we now 
underestimate $N_{\rm C}$.
(d) As panel (a), but using the estimate of $T_{\rm ex}$ given by Equation~\ref{tx_dick} 
for lines of sight with $W_{\rm CI, 1-0} > 3 \: {\rm K} \: {\rm km} \: {\rm s^{-1}}$ and adopting 
a constant value for the fainter lines of sight that is equal to the mean of the values 
determined for the bright sight-lines. \label{NCI_est}}
\end{figure*}

In Figure~\ref{NCI_est}a, we show a 2D PDF of the estimate of
$N_{\rm C}$ that we derive if we assume optically thin \ci~emission
($N_{\rm C, est}$)
versus the true value of $N_{\rm C}$ measured from the simulation. As before,
we show values only for those lines of sight with $W_{\rm CI, 1-0} > 0.1 \: {\rm K
\: km \: s^{-1}}$. To compute our column density estimate, we need to know the
mean excitation temperature along each line of sight. For this first estimate,
we set $T_{\rm ex} = T_{\rm ex, mean}$, the weighted mean excitation temperature given by 
Equation~\ref{tex_true_av} for each line of 
sight. We see from the Figure that for atomic carbon column densities of less than a few 
times $10^{16} \: {\rm cm^{-2}}$, there is very good agreement between $N_{\rm C, est}$
and $N_{\rm C}$. At higher column densities, however, $N_{\rm C, est}$ falls below
$N_{\rm C}$; i.e.\ it becomes an underestimate. This is easy to understand as
a consequence of the growing optical depth of the \ci~lines with increasing 
$N_{\rm C}$, an effect which is not accounted for in our simple estimate.

In practice, $T_{\rm ex, mean}$ is not an observable quantity. Therefore, in 
the other panels of Figure~\ref{NCI_est} 
we explore what happens if we use the different estimates of
$T_{\rm ex}$ that we derived in Section~\ref{calc_tex} to compute our estimate of $N_{\rm C}$. 
In Figure~\ref{NCI_est}b, we show the result that we obtain when we
set $T_{\rm ex} = T_{\rm ex, est}$, the excitation temperature estimate given by 
Equation~\ref{tx_sch}. We see that although there is still a good correlation between $N_{\rm C}$
and $N_{\rm C, est}$ at low column densities, our column density estimate is systematically
offset from the true value by a factor of approximately 0.7. This behaviour is consistent with our 
earlier finding that $T_{\rm ex, est}$ is an overestimate of the true excitation temperature, as this 
means that our derived value for $N_{0}$ is too small.

In Figure~\ref{NCI_est}c, we show what happens if we use 
$T_{\rm ex, est, 2}$,  the estimate of $T_{\rm ex}$ that we obtained from the optically 
thick analysis. We see that in this case, there is very poor agreement between $N_{\rm C, est}$ 
and $N_{\rm C}$. This is not unexpected, as in this approach we assume mutually contradictory
things for the \ci~emission. Our derivation of the excitation temperature assumes that \ci~is
optically thick, but our expression for $N_{\rm C}$ assumes that the emission is optically
thin. Along sight-lines where the emission is actually optically thin, we derive excitation
temperatures that are too small and hence predict values for the column density of carbon atoms in 
the $J = 0$ level that are much higher than the true value, leading to values of $N_{\rm C, est}$
that are also far too large. On the other hand, along sight-lines where the emission is
actually optically thick, our estimated values of $T_{\rm ex}$ are far more reasonable, 
but our expression for $N_{\rm C}$ is no longer accurate, because it does not account for
the effects of line opacity. In this regime, we therefore underestimate $N_{\rm C}$.

Finally, in Figure~\ref{NCI_est}d, we show the results we obtain
if we compute $T_{\rm ex}$ for lines of sight with $W_{\rm CI, 1-0} > 3 \: {\rm K} \:
{\rm km} \: {\rm s^{-1}}$ using Equation~\ref{tx_dick}, and adopt a constant value
for the fainter lines of sight that is equal to the mean of the values determined for
the $W_{\rm CI, 1-0} > 3 \: {\rm K} \: {\rm km} \: {\rm s^{-1}}$ sight-lines. We see that this strategy performs well at low $N_{\rm C}$:
estimates for $N_{\rm C}$ along individual sight-lines can differ from the true values
by as much as 50\%, but there is no systematic bias towards low or high values.
At high $N_{\rm C}$, optical depth effects once again cause this method to break 
down.

A number of studies in the literature \citep[e.g.][]{ikeda02,sch03}
attempt to correct for the effects of \ci~line opacity by including a factor 
\begin{equation}
f_{\rm corr, J \rightarrow J-1} = \frac{\tau_{\rm C, J \rightarrow J-1}}{1 - e^{-\tau_{\rm C, J \rightarrow J-1}}},  
\label{fcorr}
\end{equation}
in their conversion from integrated intensity to column density, where $\tau_{\rm C, J \rightarrow J-1}$ is
the optical depth of the transition from level $J$ to level $J-1$. For the column density of
the $J = 1$ level, one then has
\begin{equation}
N_{1} = \frac{8\i k \nu_{10}^{2}}{h c^{3} A_{10}} f_{\rm corr, 1\rightarrow 0} \int T_{\rm B, 1-0} {\rm d}v,
\label{n1_corr}
\end{equation}
where $T_{\rm B, 1-0}$ is the brightness temperature of the $1 \rightarrow 0$ line. A similar
expression can also be written down for the column density of the $J = 2$ level.
To compute this correction, it is necessary to estimate the optical depth in the
relevant transition. In the case of the $1 \rightarrow 0$ transition, this is typically done
using the expression
\begin{equation}
\tau_{\rm C} = - \log \left(1 - kT_{\rm B, 1-0, max} \frac{\exp(E_{10}/kT_{\rm ex}) - 1}{E_{10}} \right), \label{tau_est}
\end{equation}
where $T_{\rm B, 1-0, max}$ is the maximum value of $T_{\rm B, 1-0}$ observed along
the line of sight. However, the derivation of Equation~\ref{tau_est}
follows from the relationship between $T_{\rm B, 1-0}$ and $T_{\rm ex}$
that we previously encountered in Section~\ref{calc_tex}:
\begin{eqnarray}
\frac{kT_{\rm B, 1-0, max}}{E_{10}} & = & \left[\frac{1}{e^{E_{10} / kT_{\rm ex}} - 1} 
- \frac{1}{e^{E_{10} / kT_{\rm bg}} -1} \right] \nonumber \\
& & \times \left(1 - e^{-\tau} \right).
\end{eqnarray}
The procedure that gives us our best estimate of $T_{\rm ex}$ in the optically 
thin limit is based on taking the $\tau \rightarrow \infty$ limit of this expression, 
and hence does not allow us to compute a meaningful opacity correction.

\begin{figure}
\includegraphics[width=0.49\textwidth]{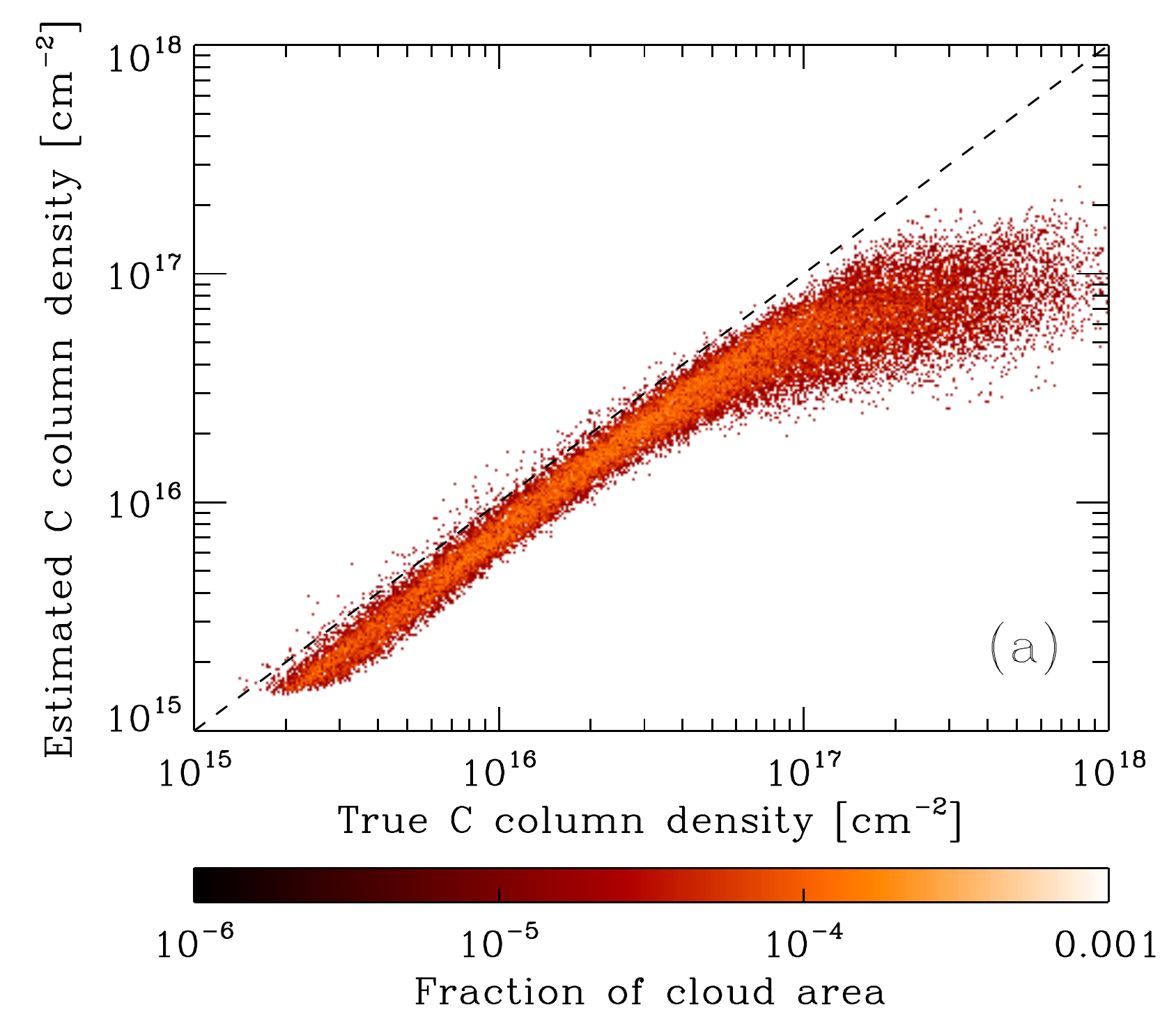}
\includegraphics[width=0.49\textwidth]{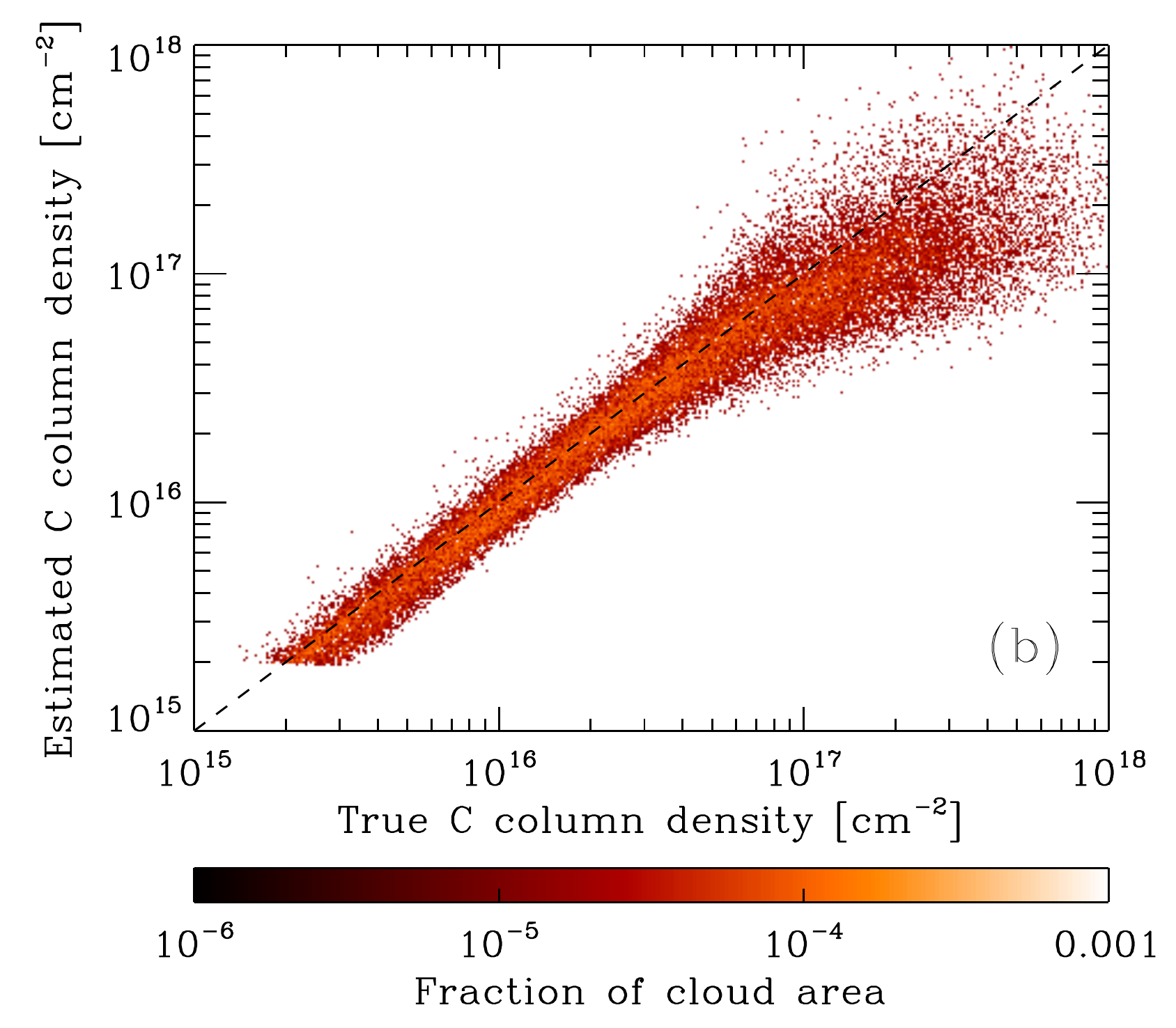}
\caption{(a) 2D PDF showing the estimated column density of neutral atomic carbon 
as a function of the true column density, computed using the estimate of $T_{\rm ex}$ 
given by Equation~\ref{tx_sch} and corrected for opacity using 
Equations~\ref{fcorr} and \ref{tau_est}. As in the previous figures, 
the colour-coding corresponds to the fraction of the total
area of the cloud represented by each point.
(b) As (a), but using an estimate of $T_{\rm ex}$ derived from the $^{12}$CO peak 
brightness temperature, as described in more detail in the text.
\label{NCI_est_tau}}
\end{figure}

If we instead use $T_{\rm ex, est}$ as our estimate of the excitation temperature,
then we can compute physically plausible values of $\tau$ using 
Equation~\ref{tau_est}. However, the resulting correction does not significantly
improve the agreement between the real and estimated column densities in the
optically thick regime, as can be seen by comparing Figures~\ref{NCI_est}b and
\ref{NCI_est_tau}a. In view of this, we have also examined the performance of
an alternative strategy that is often used to derive $T_{\rm ex}$ and $\tau$
for \ci. This makes use of the $^{12}$CO $J = 1 \rightarrow 0$ emission map 
(assuming one is available). The $^{12}$CO emission is assumed to be
optically thick, and the following expression is used to compute 
the excitation temperature of the CO molecules \citep{dick78}
\begin{equation}
T_{\rm ex, CO} = \frac{5.5}{\ln \left[ 1 + 5.5 / (T_{\rm B, 1-0, max} + 0.82) \right]}. \label{tx_co}
\end{equation}
This is directly analogous to Equation~\ref{tx_dick}, but uses parameters appropriate
for the $^{12}$CO $J = 1 \rightarrow 0$ line rather than the \ci~$1 \rightarrow 0$ line.
The excitation temperature of the carbon atoms is then simply taken to be
the same as this value. One potential issue with this approach is that the extent of the 
detectable CO emission is not necessarily the same as that of the detectable 
\ci~emission. We avoid this problem by once again computing sightline-specific
values of $T_{\rm ex}$ only for those sight-lines with $W_{\rm CI, 1-0} > 3 \: {\rm K} \:
{\rm km} \: {\rm s^{-1}}$, since this subset of sight-lines all have significant 
$^{12}$CO emission associated with them. For the remaining sight-lines, we adopt
a constant value of $T_{\rm ex}$ equal to the mean of the values determined for
the bright sight-lines.

The results that we obtain with this approach 
are plotted in Figure~\ref{NCI_est_tau}. We see that the column
density estimates yielded by this technique agree well with the true values
for column densities up to $N_{\rm C} \sim 7 \times 10^{16} \: {\rm cm^{-2}}$.
At higher column densities, this technique does not completely correct for
the effects of the opacity, and so we still underestimate the true column
density along most sightlines in this regime. Nevertheless, even here
we see significant improvement in comparison to the uncorrected case,
albeit at the cost of substantially increased scatter.

\begin{figure}
\includegraphics[width=0.49\textwidth]{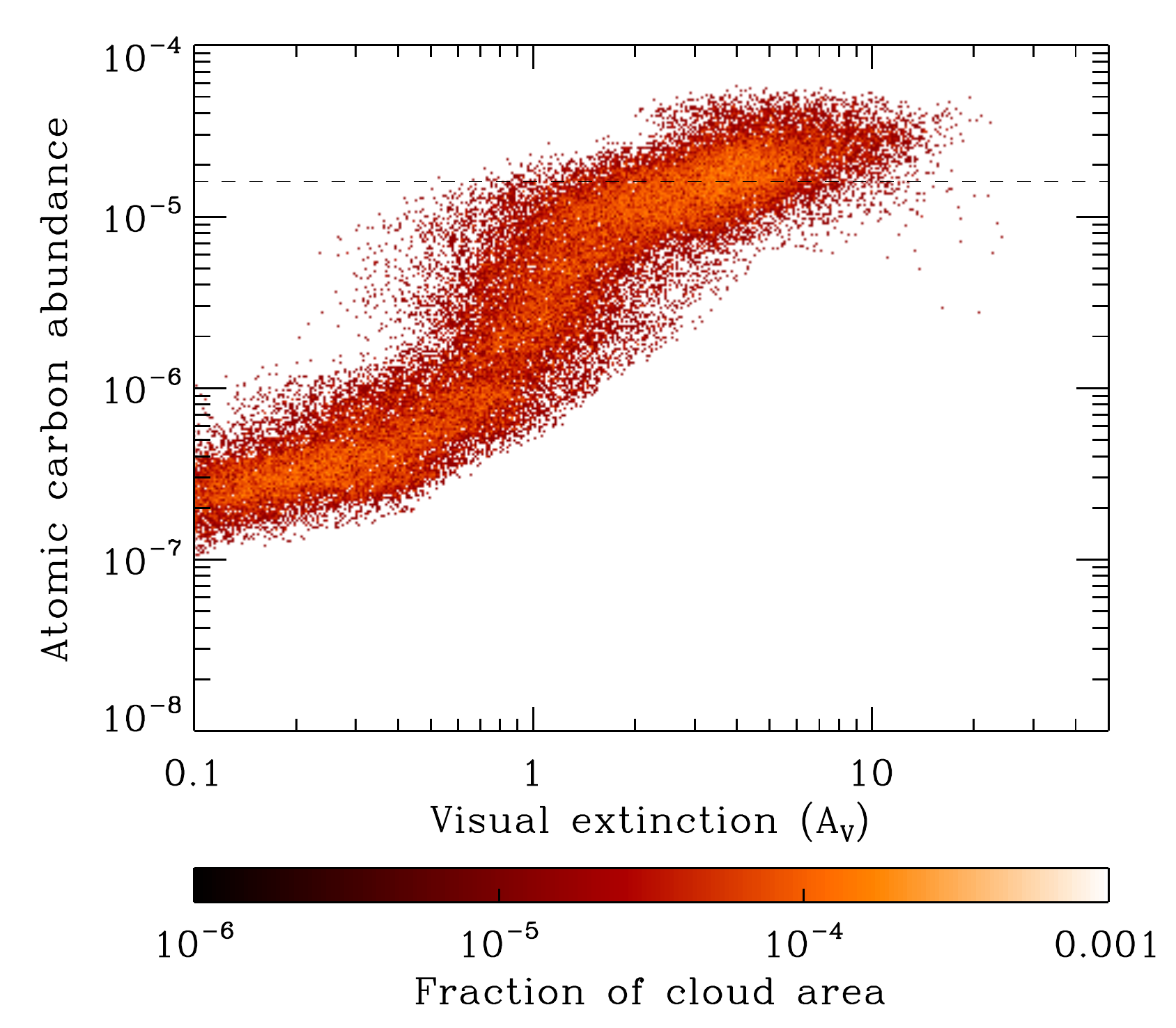}
\caption{2D PDF showing the line-of-sight averaged fractional
abundance of neutral atomic carbon as a function of the visual extinction. 
The colour-coding corresponds to the fraction of the total
area of the cloud represented by each point. The horizontal
dashed line indicates the mean fractional abundance in the
cloud as a whole. \label{XCI_AV}}
\end{figure}

We have also examined how much of the atomic carbon in the cloud we fail
to detect owing to the line opacity, and how good our various approximations
are at recovering this missing material. The total atomic carbon content that
we infer if we use the actual column-averaged excitation temperatures but
do not correct for opacity is approximately 47\% of the true value, i.e.\ we
miss around half of the atomic carbon. Using excitation temperatures derived
from the \ci~line ratio and correcting for opacity actually makes the situation
worse: the small improvement at high $N_{\rm C}$ is swamped by the
effects of the systematic underestimate at low $N_{\rm C}$, and we recover
only 37\% of the total atomic carbon. Finally, using values for $T_{\rm ex}$
estimated from the $^{12}$CO emission does improve matters, but even
in this case we recover only 65\% of the total atomic carbon.
An important conclusion that we can draw from this result is that estimates 
in the literature of the atomic carbon content of real molecular clouds that 
are based on observations of the \ci~emission will typically be too small, 
possibly by as much as a factor of two.

Given an estimate of $N_{\rm C}$, a further problem that we face is how to convert
from this to the total column density. Along a given line of sight, $N_{\rm C}$ and
$N_{\rm H, tot}$, the total hydrogen column density, are related by
$N_{\rm H, tot} = N_{\rm C} \bar{x}_{\rm C, los}^{-1}$, where $\bar{x}_{\rm C, los}$
is the fractional abundance of atomic carbon weighted by mass and averaged along
the line of sight. The mass-weighted average for the cloud as a whole is given by
$\bar{x}_{\rm C} = 1.60 \times 10^{-5}$, but as we have already seen, the local abundance
is not constant, but instead varies as a function of density and position
in the cloud.  

In Figure~\ref{XCI_AV}, we show how  $\bar{x}_{\rm C, los}$
varies as a function of the visual extinction.
We see that there is a systematic increase in the mean abundance as we move
to higher extinctions. This is particularly pronounced around $A_{\rm V} \sim 1$,
where we first start to probe regions that are well-shielded from the ISRF, but
even at much higher $A_{\rm V}$ we still see an increase. Using the mean
value for the whole cloud as a proxy for $\bar{x}_{\rm C, los}$ is a reasonable
approximation only for visual extinctions $\sim$ a few, but would lead one to
significantly underestimate the total column density along diffuse lines of
sight and overestimate it along dense sight-lines. Moreover, $\bar{x}_{\rm C}$
is not an observable quantity (unless the total column 
density of the cloud is already known), but must instead be estimated using a chemical model
of the cloud, such as that included in our hydrodynamical simulation. This
introduces yet more uncertainty into the final column density estimate, and
leading us to conclude that this technique for estimating $N_{\rm H, tot}$
is likely in practice to be much less accurate than simply using the correlation
between $W_{\rm CI, 1-0}$ and $A_{\rm V}$ discussed in the previous section.

\subsection{Comparison of \ci~and $^{13}$CO emission}
\label{13co}
We have seen in the previous sub-sections that there are some interesting similarities
between \ci~$1 \rightarrow 0$ emission and $^{13}$CO $J = 1 \rightarrow 0$ emission.
Both tracers are optically thin for a broad range of visual extinctions, and both trace a very 
similar fraction of the cloud mass when  we restrict our attention to regions where the 
integrated intensity exceeds a few K~km~s$^{-1}$. It is therefore worthwhile to directly
compare these two tracers in more detail, so that we can better understand their
strengths and weaknesses.

\begin{figure*}
\includegraphics[width=0.49\textwidth]{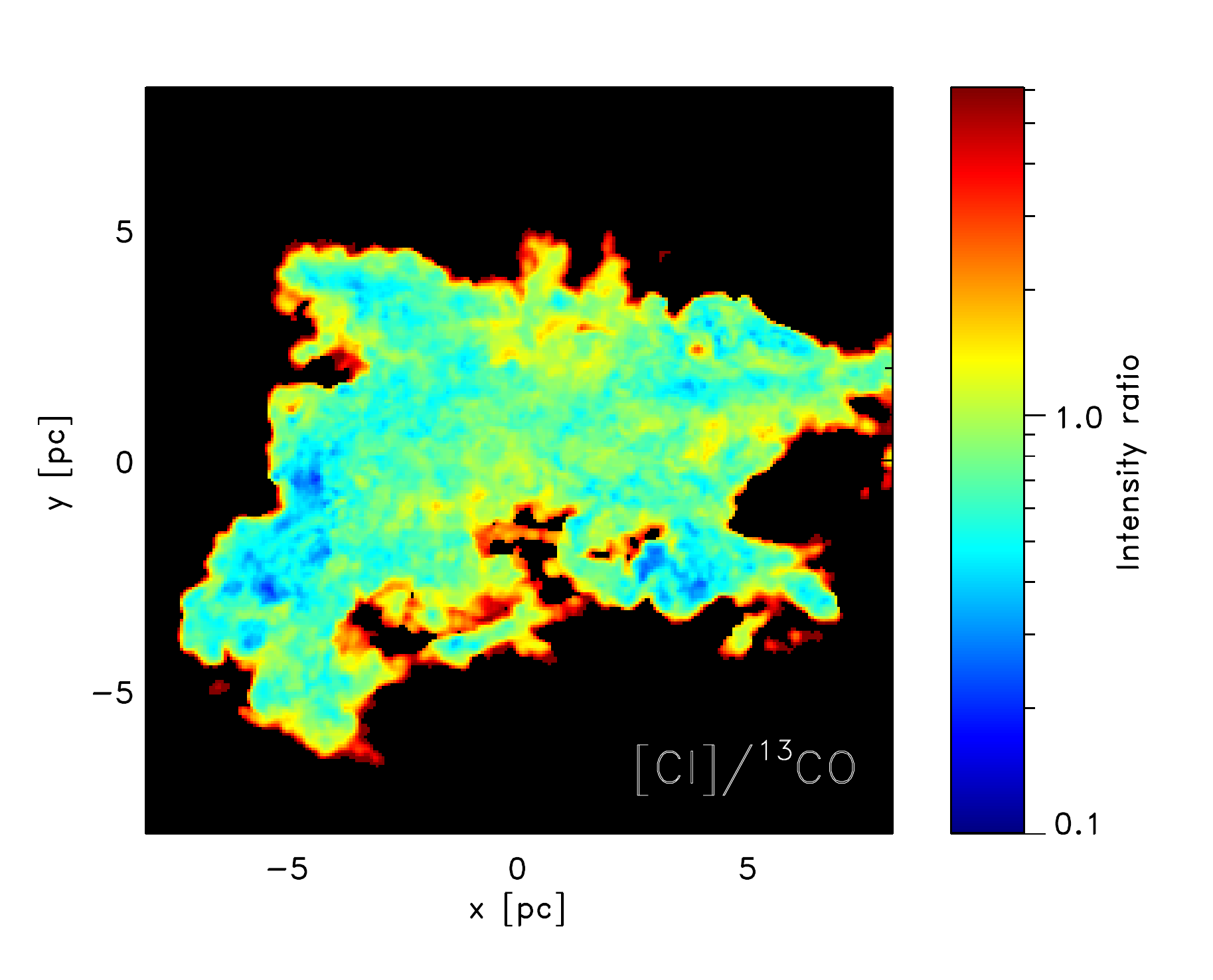}
\includegraphics[width=0.49\textwidth]{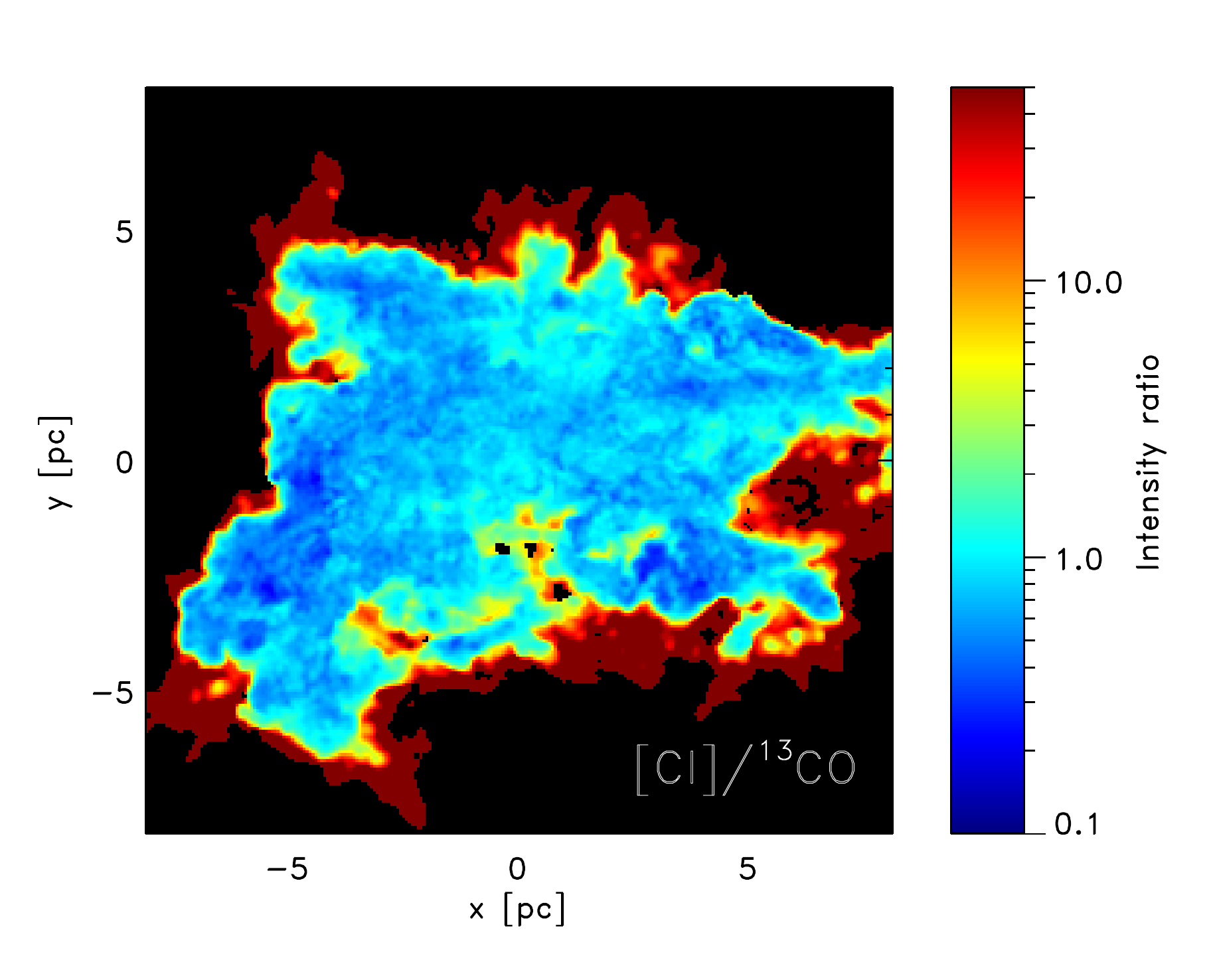}
\caption{Ratio of the integrated intensities of the \ci~$1 \rightarrow 0$
transition and the $^{13}$CO $J = 1 \rightarrow 0$ transition. In the
left-hand panel, we show the values only for those lines of sight
with $W_{\rm 13CO} \geq 0.1 \: {\rm K \: km \: s^{-1}}$, which serves
to emphasize the near uniformity of the intensity ratio within the CO-bright
region. In the right-hand panel, we show how the intensity ratio behaves
when we remove this restriction. In this case, we have chosen a different colour
scale that allows us to highlight the region at the edges of the cloud
where there is still bright \ci~emission but no significant $^{13}$CO
emission, resulting in a large value for the intensity ratio. Note that for
practical reasons, we limit the maximum displayed value of the
ratio to 50. \label{ratio_maps}}
\end{figure*}

\begin{figure*}
\includegraphics[width=0.49\textwidth]{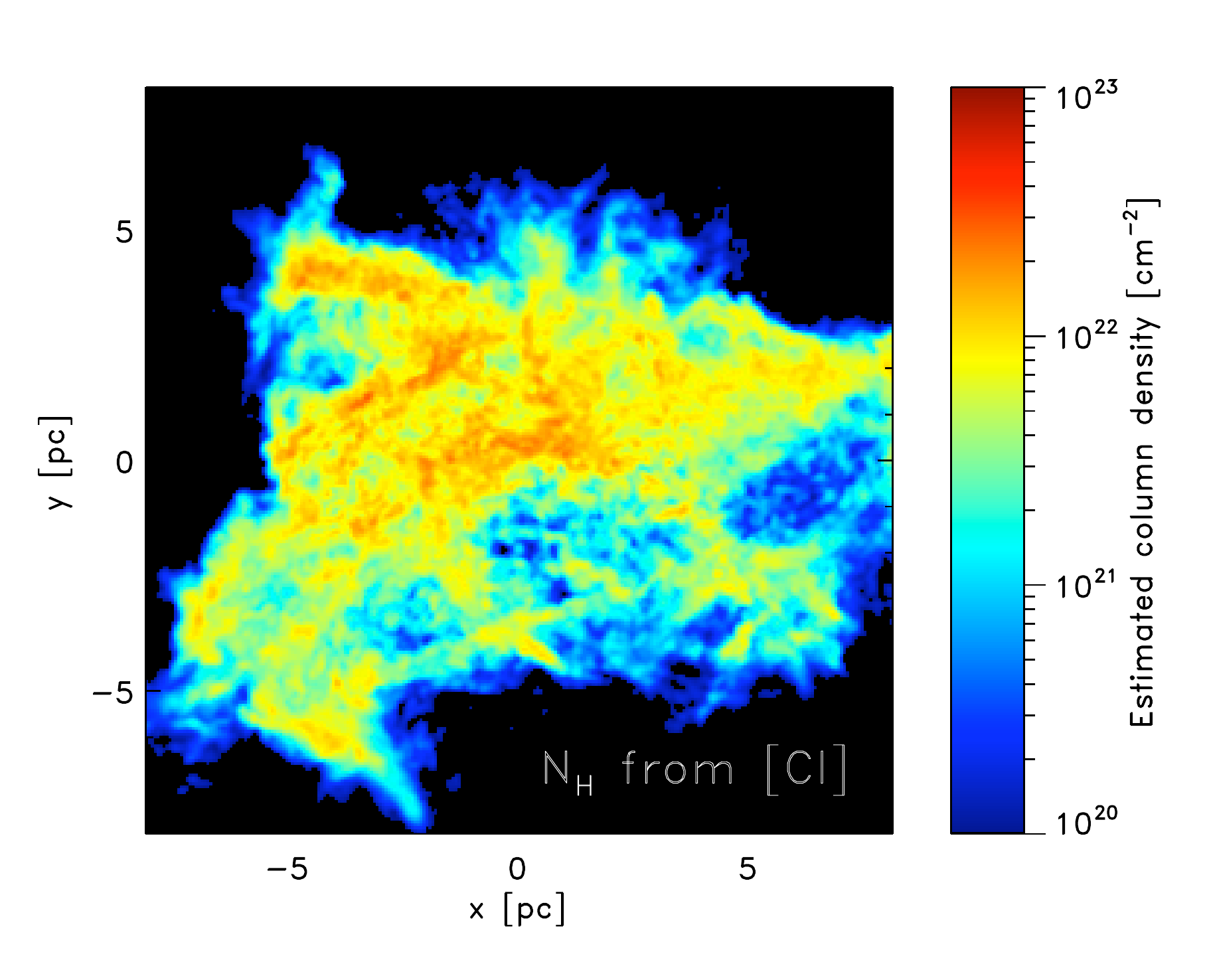}
\includegraphics[width=0.49\textwidth]{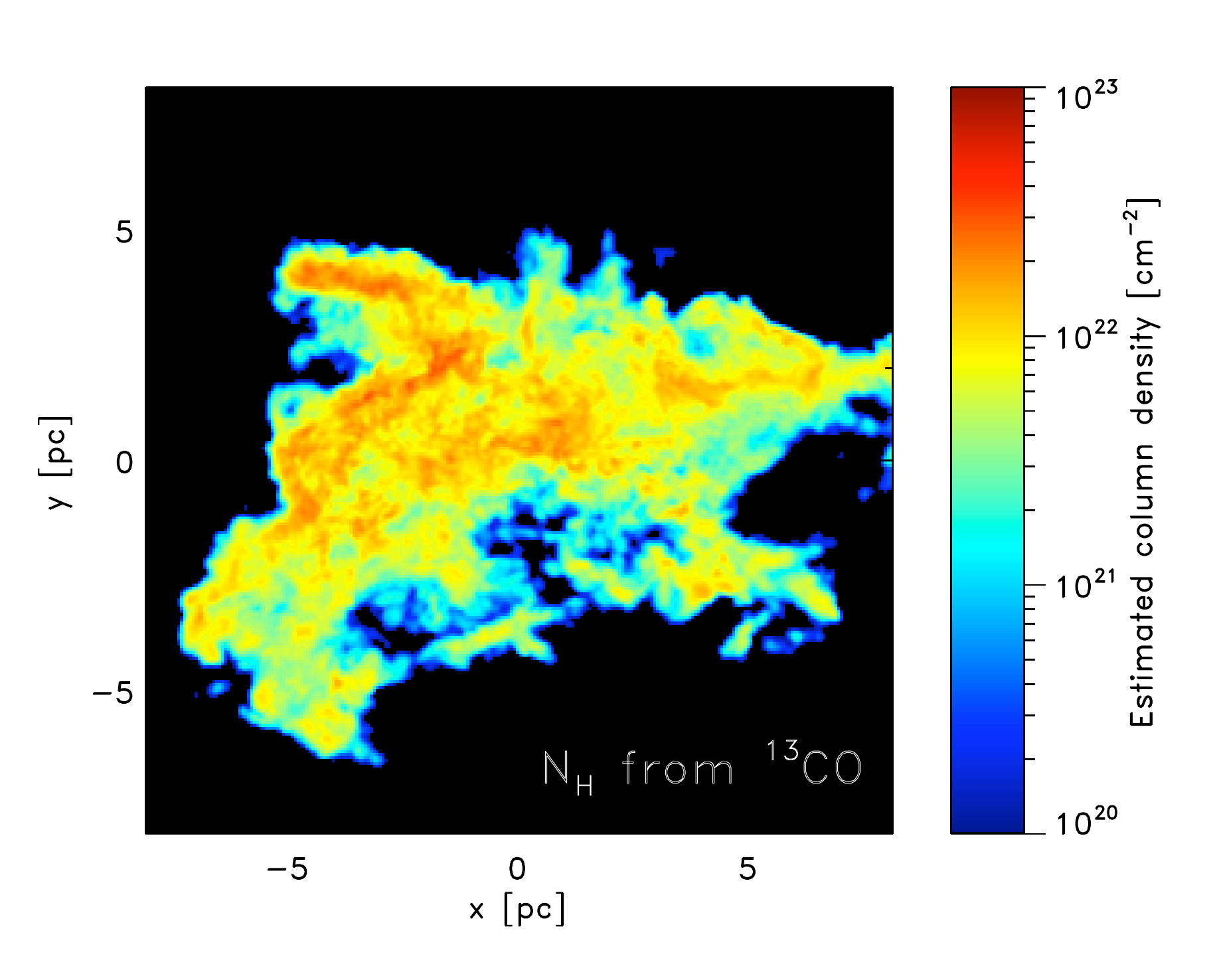}
\caption{{\it Left}: Reconstruction of the column density distribution, based on the
empirical relationship between $W_{\rm CI, 1-0}$ and $A_{\rm V}$ quantified
by Equation~\ref{emp_fit}. {\it Right}: As in the left-hand panel, but this time 
using the empirical relationship between $W_{\rm 13CO}$ and $A_{\rm V}$ given by
Equation~\ref{emp_2}. \label{recon}}
\end{figure*}

In Figure~\ref{ratio_maps}, we examine how the integrated intensity ratio
$R \equiv W_{\rm CI, 1-0} / W_{\rm 13CO}$ varies over the cloud. In the left-hand panel,
we compute this ratio only for lines-of-sight for which both $W_{\rm CI, 1-0}$ and
$W_{\rm 13CO}$ exceed $0.1 \: {\rm K \: km \: s^{-1}}$. We see that within this region,
$R$ is surprisingly uniform with a value close to one. This behaviour is in good 
agreement with observations of the \ci-to-$^{13}$CO intensity ratio in real clouds
\citep[see e.g.][]{ikeda02} that often show almost uniform ratios. It is also clear that
$R$ starts to increase towards the edge of the cloud, but the restriction of the image
to regions with $^{13}$CO emission above our $0.1 \: {\rm K \: km \: s^{-1}}$ threshold
acts to obscure how large $R$ can grow as we move towards the edge of the cloud.
Therefore, in the right-hand panel of Figure~\ref{ratio_maps}, we show how $R$ 
behaves when we remove this restriction. The colour scale is chosen to emphasize
the change in $R$ at the edge of the cloud, although for practical reasons we limit
the maximum displayed value to $R = 50$. Comparing the two maps allows one to
see very clearly the envelope of gas around the CO-bright portion of the cloud that
is well-traced by \ci~but not by $^{13}$CO.

It is also interesting to explore how well we can reconstruct the column density
distribution of the cloud using these two tracers. In Figure~\ref{recon}a, we show a column
density map of the cloud that we have constructed using the empirical relationship
between $W_{\rm CI, 1-0}$ and $A_{\rm V}$ given in Equation~\ref{emp_fit},
which we have already seen gives a good description of the relationship between
\ci~emission and dust extinction for lines of sight with $A_{\rm V} > 1.5$. In
Figure~\ref{recon}b, we show a similar map, constructed using the empirical 
relationship
\begin{equation}
W_{\rm 13CO} = 1.0 A_{\rm V}  \: {\rm K \: km \: s^{-1}},  \label{emp_2}
\end{equation}
which gives a similarly good description of the behaviour of the $^{13}$CO
integrated intensity for lines of sight with $A_{\rm V} > 3$.
We have chosen to perform this comparison using these empirical fits, rather
than making use of the more common strategy of first inferring the C (or $^{13}$CO)
column density from the observed emission, and then inferring the total column
density from the column density of our tracer, because the latter technique inevitably
introduces additional sources of error, as we have already explored in some detail.
The comparison in Figure~\ref{recon} is therefore in some sense a ``best case'' scenario --
we are unlikely to actually be able to reconstruct the column densities as well in
practice. 

Comparing these two reconstructed maps with the true column density map
shown in Figure~\ref{cloud_maps}a, we see that, as expected, \ci~allows us to better
reconstruct the column density distribution towards the edges of the cloud,
although we still miss some of the lowest column density material. On the
other hand, $^{13}$CO seems to do better at tracing the highest density peaks,
some of which are hard to pick out from the background when using 
\ci.
%~(e.g.\ the dense filament running from $x \sim +3$~pc to $x \sim +7$~pc
%at $y \sim +2$~pc).

\begin{figure}
\includegraphics[width=0.49\textwidth]{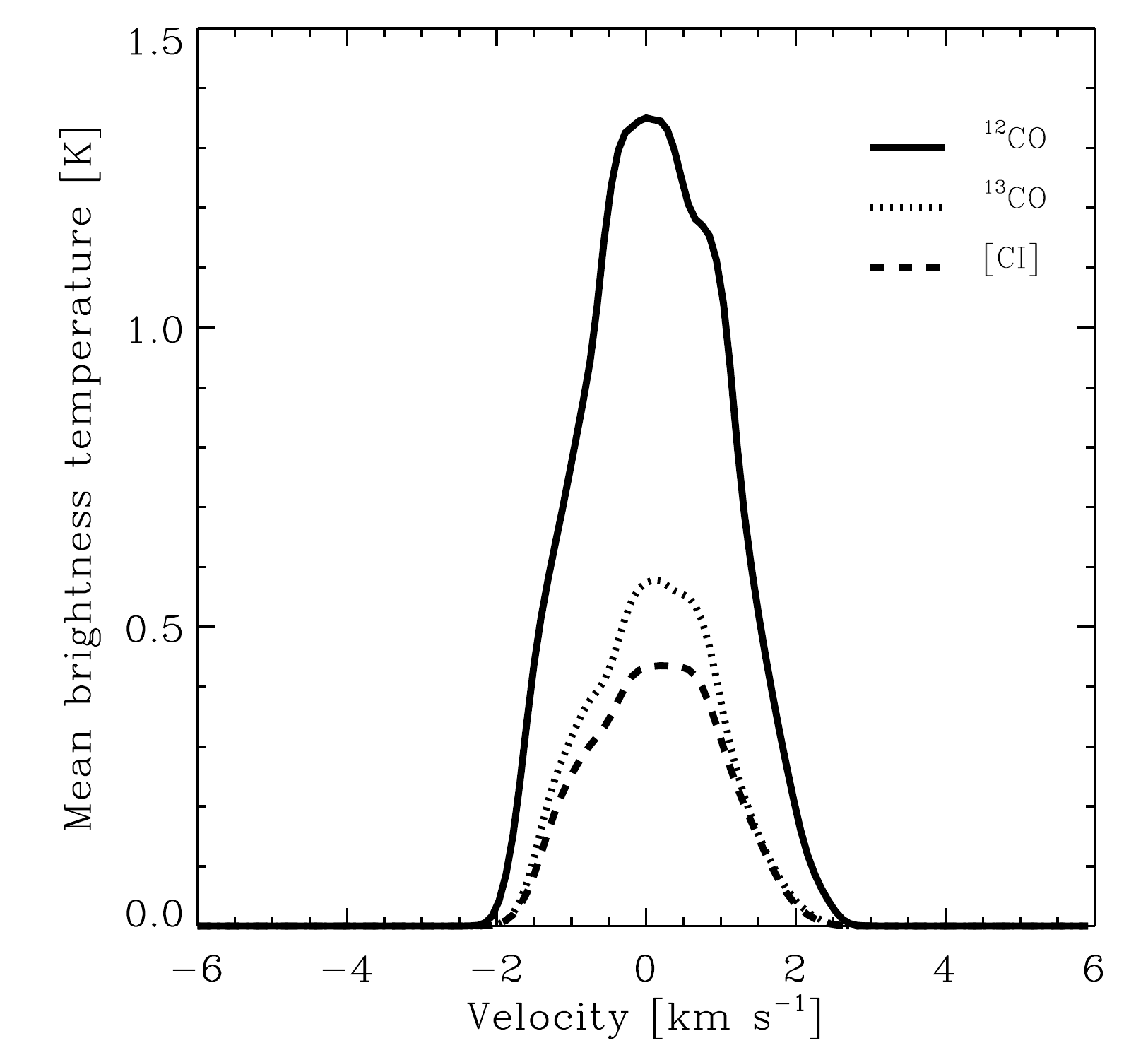}
\caption{Comparison of the mean brightness temperatures of
$^{12}$CO $J = 1 \rightarrow 0$ (solid line), $^{13}$CO $J=1 \rightarrow 0$ (dotted line) and 
the \ci~$1 \rightarrow 0$ transition (dashed line), averaged over the entire synthetic image.
\label{linewidth}}
\end{figure}

Finally, we have also examined how well \ci~and $^{13}$CO trace the cloud velocity.
We have computed cloud-averaged spectra for the \ci~$1 \rightarrow 0$ line and
for the $^{13}$CO $J=1 \rightarrow 0$ line, and have plotted these in 
Figure~\ref{linewidth}, along with a cloud-averaged spectrum for the
$^{12}$CO $J = 1 \rightarrow 0$ transition. We see that the linewidth and strength
of the \ci~and $^{13}$CO are remarkably similar. This is unsurprising, since in
both cases, the cloud-averaged spectra are dominated by the bright emission
coming from the high density regions of the cloud, and hence are largely tracing
the same material. The fact that the \ci~emission does a better job of tracing the
boundaries of the cloud does not greatly affect the average linewidth, as little of
the total \ci~emission is generated in these regions. Finally, the
much higher optical depth of the $^{12}$CO emission
means that a larger fraction of the cloud contributes to the cloud-averaged
spectrum, resulting in a much brighter line with a greater linewidth.

\section{Conclusions}
\label{conc}
In this paper, we have examined in detail the distribution of neutral atomic 
carbon within a model of a turbulent molecular cloud illuminated by the standard
local interstellar radiation field, and have also studied the \ci~fine 
structure emission produced by these carbon atoms. We show that the
density substructure created by the turbulence naturally leads to 
widespread, spatially extended \ci~emission, in good agreement with
observations of real molecular clouds. 

Most of the neutral carbon in our model is located in gas with a density
in the range $100 < n < 10^{4} \: {\rm cm^{-3}}$, and with an effective
(angle-averaged) visual extinction $A_{\rm V, eff} > 1$. This gas is relatively
cold -- around 80\% of the neutral carbon is found in regions with $T < 30$~K
-- and so \ci, like CO, is primarily a tracer of the cold, dense phase within 
molecular clouds, and not the warm, space-filling phase traced by \cii.

Our results suggest that \ci~$1 \rightarrow 0$ emission is a promising tracer 
of low column density gas in molecular clouds. Although it is easier to detect
the molecular gas using $^{12}$CO emission than using \ci~emission, the
relationship between $^{12}$CO integrated intensity and the column density
of the gas is highly non-linear, owing to the effects of CO photodissociation at
low $A_{\rm V}$ and line opacity at high $A_{\rm V}$. One can avoid the worst
of the opacity effects by using $^{13}$CO in place of $^{12}$CO, but the effects
of photodissociation are not so easily overcome. We find that  $^{13}$CO 
emission is a roughly linear tracer of column density for line-of-sight visual
extinctions in the range $3 < A_{\rm V} < 10$, although we caution that the
fact that we are neglecting CO freeze-out onto dust grains may affect our results
at high $A_{\rm V}$. In comparison, the integrated intensity of the \ci~$1 \rightarrow 0$
line is a linear tracer of column density for visual extinctions in the range
$1.5 < A_{\rm V} < 7$. Observing \ci~in place of $^{13}$CO therefore allows
one to better study the gas with visual extinctions $1.5 < A_{\rm V} < 3$, 
which in practice accounts for around 20\% of the total mass of the cloud.
However, at $A_{\rm V} > 3$, $^{13}$CO is as good a tracer of the cloud 
as \ci, while also being significantly easier to observe. 
Our results are consistent with previous work that has suggested that
\ci~emission is a good tracer of molecular gas \citep[see e.g.][]{gp00,ptv04},
although whether this is still true for clouds in other environments (e.g.\
lower metallicities, higher radiation fields) remains to be seen.

We have studied several different ways of estimating the excitation
temperature of \ci~based on the observed emission. We find that 
the best results are obtained if we use the \citet{dick78} method
for computing $T_{\rm ex}$ along sight-lines with integrated 
intensities $W_{\rm CI, 1-0} > 3 \: {\rm K} \: {\rm km} \: {\rm s^{-1}}$
and adopt a constant value of $T_{\rm ex}$ for the fainter sight-lines,
derived by averaging the estimated values for the bright sight-lines.
The resulting excitation temperatures are typically $T_{\rm ex}
\sim 11$--13~K, and are generally smaller than the kinetic 
temperature of the emitting gas, indicating that most of the
carbon atoms are not in local thermodynamic equilibrium. 

Using our estimate for $T_{\rm ex}$, we can determine the column density
of neutral atomic carbon, $N_{\rm C}$, with reasonable accuracy in the
regime where $N_{\rm C}$ is less than a few times $10^{16} \: {\rm cm^{-2}}$.
Comparison of the true and estimated
column densities shows that although the error along any particular line
of sight may be as high as a factor of two, there is no systematic offset between 
the estimated and real values. At higher carbon column densities, however, 
our estimate becomes inaccurate because it neglects the effects of line
opacity, and we start to significantly underestimate the true value of 
$N_{\rm C}$. We find that in practice, we typically miss about half of the
total atomic carbon when using this procedure.
We have also explored whether using the $^{12}$CO excitation temperature
as a proxy for the \ci~excitation temperature can allow us to better reconstruct
$N_{\rm C}$ in the optically-thick regime. We find that it does allow us to
partially correct for the effects of opacity, although we still miss as much as
a third of the total atomic carbon. We therefore recommend that if one
wants to estimate $N_{\rm C}$ based on observations of the \ci~fine
structure lines, then the following procedure should be adopted:
\begin{enumerate}
\item Determine the excitation temperature of the $1 \rightarrow 0$ line of
$^{12}$CO using Equation~\ref{tx_co}. 
\item Using this excitation temperature as a proxy for that of \ci, estimate 
the optical depth in the \ci~$1 \rightarrow 0$ transition using Equation~\ref{tau_est}.
\item Compute the column density of atomic carbon in the $J = 1$ level,
$N_{1}$, using Equations~\ref{fcorr} and \ref{n1_corr}.
\item Repeat steps 2 and 3 for the $J = 2$ level, using the integrated intensity
of the $2 \rightarrow 1$ transition in place of that of the $1 \rightarrow 0$ 
transition, and using $E_{21}$, $\nu_{21}$ and $A_{21}$ in place of 
$E_{10}$, $\nu_{10}$ and $A_{10}$.
\item Use the excitation temperature estimate together with Equation~\ref{tex_defn}
to compute the column density of carbon in the $J = 0$ level.
\item Sum the column densities of the three levels to obtain the final estimate
for $N_{\rm C}$.
\end{enumerate}

If, as is often the case, the atomic carbon column density is estimated
without correcting for opacity effects and with the assumption that the levels
have LTE populations, then our study suggests that the resulting values
could be in error by as much as a factor of two.

A major limitation of our present study is the fact that we have restricted
our attention to a single example of a turbulent cloud. In future work, we
plan to examine a wider sample of clouds, and to explore whether 
\ci~emission remains a good tracer of the H$_{2}$ column density as we 
reduce the metallicity of the cloud and/or increase the strength of the
ambient interstellar radiation field. It will also be interesting to explore 
how the ability of \ci~emission to trace H$_{2}$ changes over time as
the cloud evolves, as we have seen in previous studies that the CO
emission from clouds that are in the process of forming can change
quickly on a relatively short timescale.

\section*{Acknowledgments}
The authors would like to thank H.~Beuther, F.~Bigiel, T.~Bisbas, D.~Cormier, I.~De~Looze, S.~Madden, S.~Malhotra, R.~Plume 
and K.~Sandstrom for useful conversations on the topic of this work. They would also like to thank S.~Ragan for helpful comments on
an earlier draft of this manuscript. Special thanks also go to V.~Ossenkopf for stressing to 
us the potential importance of neutral atomic carbon as a cloud tracer, to J.~Carpenter for suggesting 
that we produce a figure along the lines of what is now Figure~\ref{cumul_emission}, and to the referee,
P.~Goldsmith, whose detailed and thoughtful referee report allowed us to greatly improve the paper.
SCOG, FM and PCC
acknowledge support from the DFG via SFB project 881 ``The Milky Way System'' (sub-projects B1, B2, B3, B5 and B8). 
FM and MM also acknowledge financial support by the International Max Planck Research School for 
Astronomy and Cosmic Physics at the University of Heidelberg (IMPRS-HD) and the Heidelberg Graduate 
School of Fundamental Physics (HGSFP). The HGSFP is funded by the Excellence Initiative of the German 
Research Foundation DFG GSC 129/1. PCC further acknowledges support from the DFG priority program
SPP 1573 ``Physics of the ISM" (grant CL 463/2-1). M.M.\  also acknowledges support from the Ministry of Education, 
Science and Technological Development of the Republic of Serbia through the project No.\ 176021, ``Visible and 
Invisible Matter in Nearby Galaxies: Theory and Observations" and support provided by the European Commission 
through FP7 project BELISSIMA (BELgrade Initiative for Space Science, Instrumentation and Modelling in 
Astrophysics, call FP7-REGPOT-2010-5, contract no.~256772). The simulations described in this paper were 
performed on the {\em kolob} cluster at the University of Heidelberg, which is funded in part by the DFG via 
Emmy-Noether grant BA 3706, and via a Frontier grant of Heidelberg University, sponsored by the German 
Excellence Initiative as well as the Baden-W\"urttemberg Foundation.

\appendix
\section{Resolution study}
Any numerical study of the physics of molecular clouds inevitably has to grapple with the issue
of limited numerical resolution. Computational limitations prevent us from running our models
at arbitrarily high resolution, and so it is important to establish whether this limitation significantly
affects the conclusions that we draw from our model clouds. In the present study, there are three
potential sources of resolution-dependence: the numerical resolution of the hydrodynamical model
itself, the size of the Cartesian grid used during our radiative transfer post-processing step, and the
discretization of column densities in the column density maps generated by our {\sc TreeCol} 
algorithm. We examine each of these in more detail below.

\subsection{Hydrodynamical resolution}
\label{hydro_res}
The simulation that we have studied in detail here was performed using two million SPH particles
to simulate $10^{4} \: {\rm M_{\odot}}$ of gas. This gives a particle mass of $5 \times 10^{-3}
\: {\rm M_{\odot}}$, which corresponds to a mass resolution of $0.5 \: {\rm M_{\odot}}$, if we
use the standard convention that any ``resolved'' structure must consist of at least 100 SPH 
particles. We have previously shown that this resolution is sufficient to model the star formation
rate within our turbulent cloud \citep{gc12c}, but that the low mass end of the stellar initial mass
function is not properly resolved. This suggests that the resolution is sufficient to model the
formation of pre-stellar cores, but not sufficient to properly follow the fragmentation that occurs
within them. As most of the \ci~emission in our model comes from gas that is at lower densities
than the density of a typical pre-stellar core, it is reasonable to assume that two million
particles will give us sufficient resolution to accurately model this emission. 

To verify this assumption, we have run a higher resolution simulation of the
same cloud using 20~million SPH particles. The higher computational cost of this 
simulation makes it impractical to run it for as long as our lower resolution simulation, 
and so we have compared the results from the two simulations at a time 
$t = 0.72$~Myr, corresponding to roughly 30\% of a free-fall time. Turbulence has 
already created significant density substructure by this point in the evolution of the
cloud and substantial amounts of neutral atomic carbon are present. 

\begin{figure}
\includegraphics[width=3.3in]{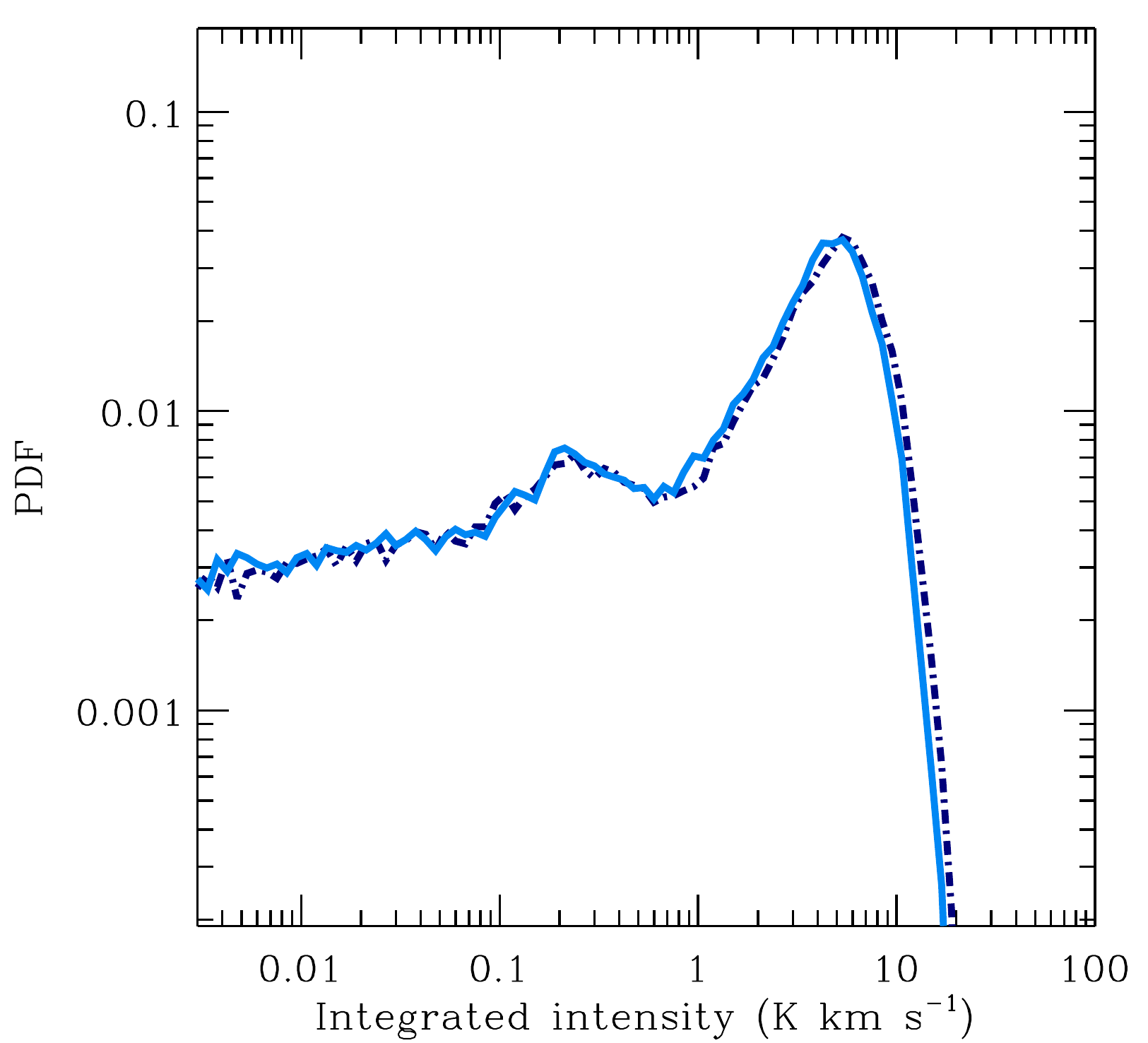}
\caption{PDF of the velocity-integrated emission in the \ci~$1 \rightarrow 0$ line,
computed at a time $t = 0.72$~Myr after the beginning of the simulation. The 
solid line is for a run with two million SPH particles, while the dot-dashed line
is for a higher resolution run with 20 million SPH particles. There is slightly more
emission from regions with very high integrated intensities in the high resolution
run, but otherwise the two PDFs are very similar. \label{WCI_hydro}}
\end{figure}

In Figure~\ref{WCI_hydro}, we compare the probability density function of 
$W_{\rm CI, 1-0}$ that we obtain in the two different simulations. The only
real difference that is apparent is that the high resolution simulation produces 
slightly more emission at high values of $W_{\rm CI, 1-0}$ than the low resolution 
simulation. This may be a result of the fact that in the high resolution simulation,
we are better able to resolve the wings of the density PDF, and hence are 
capturing slightly more of the emission from dense, \ci~bright gas. However,
the effect is small, and the mean \ci~integrated intensity changes by no more
than around 5\%. It is therefore safe to conclude that the results presented in 
this study are not sensitive to the limited hydrodynamical resolution of our simulation.

\subsection{RADMC-3D post-processing}
\label{radmc-resolution}
\begin{figure}
\includegraphics[width=3.3in]{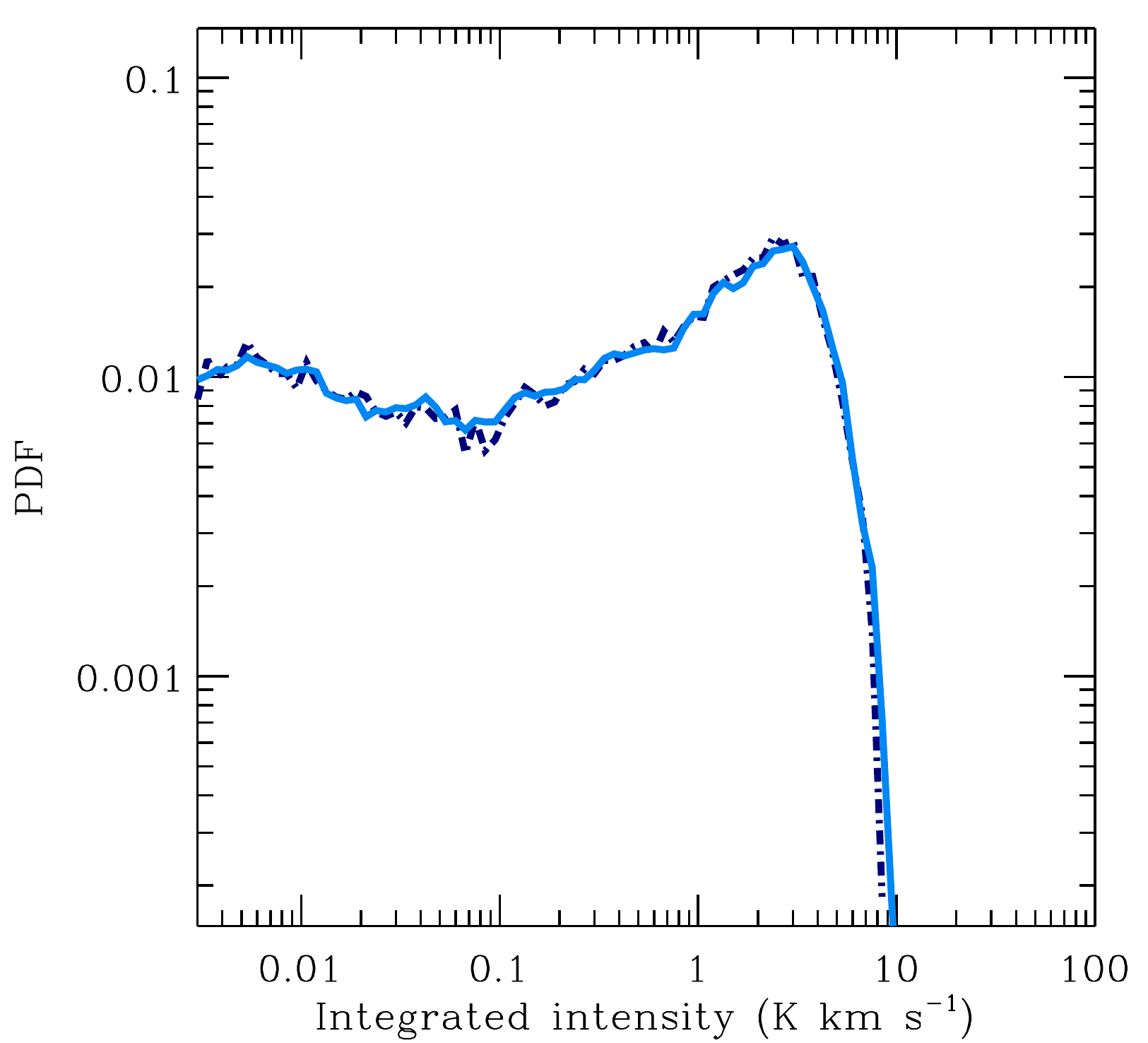}
\caption{PDF of the velocity-integrated emission in the \ci~$1 \rightarrow 0$ line. The two
sets of results shown here were produced using different grid resolutions during the
{\sc radmc-3d} post-processing step. The solid line corresponds to a grid resolution
of $256^{3}$ zones, while the dot-dashed line corresponds to a grid resolution of
only $128^{3}$ zones. The PDF produced using the lower resolution grid is somewhat
noisier, particularly in the tail of the distribution at low intensities, but other than this, 
there is not a significant difference between the two distributions. \label{radmc-test}}
\end{figure}

A second possible source of resolution dependence comes from our post-processing procedure.
As described in Section~\ref{post}, {\sc radmc-3d} cannot presently deal directly with SPH data, and 
hence we must interpolate our data onto a Cartesian grid before running the code. The results
presented in Section~\ref{res} are all based on maps produced using a cubical grid with a side length
of $5 \times 10^{19} \: {\rm cm}$ and a grid resolution of $256^{3}$ zones. We have examined
the sensitivity of our results to the choice of grid resolution by performing a similar post-processing
step using a resolution of only $128^{3}$ grid zones. In Figure~\ref{radmc-test} we compare the probability 
density function of $W_{\rm CI, 1-0}$ that we obtain with the two different grid resolutions. We see that they are
almost identical, suggesting that our results are insensitive to our choice of grid resolution. In fact, 
it is easy to understand why this is the case. The size of a zone in the $256^{3}$ case is around 
0.06~pc. We can estimate the density at which our SPH smoothing length becomes smaller than a
single grid zone by writing it as $\rho = N_{\rm neigh} M_{\rm part} / V$, where $N_{\rm neigh}$ is
the number of neighbouring SPH particles smoothed over, $M_{\rm part}$ is the particle mass and
$V = 4\pi h^{3} / 3$, where $h$ is the SPH smoothing length. In our calculations, $N_{\rm neigh}
= 50$, $M_{\rm part} = 5 \times 10^{-3} \: {\rm M_{\odot}}$, and hence $\rho \simeq 4 \times 10^{-24}
(h / 1 \: {\rm pc})^{-3} \: {\rm g \: cm^{-3}}$. If we now set $h = 0.06 \: {\rm pc}$, we find that $\rho 
\simeq 2 \times 10^{-20} \: {\rm g \: cm^{-3}}$, corresponding to a number density $n \simeq 
8000 \: {\rm cm^{-3}}$. This value is significantly higher than the \ci~critical density, and 
Figure~\ref{cumul_dense} demonstrates that not only is most of the gas in our simulation 
found at densities below this, but so is most of the neutral atomic carbon. Therefore, the dense 
regions that are not well-resolved by our $256^{3}$ grid contribute only a small fraction of the 
total \ci~emission, and our failure to resolve these regions has little impact on our 
results.

\begin{figure}
\includegraphics[width=3.3in]{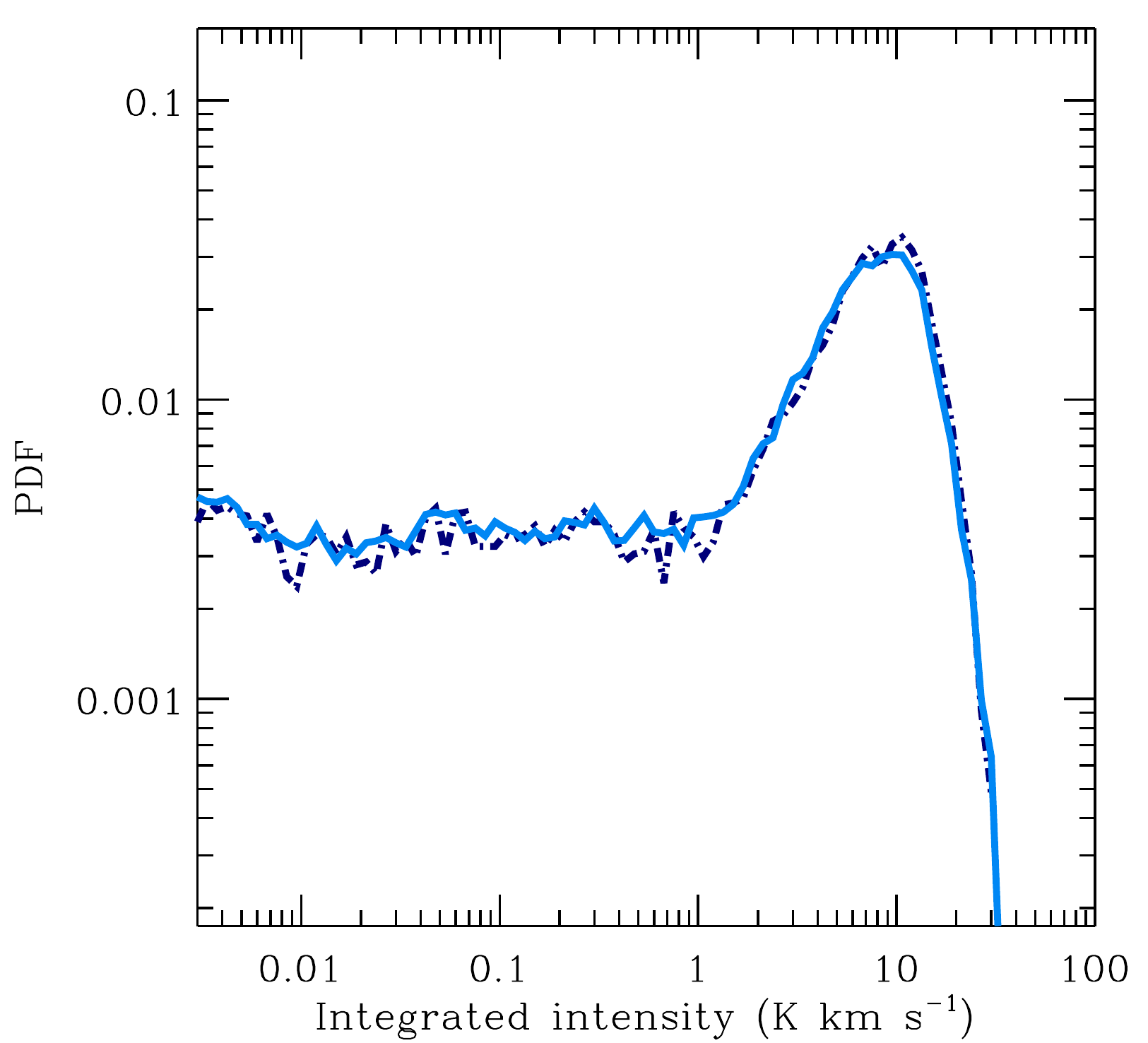}
\caption{As Figure~\ref{radmc-test}, but for the $^{12}$CO $J = 1 \rightarrow 0$ line.
Once again, we see that our grid resolution is sufficient to capture the bulk of the
emission. \label{radmc-test2}}
\end{figure}

We have also performed a similar analysis for our CO emission maps (see Figure~\ref{radmc-test2}). 
Once again, we find that our failure to resolve the highest density structures with our Cartesian grid
has little impact on our results. 

\subsection{TREECOL}
\label{treecol_res}

\begin{figure}
\includegraphics[width=3.3in]{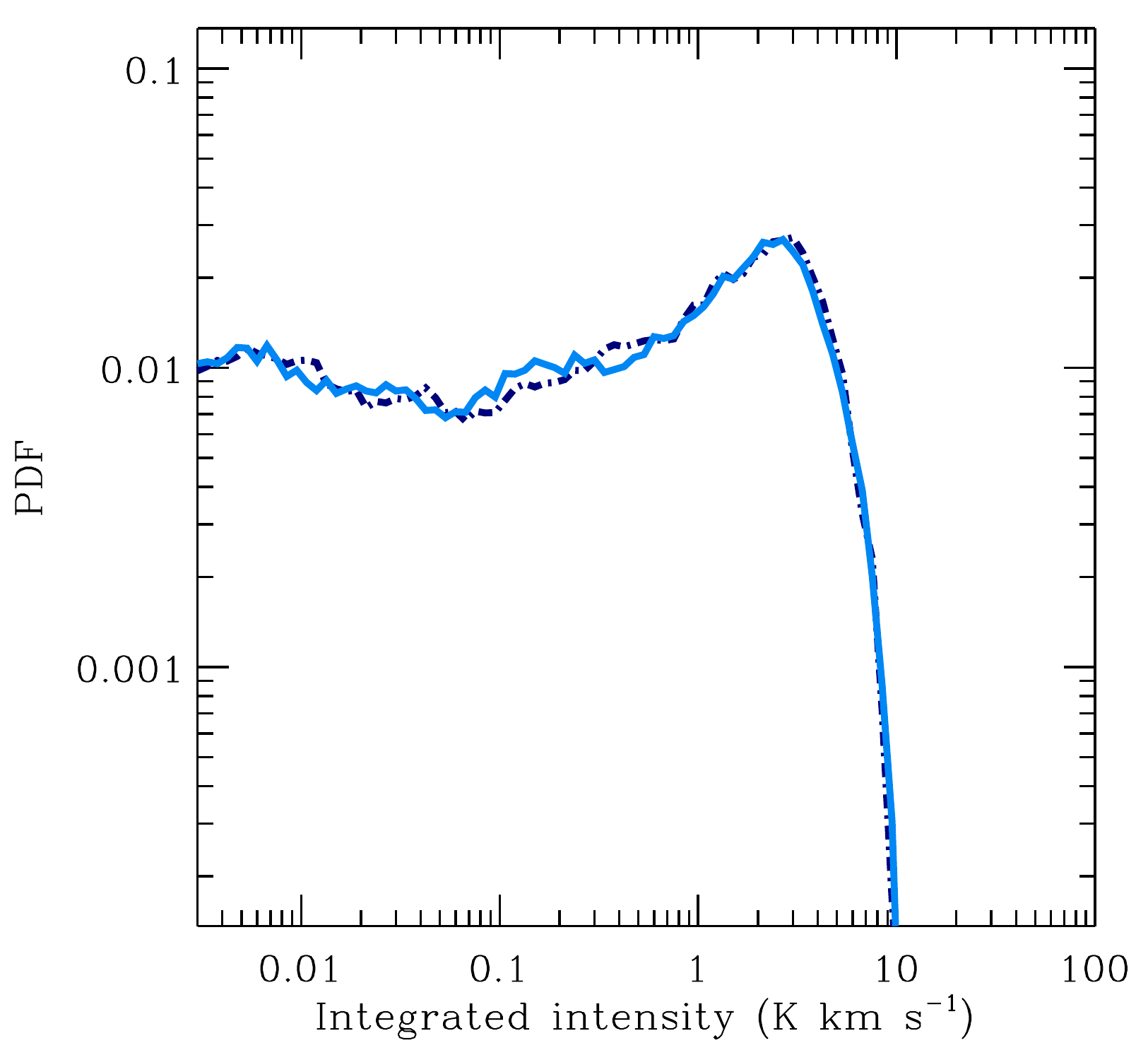}
\caption{PDF of the velocity-integrated emission in the \ci~$1 \rightarrow 0$ line. The solid blue 
line shows results from a simulation performed using $N_{\rm pix} = 48$ {\sc Healpix} pixels in 
the {\sc TreeCol} algorithm, while the dark blue dash-dotted line shows results from a simulation
performed using $N_{\rm pix} = 192$ {\sc Healpix} pixels. We see that the results are essentially
indistinguishable.
\label{treecol-test}}
\end{figure}

The final area in which the effects of limited numerical resolution may make themselves felt
is in our treatment of the attenuation of the interstellar radiation field within the cloud. We model
this using the {\sc TreeCol} algorithm \citep{cgk12}. This algorithm provides us with an approximate 
4$\pi$ steradian map of the column densities of hydrogen nuclei, H$_{2}$ and CO seen by each 
SPH particle, discretized into a number of equal-area pixels using the {\sc Healpix} pixelation
scheme \citep{healpix}. The {\sc Healpix} scheme involves splitting up the sphere into 12 equal-area
base pixels, each of which can then be further sub-divided. The amount of sub-division can be
quantified by $N_{\rm side}$, the number of pixels along each side of one of the base pixels;
the total number of pixels then follows as $N_{\rm pix} = 12 N_{\rm side}^{2}$. In {\sc TreeCol},
by default we sub-divide each of the {\sc Healpix} base pixels into four sub-pixels, and hence
$N_{\rm side} = 2$ and $N_{\rm pix} = 48$. We have found in testing that this choice appears
to give us the best balance between computational efficiency (which argues for a small value
for $N_{\rm pix}$) and accuracy (which argues for a larger value). However, we have also 
investigated the extent to which our results depends upon this choice. In Figure~\ref{treecol-test}, 
we compare the PDF of integrated intensity in the $1 \rightarrow 0$ line of \ci~that we obtain from
our default run with the corresponding distribution produced by a run with $N_{\rm side} = 4$ 
and $N_{\rm pix} = 192$. We see that the results are essentially indistinguishable, justifying
our decision to use a small value for $N_{\rm pix}$.

\section{Relative optical depths of the \ci~lines}
\label{op_depths}
In this Appendix, we examine whether there are any physical conditions in which the $2 \rightarrow 1$
 transition of \ci~is optically thick while the $1 \rightarrow 0$ transition remains optically thin.

 We can write the optical depth at line centre in a transition with upper state $u$ and lower state $l$ due 
 to an absorber with physical size $L$ as
 \begin{equation}
 \tau_{ul} = \int_{0}^{L} \alpha_{ul} \, {\rm d}s,
 \end{equation}
 where $\alpha_{ul}$ is the absorption coefficient for the transition and we integrate along the path of
 the light-ray through the absorber. The value of the absorption coefficient at the centre of the line can
 be written as
 \begin{equation}
  \alpha_{ul}(\nu_{ul}) = \frac{h \nu_{ul}}{4 \pi} B_{ul} \left[\frac{g_{u}}{g_{l}} n_{l} - n_{u} \right] \phi(\nu_{ul}),
 \end{equation}
 where $B_{ul}$ is the Einstein coefficient for stimulated emission for the transition, $n_{l}$ and $n_{u}$
 are the number densities of absorbers in the lower and upper states, $g_{l}$ and $g_{u}$ are the statistical
 weights of these two states, $\nu_{ul}$ is the frequency of the transition at the centre of the line, and
 $\phi(\nu)$ is the line profile function. In general, for weak transitions such as the fine structure lines of
 \ci, Doppler broadening provides the dominant contribution to $\phi(\nu)$, allowing us to write its
 functional form as
\begin{equation}
\phi(\nu) = \frac{1}{\sqrt{\pi} \Delta \nu_{\rm D}} e^{-(\Delta \nu / \Delta \nu_{\rm D})^{2}}, \label{dopp}
\end{equation}
where $\Delta \nu = \nu - \nu_{ul}$, $\Delta \nu_{\rm D} = b (\nu_{ul} / c)$ is the Doppler width of the line 
and $b = (2kT / m_{\rm C})^{1/2}$ is the Doppler broadening parameter. Its value at line centre is then 
simply
\begin{equation}
\phi(\nu_{ul}) = \frac{c}{\sqrt{\pi} b \nu_{ul}}.
\end{equation}

If we assume, for simplicity, that the gas is isothermal, then $b$ is independent of our position along the
light-ray and can be taken out of the integral, allowing us to write the optical depth of the transition as
\begin{eqnarray}
\tau_{ul} & = & \frac{h \nu_{ul}}{4 \pi} B_{ul}  \frac{c}{\sqrt{\pi} b \nu_{ul}} \int_{0}^{L}  \left[\frac{g_{u}}{g_{l}} n_{l} - n_{u} \right] {\rm d}s, \\
& = & \frac{hc B_{ul}}{4 \pi^{3/2} b}  \left[\frac{g_{u}}{g_{l}} N_{l} - N_{u} \right],
\end{eqnarray}
 where $N_{l}$ and $N_{u}$ are the column densities of absorbers in the lower and upper states, respectively.
Using this expression, we can then write the ratio of the optical depth in the  $2 \rightarrow 1$ transition,
 $\tau_{21}$, to the optical depth in the  $1 \rightarrow 0$ transition, $\tau_{10}$, as
 \begin{equation}
 \frac{\tau_{21}}{\tau_{10}} = \frac{B_{21}}{B_{10}} \frac{g_{0}}{g_{1}} \left[\frac{g_{2} N_{1} - g_{1} N_{2}}{g_{1} N_{0} - g_{0} N_{1}} \right].
\end{equation} 
The statistical weights for the $J = 0, 1$ and 2 fine structure levels of atomic carbon are
$g_{0} = 1$, $g_{1} = 3$ and $g_{2} = 5$, and so we can also write this as
 \begin{equation}
 \frac{\tau_{21}}{\tau_{10}} = \frac{1}{3} \frac{B_{21}}{B_{10}}  \left[\frac{5 N_{1} - 3 N_{2}}{3 N_{0} - N_{1}} \right].
\end{equation} 
Finally, since 
\begin{equation}
B_{ul} = \frac{c^{2}}{2h \nu_{ul}^{3}} A_{ul},
\end{equation}
we have
\begin{equation}
\frac{\tau_{21}}{\tau_{10}} = \frac{1}{3} \left(\frac{\nu_{10}}{\nu_{21}} \right)^{3} \frac{A_{21}}{A_{10}}   
\left[\frac{5 N_{1} - 3 N_{2}}{3 N_{0} - N_{1}} \right].
\end{equation} 
Evaluating this using data on the frequencies and spontaneous transition rates for the two transitions
taken from the LAMDA database \citep{sch05}, we find that
\begin{equation}
\frac{\tau_{21}}{\tau_{10}} \simeq 0.25 \left[\frac{5 N_{1} - 3 N_{2}}{3 N_{0} - N_{1}} \right].
\end{equation}

If we assume that all three fine structure levels of the carbon atom have their LTE level populations, 
then using this expression, it is easy to show that $\tau_{21}/\tau_{10} < 1$ for all temperatures 
$T < 40$~K, and that in the limit of large temperatures, $\tau_{21}/\tau_{10} \sim 2$. If the level
populations are instead significantly sub-thermal, then the optical depth ratio is much smaller.
We therefore see that it is indeed possible for the $2 \rightarrow 1$ transition of \ci~to be optically thick 
while the $1 \rightarrow 0$ transition remains optically thin, but that it requires rather unusual conditions:
the gas must be hot, sufficiently dense that the level populations are close to LTE, and must have a total
column density of carbon such that $0.5 < \tau_{10} < 1.0$. The last condition is particularly restrictive,
as it implies that for any given combination of temperature and density, the range of carbon column densities
that gives rise to a situation with $\tau_{10} < 1$ and $\tau_{21} > 1$ spans only a factor of two.  For
comparison, in our simulations, we find values of $N_{\rm C}$ spanning almost three orders of magnitude.
Therefore, in most cases, either both lines will be optically thin, both will be optically thick, or the 
$1 \rightarrow 0$ transition will be optically thick while the $2 \rightarrow 1$ transition remains optically thin,
with the last possibility being particularly likely when the excitation of the fine structure levels is sub-thermal.

In the particular case of the cloud modelled in this paper, we can be confident that we are never in a situation
with $\tau_{10} < 1$ and $\tau_{21} > 1$. As Figure~\ref{rhotemp} demonstrates, almost all of the atomic carbon 
is found in gas with $T < 40$~K, and the little that is not is found at densities $n \ll n_{\rm crit}$, and hence will be
sub-thermally excited.

\end{document}